\documentclass{thesis}
\usepackage{color}
\usepackage{graphicx}
\linespacing{1.33}

\newif\ifshowprogram
\showprogramtrue

\degree{Doctor of Philosophy}
\department{Arts and Sciences}
\gradyear{2000}
\author{Michael K. Rivera}
\title{The Inverse Energy Cascade of Two-Dimensional Turbulence}

\begin{document}
\begin{preliminary}
\maketitle
\begin{abstract}
This thesis presents an experimental study of the inverse energy cascade as it occurs in an electromagnetically forced soap film. It focuses on characterizing important features of the inverse cascade such as it's range, how energy is distributed over the range and how energy flows through the range.  The thesis also probes the assumption of scale invariance that is associated with the existence of an inverse cascade.  These investigations demonstrate that the extent of the inverse cascade range and the behavior of the energy distribution are in agreement with dimensional predictions.  The energy flow in the inverse cascade range is shown to be well described by exact mathematical predictions obtained from the Navier-Stokes equation.  At no time does the energy flow in the inverse cascade range produced by the e-m cell behave inertially or in a scale invariant manner.  Evidence that the cascade could become scale invariant should an inertial range develop is presented, as are the requirements that a system must satisfy to create such an inertial range.
\end{abstract}

\tableofcontents
\listoftables
\listoffigures
\end{preliminary}

\chapter{Introduction \label{intro}}

One of the curious predictions in turbulence theory is that there might possibly exist a range of length scales in a two-dimensional (2D) turbulent fluid over which kinetic energy is transferred from small to large length scales.  That this range could exist was first predicted in the late 1960's by Kraichnan\cite{Kraichnan:PFL67}.  Numerical simulations that followed yielded varying degrees of agreement with this prediction\cite{Siggia:PFL81,Frisch:PFL84,Maltrud:JFM91}.  Experimental verification of the existence of such a range did not come about until much later due to the difficulty inherent in building and maintaining a system which approximates a 2D fluid\cite{Paret:PFL98,Rutgers:PRL98,Rivera:PRL00}.  This thesis presents an investigation of the 2D inverse energy cascade in a new apparatus, the electromagnetically forced soap film.  What follows in this chapter is a description of the phenomenology surrounding the inverse energy cascade and a discussion of the experiments that have attempted to probe it's properties.

\section{The Inverse Energy Cascade}

\begin{figure}
\hskip 1in
\includegraphics*[width=4in,height=1.75in]{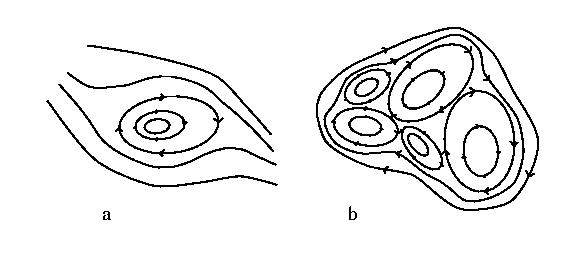}
\caption{Two pictures of the ``eddy" concept for a 2D fluid:  (a) a single large eddy and (b) a large eddy made from many interacting smaller eddies.}
\label{fig: eddies}
\end{figure}

The phenomenology of turbulence, in three-dimensions (3D) or 2D, is usually phrased in terms of ``eddies".  An eddy itself is not a well defined object, though there have been many recent attempts using wavelets to better define the concept\cite{Farge:PFL99}.  Loosely speaking it is a region in a fluid that is behaving coherently.  The extent of an eddy is dictated by boundaries within which an arbitrary determination is made that some sort of structure exists.  Thus an eddy can be a single large region of rotation, such as the whirlpool which forms above a bathroom drain.  Or an eddy can be a large region containing many smaller eddies which are interacting with one another while behaving distinctly (again by an arbitrary determination) from other neighboring clusters of eddies.  These two ideas are drawn in Fig. \ref{fig: eddies} for the case of a 2D fluid. Eddies in 3D are much more difficult to picture. Two important properties that are associated with an eddy are size and energy. These two properties allow predictions about energy motion in fluids to be made if some knowledge of how eddies interact in the system is known.

In 2D fluids, one way in which eddies (which assume the more familiar label ``vortices" in 2D) interact with each other is through a process known as ``vortex cannibalization".  A cannibalization event is when two neighboring eddies of like rotational sense merge to form a single larger eddy.  When cannibalization occurs energy flows out of the length scales of the initial eddies and into the length scale of the final eddy.  Since the final eddy is larger than the initial ones, cannibalization results in the flow of energy from small to large length scales.

In a 2D turbulent fluid, many eddies are generally created at a small length scale called the energy injection scale, $r_{inj}$.  The expectation is that through interaction by cannibalization these small eddies cluster and merge into larger eddies.  These larger eddies are also expected to cluster and merge to form even larger eddies and so on.  This means that energy, initially injected into the turbulence at the length scale $r_{inj}$ should gradually be moved by consecutive cannibalization events to larger length scales.  This type of energy motion constitutes an inverse energy cascade\cite{Kraichnan:PFL67}.

Using the eddy concept has the advantage of highlighting two important features associated with the existence of an inverse energy cascade: scale invariance and locality of interaction.  The first of these can be understood by looking again at Fig. \ref{fig: eddies}(b) which shows many smaller like signed eddies clustering to form a single larger eddy.  Presumably, the small eddies in the figure are themselves formed by the clustering of even smaller eddies, which in turn are formed by even smaller eddies.  Likewise, the large eddy cluster in the figure is most likely interacting with other eddy clusters in the system.  As long as the eddies at the very smallest scale, the injection scale, are being continuously created to replenish those which are cannibalized, the inverse cascade range is scale-invariant.  That is to say that no length scale in the inverse cascade range can be distinguished from any other length scale that is also in the range.  Scale invariance is exceedingly important from a theoretical stand point.  The assumption of scale invariance of fields, such as the probability of velocity difference on a length scale $r$, allow important predictions about turbulence to be made (see chapter \ref{himom})\cite{Frisch}.

Before discussing locality, a delicate point must be made.  If the eddies at the injection length scale are not being continuously replenished then the number of eddies at the smallest scales gradually begins to decrease as more and more eddies are lost to cannibalization events.  To maintain an inverse cascade range, then, the turbulence has to be continuously forced.  That is, eddies must be continuously created at the energy injection scale.  If the turbulence is not forced then the cascade range will eventually consume itself from small scales up, ultimately leading to a state which can be described as a diffuse gas of large individual eddies (eddies not made of clusters of smaller eddies)\cite{Chasnov:PFL97,Rivera:PRL98}.  The term ``coarsening" is used to describe decaying 2D turbulence's behavior in order to distinguish it from the inverse energy cascade.

The second property assumed to hold in the inverse cascade is locality of interaction.  This property refers to constraints on the manner in which eddies interact.  If an eddy of very small size is close to, or embedded in, an eddy of exceedingly large size, the small eddy will merely be swept along by the large eddy and not strongly deformed.  Likewise the large eddy will not be significantly effected by it's small companion.  Since neither of the eddies is strongly deformed, the cannibalization process is expected to happen over a long period of time, if at all\cite{Frisch}.  On the other hand, two neighboring  eddies of similar size interact and deform one another strongly, and thus the cannibalization happens swiftly.  Energy transfer by cannibalization is therefore most efficient when occurring between similarly sized length scales; this is what is meant by locality.  Due to locality, the kinetic energy at small scales in the inverse cascade is expected to be moved to large scales in a continuous manner, stepping through the intervening length scales by local interactions rather than making large length scale jumps by the merger of a small and large eddy.  Hence the term cascade.

The picture of 2D turbulence and it's inverse cascade is now almost complete.  Energy is continually injected into a fluid in the form of small eddies.  These small eddies cluster to form large eddies moving energy to larger scales. In turn the eddy clusters themselves cluster to form larger clusters of eddy clusters, etcetera.  It is the etcetera that is of concern.  At what point does the vortex merger process and growth of larger and larger eddies stop?  That is, how is the energy injected into a 2D turbulent fluid thermostated?

Consider first the thermostating mechanism in 3D turbulence.  In 3D turbulence there exists a direct energy cascade, where instead of energy being moved from small to large scales by eddy merger the opposite happens; energy is moved from large to small scales by eddy stretching (commonly called vortex stretching).  Eventually, through continuous vortex stretching, a smallest eddy scale is reached, at which point the kinetic energy contained in these small eddies is dissipated into heat by the fluids internal viscosity.  All of the energy that is injected into the large length scales of a 3D turbulent fluid is eventually exhausted by viscosity at small length scales\cite{Frisch}.

Internal viscosity is a short range force, only becoming a good thermostat when the kinetic energy reaches small length scales\cite{Frisch}.  Thermostating is not an issue in 3D where the direct cascade takes energy down to such small scales.  In 2D, however, the inverse cascade moves energy away from small scales.  Therefore viscosity has no chance to exhaust the injected energy.  An ideal 2D turbulent fluid driven to a state of turbulence with a continuous forcing would never be in a steady state since the total energy in the flow would continue to build up as larger and larger eddies form\cite{Yakhot:PRE99}.  What is needed to maintain 2D forced turbulence in a steady state and stop the inverse cascade process is some sort of external dissipation mechanism which is an effective thermostat at large length scales.  In other words, some sort of dissipation mechanism that is not internal to the fluid itself must exist to take energy out of large length scales and dictate the largest size eddies that can be formed by the cascade.

Fortunately, 2D experiments are almost always coupled to the surrounding 3D environment by frictional forces\cite{Paret:PFL97,Rutgers:PFL96,Rivera:PRL00}.  In these experiments this external frictional force provide the turbulence with an effective large scale thermostat and sets the largest length scales which can be reached by the inverse cascade process.  The inclusion of an external thermostat completes this phenomenological description of the inverse energy cascade.

\section{History of Experiments}

Laboratory experiments which have attempted to probe the inverse energy cascade of 2D turbulence fall into two major categories: soap films and stratified shallow layers of fluid.  The soap film experiments in 2D fluid mechanics were initiated by Couder in the early 1980's\cite{Couder:JFM86}.  This early work investigated coarse features of both 2D turbulence and 2D hydrodynamics.  Further attempts at using the soap film to measure 2D turbulence in the search for an inverse cascade were done by Gharib and Derango a few years later\cite{Gharib:PAD89}.  The experimental system that was used by Gharib and Derango, called a soap film tunnel, was later perfected by Kellay {\em et al.}\cite{Kellay:PRL95} and Rutgers {\em et al.}\cite{Rutgers:PFL96}.

\begin{figure}
\hskip 1in
\includegraphics*[width=4in,height=2.85in]{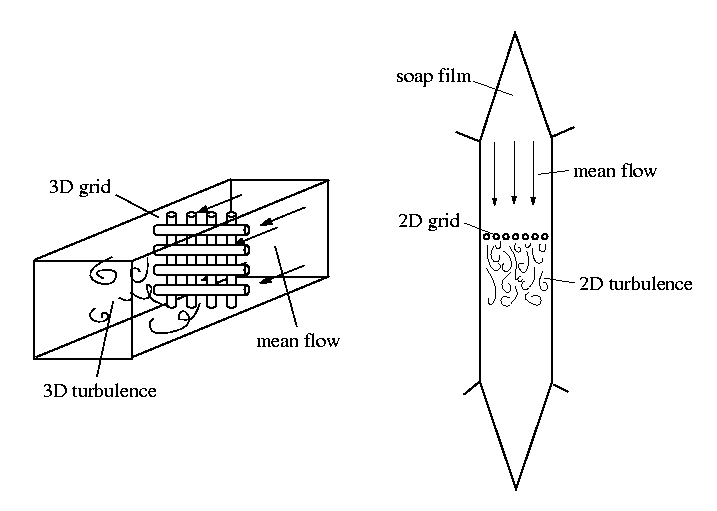}
\caption{A 3D wind tunnel creating turbulence and it's 2D equivalent: the soap film tunnel}
\label{fig: windtunnel}
\end{figure}

The soap film tunnel is a 2D equivalent of the wind tunnel, which is the mainstay of 3D turbulence research.  The manner in which turbulence is created in each is identical.  A grid, or some other obstacle, is placed in the path of a swiftly moving mean flow.  If the flow speed is fast enough, the fluid becomes turbulent downstream from the grid.  Such 2D and 3D tunnels are shown in Fig. \ref{fig: windtunnel}.  Soap film tunnels would seem to be ideal 2D fluids for performing turbulence research in because their aspect ratios are exceedingly large (many cm across to a few micrometers thick) and thus the fluid flow is almost entirely two-dimensional.  There are, however, difficulties inherent in the use of soap films.  For example thin films couple strongly to the air and the magnitude of their internal viscosity is large compared to that of water.  For the most part these difficulties are thought to be mitigated by clever experimental techniques, such as the use of vacuum chambers\cite{Rutgers:PFL96} or by using thick ($\approx 10 \mu$m) films\cite{Rivera:PRL98}.

Every attempt to study the inverse energy cascade in a 2D soap film tunnel with the configuration shown in Fig. \ref{fig: windtunnel} has met with failure.  This is not a disparaging comment about the researchers involved in the effort.  Indeed their considerable skill eventually tamed the delicate and whimsical soap films into a useful experimental system.  The lack of inverse cascade in these systems reflects the fact that the configuration shown in Fig. \ref{fig: windtunnel} creates decaying 2D turbulence.  The eddies that are injected at the grid are not replenished as the fluid moves downstream.  By the discussion in the last section this means that the system does not have the ability to form the eddy clusters that is expected of an inverse cascade.

\begin{figure}
\hskip 1in
\includegraphics*[width=4in,height=2.74in]{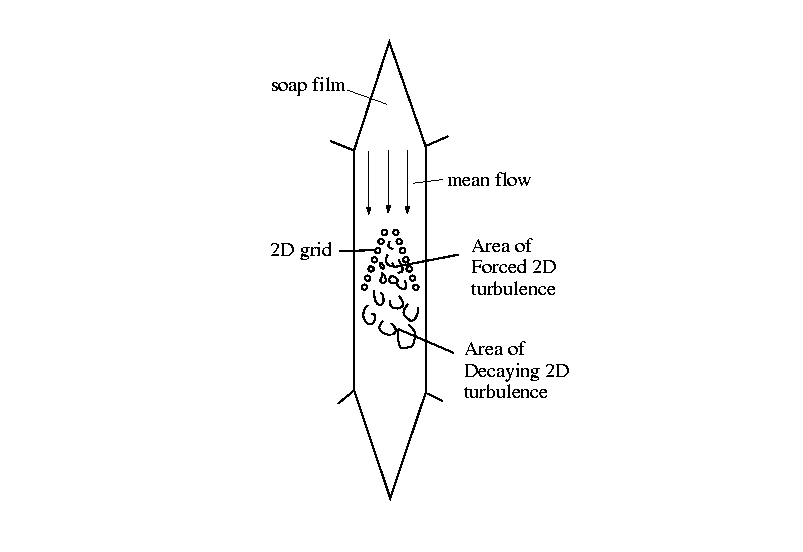}
\caption{A 2D soap film tunnel creating forced 2D turbulence.}
\label{fig: tunnel2}
\end{figure}

Once the understanding that 2D turbulence needs to be forced for an inverse cascade to be present was reached, the film tunnel design was modified to create forced turbulence\cite{Rutgers:PRL98}.
The design of the film tunnel is identical to that shown before except for the orientation of the turbulence producing grid.  Instead of having a single grid oriented perpendicular to the flow direction, two grids were oriented at angles to the flow direction so that they formed two sides of a triangle with the tip of the triangle oriented upstream.  This modified form is shown in Fig. \ref{fig: tunnel2}.  An area of forced turbulence exists in the interior between the two grids since it is here that vortices created at the grids are able to diffuse into the interior and replenish those lost to cannibalization.

Though certain properties of the inverse cascade can be investigated with such a modified film tunnel, the setup is not ideal because the turbulence it creates is inhomogeneous.  One can imagine that the fluctuation in the driven turbulence area near the grids are quite large compared to those in the interior.  Homogeneity is a critical simplifying assumption in almost all areas of turbulence theory.  The inhomogeneity of the modified film tunnel, then, has devastating consequences with regard to comparing results with theory.  A more ideal setup would involve the injection of vortices directly into the body of the fluid by some sort of external force, rather than injecting from the boundaries.  This is  a difficult task to do in soap film tunnels.

A system which does achieve such an injection of vortices falls into the second class of 2D turbulence experiments: stratified shallow layers of fluid.  For the most part, the use of such layers in turbulence research has been pioneered by Tabeling and others in the early 1990's.  A successful observation of an inverse cascade regime was reached a few years later by the same group\cite{Paret:PRL97}.  The stratified shallow layer apparatus, as it's name suggest, suspends a layer of pure water, above a higher density layer of salt water.  The layers in question are only a few millimeters thick, and the area tends to be on the order of ten centimeters so that the aspect ratio, while not nearing that of soap films, is still large.  The salt water layer is subject to a current flowing in it's plane, and placed in a spatially varying magnetic field.  The resultant Lorentz force acts directly on the fluid layer driving it to a state of turbulence.  Stratification helps to impose a measure of two-dimensionality to the fluctuations, thus one has 2D forced turbulence.

Note that in this system the force is acting directly on the fluid, unlike the modified film tunnel where forcing happened near the grid.  This restores homogeneity to the system, allowing accurate comparison with theory.  The place that shallow layers suffer is in their approximation of a two-dimensional fluid.  The bottom of the fluid container in shallow layer systems enforces a no slip boundary on the lower surface of the fluid.  If the velocity in the fluid becomes large, a strong shearing can develop between the upper and lower layers of fluid.  This inevitably causes mixing which destroys the stratification and sacrifices two-dimensionality.  Thus, shallow layer systems are severely limited in the strength of the turbulent fluctuations which they can successfully explore.

This brief experimental history, then, shows that there have been two systems used to investigate the inverse cascade, neither of which are ideal.  However, each system complements the others difficulties almost perfectly.  Film tunnels are great 2D fluids, but not easily forced homogeneously.  Stratified layers are marginal 2D fluids, but easily forced homogeneously.  What is needed then, as an ideal test apparatus for the inverse cascade, is a combination of these two systems that takes advantage of their benefits without inheriting their difficulties.  What is needed is the electromagnetically forced soap film, simply called the e-m cell.

\section{Thesis Overview}

This thesis presents an experimental study of the inverse energy cascade as it occurs in an electromagnetically forced soap film.  In particular it focuses on characterizing important features of the inverse cascade such as it's range, how energy is distributed over the range and how energy flows through the range.  The thesis also probes the assumption of scale invariance that is associated with the existence of an inverse cascade.

Chapter \ref{experimental} describes the workings of the e-m cell and measurement apparatus.  The basic design and implementation of the e-m cell is reviewed in the first two sections.  How the frictional coupling of the soap film to the air is controlled is described in the following section.  The fourth section explains certain limitations on apparatus size that are imposed by the existence of gravity.  This size constraint is one of the chief difficulties that limit results throughout the rest of the thesis.  The fifth section in the chapter is an overview of the measurement system that was developed to extract velocity information from the e-m cell.  The chapter is concluded with a brief overview of the operation of the e-m cell.

Chapter \ref{ebudg} is a systematic check that the e-m cell does behave as a 2D fluid.  The first section motivates the need for this test and the second derives the relevant mathematical relationships necessary for such a test.  The third section compares the data from the cell to these derived relationships to verify that the e-m cell behaves as a 2D fluid.  This section also establishes a model for the external dissipation mechanism in the e-m cell.  A self consistency check is also performed to help strengthen this verification.

Chapter \ref{eflux} begins the systematic investigation of the properties of the inverse energy cascade.  The extent of the range and how energy is distributed over this range is measured and presented in the first section.  The manner of energy transfer is described in the following section.  In particular, this second section attempts to determine if the energy flow is ``inertial", a property which is necessary if one wants results which are universal to all 2D turbulence systems.

Finally, chapter \ref{himom} attempts to determine if the inverse cascade is scale invariant.  Predictions for structure functions in a scale invariant fluid are presented in the first section. The next two sections attempt to extend results from the e-m cell to test these predictions, as well as explain why one should be critical of such extensions.  While conclusions presented in earlier chapters are quite strong, the conclusions drawn in this chapter are weak.  This weakness stems from the size limitations presented in chapter \ref{experimental} and can not be overcome in the current apparatus.

\chapter{The E-M Cell \label{experimental}}

A fluid which carries a current density, $\vec{J}$, in the presence of a magnetic field, $\vec{B}$, is subject to a force per unit mass $\vec{F} = \vec{J} \times \vec{B}$ which drives fluid motion.  This principle has been used in earlier experiments to excite motion in shallow layers of electrolytic fluid\cite{Paret:PFL98,Sommeria:JFM86}, and the techniques used in these experiments may be readily adapted for use in soap films.  The result of such adaptation is the electromagnetically forced soap film, called briefly an e-m cell.  The e-m cell is a useful tool in the study of 2D hydrodynamics, and in particular 2D turbulence\cite{Rivera:PRL00}.  This chapter reviews the design and operation of the e-m cell, as well as the measurement technique used to obtain velocity data from it.

\section{The E-M Cell}

\begin{figure}
\hskip 1in
\includegraphics*[width=4in,height=1.81in]{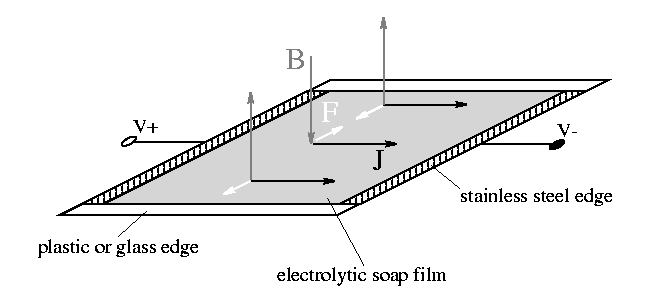}
\caption{Basic operation of the e-m cell.}
\label{fig: experimental1}
\end{figure}

The main component of the e-m cell is a free standing soap film drawn from an electrolytic soap solution across a square frame.  The frame has two opposing sides made from stainless steel, while the remaining sides are constructed from plastic or glass.  A voltage difference applied to the stainless steel sides results in a current which lies in the plane of the film. A spatially varying external magnetic field is then created and oriented so that it penetrates the film plane perpendicularly.  The resulting Lorentz force lies in the plane of the film and drives the fluid motion.  A diagram of this is shown in Fig. \ref{fig: experimental1}.

The electrolytic soap solution which the film is drawn from is made from $400$ ml distilled water, $80$ g ammonium chloride salt, $40$ ml glycerol and $5$ ml commercial liquid detergent (regular Dawn or Joy).  To this solution particles are added up to a volume fraction of about $10^{-3}$.  These particles, either $10$ $\mu$m hollow glass spheres or lycopodium mushroom spores which appear as $\sim 40$ $\mu$m particles, are of a density comparable to the soap solution so that they closely follow the surrounding flow.  These particles will be used in the measurement technique to be described later.  The salt used to make the solution electrolytic, ammonium chloride, was chosen after numerous trials which included sodium chloride and potassium chloride.  Ammonium chloride was chosen because high concentrations could be reached with little effect on the stability of the film. This was found not to be the case with potassium or sodium chloride.  High concentrations of salt are necessary to keep the films electrical resistance as small as possible.  This limits film evaporation due to Joule heating caused by the driving current, a necessity if a film is to be studied for long periods of time.

The soap film, once drawn, is maintained at a thickness of around $50$ $\mu$m.  Earlier soap film work tended to use thin films with thicknesses of $5$ $\mu$m or less.  There are a number of advantages that come from using thick films.  First, for the e-m cell, it is desirable to keep the electrical resistance low, again to limit evaporation due to Joule heating.  The larger the cross sectional area through which the current is passed, then, the better.  Another reason to use thick films is that they lose a smaller percentage of their energy than thin films to frictional rubbing against the surrounding air.  Recall from chapter \ref{intro} that 2D driven turbulence requires an external dissipation mechanism to maintain energy balance.  Air friction plays this role in the e-m cell.  If the air friction is strong then it can easily dissipate large amounts of energy. Therefore the energy injected from the electromagnetic force must be exceptionally high to maintain a state of strong turbulence.  This would require large currents which would enhance the Joule heating and evaporation of the film.  Using thick films reduces this coupling to the air and allows strong turbulence to be maintained for reasonable values of the driving current.   A final reason for using thick films arises from the observation that a soap films kinematic viscosity, $\nu$, is dependent on it's thickness.  Trapeznikov predicted that the effective kinematic viscosity should depend on the thickness, $h$, as $\nu = \nu_{bulk} + 2\nu_{surf}/h$, where $\nu_{bulk}$ is the kinematic viscosity of the soap solution and $\nu_{surf}$ is the 2D viscosity due to the soap film surfaces\cite{Trapeznikov:SICSA57}.  The surface viscosity was recently measured\footnote{This measurement utilized a flowing soap film tunnel to analyze the shedding of vortices from a cylinder placed in the flow.  The vortex shedding frequency is sensitive to changes in the fluids viscosity and through proper normalization allows $\nu_{surf}$ to be approximated.} to be around $\nu_{surf} \approx 1.5 \times 10^{-5}$ cm$^3$/s for soap films similar to that in the e-m cell\cite{Vorobieff:PRE99}. Since the soap solution's viscosity is $.01$ cm$^2/$s the effective viscosity of a $50$ $\mu$m thick soap film is approximately $0.016$ cm$^2/$s.  Using thick films, then, reduces the amount of energy lost to viscous dissipation, and by the same argument given above, allows strong turbulence to be maintained for reasonable values of driving current.

There are, of course, disadvantages to the use of thick soap films.  One is that the speed of a 2D density wave in the soap film is dependent on the thickness of the film\footnote{A 2D density wave in a soap film is a thickness wave, where the film bulges or shrinks in the third dimension.}.  Thicker films contain more interstitial fluid than thin films, and therefore have a lower wave velocity due to their increased mass.  Therefore, as the film becomes thicker it begins to be more easily compressed.  That is to say that it's 2D density couples more strongly to the velocity field in the film.  A measure of the importance of such compressibility effects is the Mach number $M = u_{rms}/c$, where $c$ is the density wave speed in the medium.  In the case of a $50$ $\mu$m thick film the wave velocity is approximately $2$ m/s.  Velocity fluctuations in the e-m cell are therefore kept less than $20$ cm/s so that $M < 0.1$.  With such a small Mach number the system does not develop shock waves and behaves approximately as an incompressible fluid.

\begin{figure}
\hskip 1in
\includegraphics*[width=4in,height=1.63in]{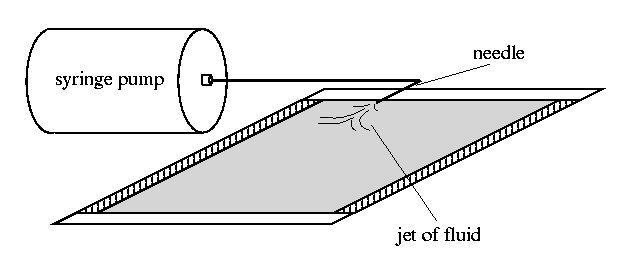}
\caption{Replenishing fluid lost to evaporation.}
\label{fig: experimental2}
\end{figure}

As noted earlier, Joule heating evaporates fluid from the film.  This causes the average thickness of the soap film to change over time.  Since thick films were used to minimize this effect, thickness changes due to evaporation happen at a slow rate and can be balanced by injecting small amounts of fluid into the film.  In the e-m cell, soap solution is injected by a syringe pump through a small needle inserted through the plastic side of the frame as in Fig. \ref{fig: experimental2}.  It is important that the fluid be burst in, as a brief jet, and not slowly injected.  Without initial momentum, fluid builds up near the injection point forming a droplet which eventually pops the film.  Fluid injected with a burst shoots to the center of the e-m cell where it is quickly mixed into the film replenishing the lost fluid.  Though not used in the experiments reported here, this process can be automated by monitoring the film resistance.  When the resistance becomes to high a small burst of fluid is shot into the film raising the thickness and lowering the resistance to an acceptable level.

\begin{figure}
\hskip 1in
\includegraphics*[width=4in,height=2.50in]{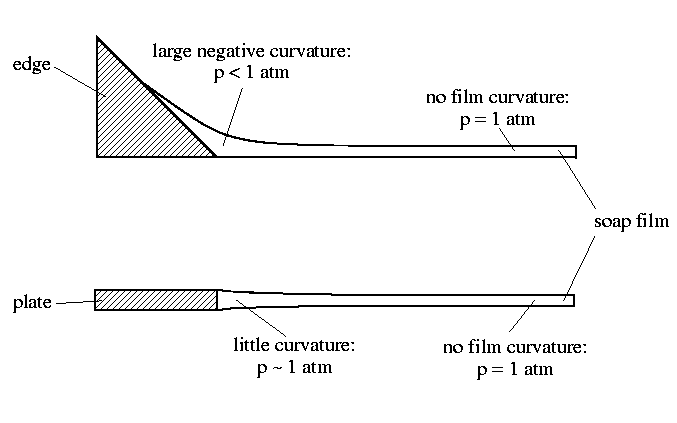}
\caption{Film curvature near an edge and a plate.}
\label{fig: filmcurve}
\end{figure}

The square frame that the film is drawn across is limited in size to $7 \times 7$ cm$^2$ in all of the experiments reported here for reasons which will be made clear later.  Both the stainless steel and plastic sides are milled to a sharp edge of about $45^\circ$, though the stainless steel edge generally dulls over time by corrosion.  Recent efforts have attempted to replace these edges with sheets of stainless steel and glass that are less than $100$ $\mu$m thick.  The intent of this is to limit the size of the wetting region near the edge of the frame, as shown in Fig. \ref{fig: filmcurve}.  Since the film has a finite surface tension, there must be a pressure jump across the film surface if it is curved.  A wetting region induces a negative curvature near the edge of the frame, therefore the pressure inside the film near the frame edge is less than at the film center where there is no curvature.  This causes fluid to be forced through the film to the edges, causing the center of the film to drain.  Elimination of these wetting regions eliminates such drainage.  This effect is relatively weak compared to other effects which cause drainage so using sheets instead of edges is a low priority, sheets being difficult to work with due to their delicacy.  All experiments reported in this thesis use edges instead of thin sheets.  However, the preliminary work with sheets which is not reported here has provided encouraging early results.

Using a $50$ $\mu$m film made from the above ammonium chloride solution on the $7 \times 7$ cm$^2$ frame results in resistance values of around $1$ k$\Omega$.  Driving currents of between $10$ and $45$ mA are used depending on the strength of turbulence desired, strength of the air dissipation, and strength of externally applied magnetic field.  In all turbulence experiments, the applied voltage oscillates at $3$ Hz with a square waveform. There are several reasons for this, one is that the current in the film causes chlorine gas bubbles to  accumulate on the positive electrode.  By oscillating the current, the bubbles form at both electrodes, slowing the inevitable bubble build up which eventually invades the film and renders it useless.  Another reason for current oscillation is to eliminate the formation of large vortices which form and dominate regions of the fluid containing a force field which is  sympathetic to it's motion.  Once formed, such structures induce a spatially varying mean flow, and thus a mean shear, rendering the turbulence in the e-m cell inhomogeneous.\footnote{ Oscillating the current may also eliminate any polarization of charge in the e-m cell caused by the net motion of ions in the fluid.  At this time it is unclear whether this polarization is important in the e-m cell.  Experiments done in both laminar and turbulent flow with D.C. forcing do not exhibit signs of charge polarization, such as the gradual decrease in applied force due to shielding of the electrodes.  However, these experiments where done over short (minutes) time scale. What effect polarization may have over a typical experiment time scale (around fifteen minutes) is unknown since such effects are masked by the effects of evaporation.}

A final note: due to the configuration of the e-m cell, the current which is driven in the plane of the film is unidirectional, namely parallel to the plastic edges of the frame running from one electrode to another.  For the rest of the paper the current direction will be assumed to lie along the $y$ axis.  By the Lorentz law, this unidirectional current creates a unidirectional electromagnetic force which lies on the $x$ axis.  Due to this unidirectionality, the e-m cell can not behave isotropically, that is the system is not rotationally invariant, regardless of the symmetry of the spatial variation of the magnetic field.  This lack of isotropy will be an exceedingly important consideration later in the thesis.

\section{The Magnet Array}

\begin{figure}
\hskip 1in
\includegraphics*[width=4in,height=1.68in]{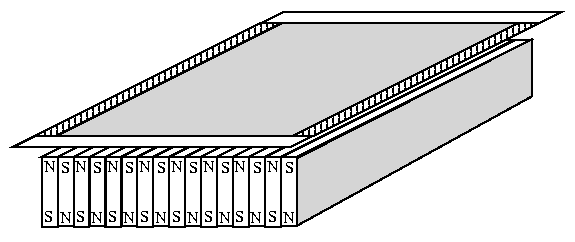}
\caption{Array of magnets creating the spatially varying external magnetic field.}
\label{fig: magnetarray}
\end{figure}

The spatially varying magnetic field used in the creation of the Lorentz force is generated by an array of Neodymium rare earth magnets (NdFeB) placed just below the film as shown in Fig. \ref{fig: magnetarray}.  These magnets produce quite powerful fields, typically on the order of $0.1$ T at their surface.  This field decays away from the surface, however, with a typical length scale of the order of the magnet size.  Since the magnets are small, generally less than $0.5$ mm, they must be brought very close to the film surface to generate fields strong enough to drive flow.  One could use larger magnets to induce motion.  This is undesirable, though, since the magnet size dictates the energy injection scale, $r_{inj}$.  Recall from the discussion in chapter \ref{intro} that $r_{inj}$ must be significantly smaller than the system size for an inverse cascade region to exist in an experimental 2D turbulence system.  For the e-m cell described above, the system size is the frame size, which is limited to $7$ cm.  Therefore magnet arrays with injection scales less than $0.7$ cm must be used to allow for a decade of inverse cascade range.

\begin{figure}
\includegraphics*[width=6in,height=5.84in]{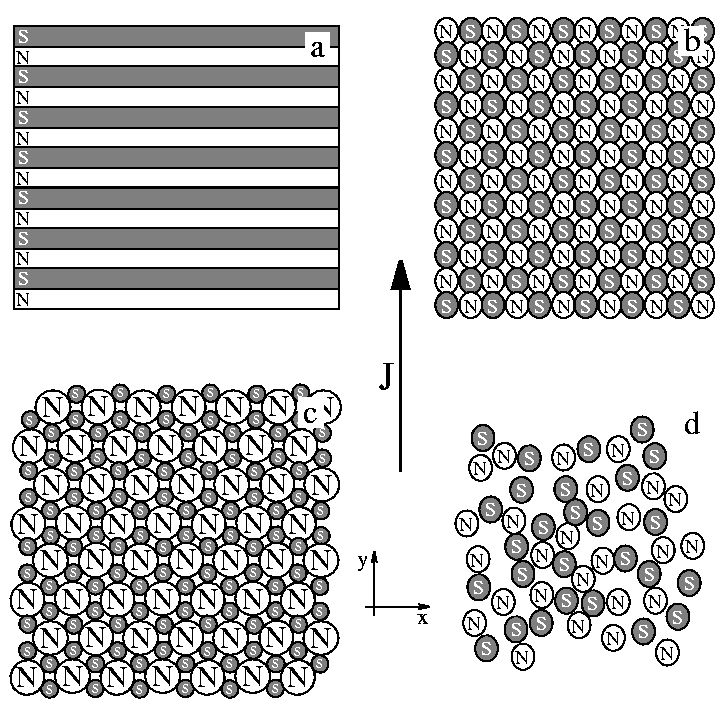}
\caption{Top view of the magnet arrays which create the spatially varying external magnetic field in the e-m cell: (a) the Kolmogorov array, (b) the square array, (c) stretched hexagonal array, (d) pseudo-random array.  The direction of the current $J$ is shown as is the coordinate axis.}
\label{fig: arrays}
\end{figure}

Several types of magnet arrays are used in the e-m cell, each distinguished by the type of flow it induces at small Reynolds number (it's laminar behavior) or equivalently by the characteristics of the force field it produces in the e-m cell.  The first type of array, and by far the most important, is the Kolmogorov array.  This array is made of alternating north-south layers of long rectangular magnets of approximately $0.3$ cm in width, as shown in Fig. \ref{fig: arrays} a.  The magnetic field it produces in the film varies approximately sinusoidally in space.  The magnets are oriented so that this variation lies in the direction of the current, causing the Lorentz force in the film to have the form $F_x = f_0 sin(k_0 y)$ (Recall $F_y=0$ because the force is unidirectional).  Two magnets make a single north-south oscillation, therefore the wavelength of the sinusoidal variations in the force field (which is the energy injection scale) is $ r_{inj}= 0.6$ cm.  The associated injection wavenumber is $k_0 = 2\pi/r_{inj} = 10$ cm$^{-1}$.  The laminar flow this forcing produces is one of alternating shear layers, a flow which Kolmogorov proposed investigating as an interesting model to study a fluids transition from laminar to turbulent motion (hence the name).  The Kolmogorov array has several properties which will be of importance in later analysis.  The first is that the forcing is divergence free, i.e. $\vec{\nabla} \cdot \vec{F} = 0$.  This property will allow pressure fields to be approximated from velocity fields using Fourier techniques.  Another property of importance is that the forcing is invariant to translation along the $\hat{x}$ direction.  Chapter \ref{ebudg} will demonstrate how this property may be used to obtain the energy injection rate from the forcing without having to explicitly measure the magnitude of the forcing $f_0$.

The next type of magnet array, shown in Fig. \ref{fig: arrays} b, is called the square array.  It is made of round magnets oriented in a tick-tack-toe arrangement with like poled magnets along diagonals.  There are two such arrays in the e-m cell arsenal made from $0.3$ cm and $0.6$ cm magnets.  The force field created by these arrays has the form $F_x = f_0(sin(k_0(x+y)) + sin(k_0(x-y)))$.  Here the wavelength along the diagonals is $r_{inj} = \sqrt{2}w$ where $w$ is the magnet diameter, and again $k_0 = 2\pi/r_{inj}$.  The $0.3$ cm square array has the smallest injection scale of all the arrays used with $r_{inj} = 0.42$ cm.  Unfortunately these magnets are exceptionally weak, limiting the magnitude of the forcing one can create with reasonable currents.  Also, the forcing created by these arrays are not divergence free.  Without the ability to account for pressure much the usefulness of this array is limited to testing ideas of universality (Chapter \ref{himom}).

The final two arrays are seldom used and are mentioned here only for completeness.  The first, shown in Fig. \ref{fig: arrays} c, is a stretched hexagonal array, called the hex array, and the second, shown in Fig. \ref{fig: arrays} d, is a pseudo-random array.  The former is constructed from a mixture of both $0.3$ cm round and $0.6$ cm round magnets.  The result could be though of as a Kolmogorov flow with a periodic variation along every other magnet.  The injection wavenumber associated with this array is quite large, limiting it's usefulness.  The laminar behavior a hex array produces is a triangular vortex crystal.  The pseudo-random array is constructed from $0.3$ cm round magnets placed at random on an iron sheet.  The positions of the magnets where generated via a random number generator which attempted to maximize the magnet density.  It was constructed in a naive attempt to obtain more homogenous turbulence in the e-m cell.  The pseudo-random flow has no well-defined laminar flow behavior.

\section{External Dissipation: The Air Friction}

The air friction has already been described as the external dissipation mechanism in the e-m cell.  
Chapter \ref{ebudg} will demonstrate that the air friction can be adequately modeled as a linear friction, that is if an element of the film is moving with velocity $\vec{u}$ it experiences a drag force from the air of the form $\vec{F}_{drag} = -\alpha \vec{u}$.   The following discussion sets the basis for this linear drag model.

\begin{figure}
\hskip 1in
\includegraphics*[width=4in,height=1.40in]{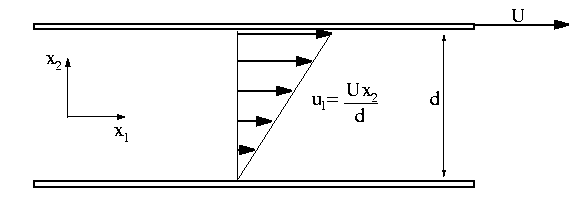}
\caption{Fluid between a top plate moving with velocity $U$ and fixed bottom plate produces a linear velocity profile.}
\label{fig: airdrag}
\end{figure}

A fluid between two parallel plates separated in the $x_2$ direction by a small distance $d$, one fixed and the other moving with velocity $U$ in the $x_1$ direction, assumes a linear profile of the form, $u_1 = Ux_2/d$ (This assumes there are no other forces acting on the fluid, the plates are infinite and boundary conditions are no slip at either plate).  This is shown in Fig. \ref{fig: airdrag}.  The drag force per unit area exerted on the top plate by the fluid is $f_{drag} = \eta \frac{\partial u_1}{\partial x_2}|_{d} = \frac{\eta U}{d}$ where $\eta$ is the ordinary 3D shear viscosity of the fluid between the plates.  In the e-m cell the role of the top plate is played by the soap film, while the role of the lower plate is played by the magnet array.  The fluid is the surrounding air and the length $d$ is just the distance of the magnets to the film.  Thus raising the magnets closer to the film increases the strength of the air drag in the e-m cell.

The drag force must be normalized by the films 2D density so that the force per unit mass can be considered.  The 2D density of the film is given by $\rho h$ where $\rho$ is the density of the soap solution (about that of water) and $h$ is the films thickness.  The force per unit mass caused by the air drag on the film is then $\vec{F}_{drag} = -\frac{\eta}{\rho h d} \vec{u}$.  This demonstrates a point that was made earlier in this chapter, that the frictional coupling to the air depends not only on the magnet-film distance but on the thickness of the soap film as well. Keeping the film as thick as possible, then, minimizes this coupling.

There are a couple of reasons to worry about the applicability of this linear friction model.  First is that the soap film in the e-m cell is not an infinite flat plate moving with a constant velocity, but has a velocity which fluctuates from point to point in the flow.  As long as the size of these velocity fluctuations, that is the size of the vortices, is larger than both the magnet-film distance, $d$, and the film thickness, $h$, the approximation as an infinite plate should apply.  Recall from the discussion above that the magnetic field created by the array quickly decays within the width of a single magnet.  Therefore the magnets are always kept within one magnet width of the film, otherwise the magnetic field would be too weak to drive turbulence.  Since the magnet width also dictates the smallest vortex size, $r_{inj}$, the requirement that $d < r_{inj}$ is always met in the e-m cell.

The second complication is the fact that there is more than one side to the film.  The linear drag model accounts for the drag force exerted on the lower surface of the soap film.  The upper surface of the film is also dragging along a layer of air.  The velocity profile of the air above the film is not a simple linear shear as it is below the film, but a more complex Prandtl-like boundary layer due to the absence of a second fixed plate.  Chapter \ref{eflux} will show that this causes a non-negligible correction to the above linear drag model.

\section{Gravity}

To this point there has been no discussion about the orientation of the soap film with respect to the earth's gravitational field.  Vertical orientation, that is the film plane lying along the gravitational field direction, is not desirable because it would strongly stratify the soap film, making it thinner on top than on the bottom.  This is tantamount to both a severe change in the 2D density and a change in the films kinematic viscosity from the top of the film to the bottom.  In other words the film becomes an inhomogeneous fluid.  Vertical orientation must therefore be discarded and a horizontal orientation used to allow the film to approximate a homogenous fluid.

\begin{figure}
\hskip 1in
\includegraphics*[width=4in,height=1.51in]{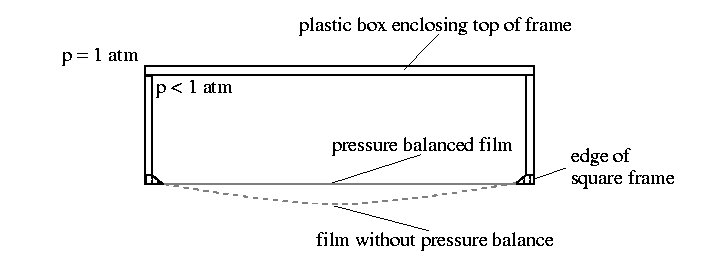}
\caption{A thick film droops under the action of gravity, as shown by the dotted line.  A box enclosing the top of the e-m cell frame is brought to a lower pressure than the surrounding environment to balance gravity.}
\label{fig: experimental4}
\end{figure}

A horizontally oriented soap film droops under the force of the earth's gravitational field.  This effect is exacerbated by the fact that the film is $50$ $\mu$m thick.  To balance gravity, a box enclosing the region on the top surface of the soap film is evacuated of a small amount of air, as in Fig. \ref{fig: experimental4}.  This lowers the pressure in the box causing a pressure gradient across the film plane and thus a force opposing gravity.  Enough pressure is drawn to almost exactly balance the gravitational field.  Unfortunately the larger the soap film, the more delicate this balance becomes.  This is the reason that soap films used in the e-m cell are limited to sizes under $7 \times 7$ cm.

This pressure balance is delicate and can be disrupted by the evaporation of fluid from the soap film into the container above the film.  It must be constantly monitored to make sure that the film stays at the same level.  This is done in the e-m cell by reflecting a laser light off a portion of the film near the middle of one of the edges.  A drooping film deflects the beam, and this deflection can be monitored by various techniques, for example using a position sensitive detector or a linear CCD array.  A feedback loop based on the deflection measurement can be easily constructed and film level kept steady, even for high current and large amounts of evaporation.

\section{Particle Tracking Velocimetry}

Velocity information is obtained from the e-m cell using a particle tracking method (PTV) which is similar to the standard technique known as particle imaging velocimetry (PIV).  The PIV technique uses a camera to capture images of small particles that seed the fluid flow.  Two consecutive images are separated in time by a small amount $\Delta t$.  These images are sectioned into small regions by a discrete grid.  Corresponding regions from the two consecutive images are compared to obtain the average motion $\vec{\Delta x}$ of the particles in that region over the time $\Delta t$.  Each region is then assigned a velocity vector $\vec{u} = \vec{\Delta x} / \Delta t$, yielding an entire velocity field on a grid.  There are many papers and review articles which describe the PIV technique \cite{Rivera:PRL98,Grant:IME97,Wernet:EIF93,Prasad:EIF92}, interested readers should refer to these for a complete description.

Where PIV attempts to track the average displacement of a number of particles (usually around 10) in a square region formed by a grid, PTV attempts to determine the displacement of individual particles.  This allows PTV to obtain finer spatial resolution than PIV.  Equivalently one could say that PTV has higher vector density.  Since the algorithms used in PTV and PIV to determine translation are identical it might be expected that this increased resolution comes only with increased noise.  This is not the case, however,  because PTV has a built-in self correction that allows noise to be suppressed.  There is one sacrifice though.  Since PTV attempts to track individual particles the technique is much more sensitive to particles leaving the measurement volume than PIV.  This is not a issue in 2D flows such as are studied here (except near the boundaries of the images), though in 3D it could be a significant problem.  A copy of the PTV program is listed in Appendix \ref{program}.

\begin{figure}
\hskip 1in
\includegraphics*[width=4in,height=2.11in]{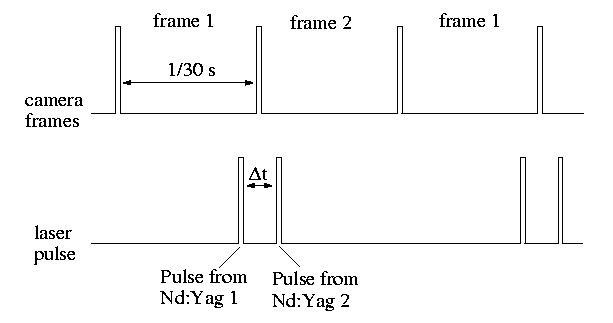}
\caption{Timing of the CCD camera frames and laser pulses used in PTV.  Frame 1 and Frame 2 denote a single PTV image pair from which velocity fields are extracted.}
\label{fig: timing}
\end{figure}

There are three main steps to PTV by which one goes from CCD images of particles to velocity information: particle identification, neighborhood comparison, and matching.  Before going into these, the manner in which images are acquired should be described.  As stated earlier the soap film in the e-m cell is seeded with small particles.  These particles are illuminated by two pulsed Nd:Yag lasers that yield $12$ mJ of energy per nano-second pulse.  Images of the particles illuminated by the lasers are obtained using a $30$ Hz, $8$ bit CCD camera of resolution $768 \times 480$ rectangular pixels with aspect ratio $1:1.17$.  The lasers are slaved to the camera so that the first Nd:Yag laser pulses at the end of the first image and the second laser pulses a time $\Delta t$ later in the second image.  In this manner two images of particles separated by time $\Delta t$ are obtained on a camera with a fixed time resolution of $.03$ ms.  This timing is shown in Fig. \ref{fig: timing}.

To determine the positions of the particles in the flow the background of each image must be subtracted off.  There are many routines by which one can do this; in this thesis a high pass filter is used since the particles are spatially small compared to the image size.  Once the background is subtracted the particles in each image are found by an exhaustive nearest neighbor searching algorithms.  If the $(i,j)$ pixel is found to be non-zero then this serves as the base of a particle group. The four pixels at $(i+1,j)$, $(i-1,j)$,$(i,j+1)$, and $(i,j-1)$ are said to neighbor the base pixel, and any of these which are non-zero are added to the group.  Any neighbor of a pixel in the group is then considered, and if it is non-zero and not already part of the group, it is added to the group.  This process continues until no pixels are being added to the group.  The final group of non-zero pixels is called a particle.  This process is performed until all non-zero pixels in each image are accounted for in a particle.  In what follows the particles in the first image will be indexed by $a$ and those in the second image by $b$.

This manner of finding a particle does not distinguish between an individual particle and particles that are so close together that they form a continuous image on the CCD camera.  Since the turbulence in the e-m cell is only mildly compressible, particles which start initially very close should stay close over the short period of time, $\Delta t$, between laser flashes.  The indistinguishability, then, should not be an issue.

There are three quantities which need to be determined for each particle: it's centroid, pixel centroid and size.  The centroid is the center of the group of pixels which make the particle weighted by the intensity, $I^{(z)}_{(i,j)}$, of the pixels in the group (z denotes the image).  If there are N pixels, indexed by $n$, in the pixel group of particle $a$ in image one, then the centroid, $(x^{(a)}_c,y^{(a)}_c)$, is given by:
\[
(x^{(a)}_c,y^{(a)}_c) \equiv \frac{\sum_{n=1}^{N} (i(n),j(n))I^{(1)}_{(i(n),j(n))}}{\sum_{n=1}^{N}I^{(1)}_{(i(n),j(n))}}.
\]
Note that where the pixels themselves are discretized on a finite grid, the center of mass of a particle need not be if there is more than one pixel contained in it's group.  This phenomenon is called sub-pixel resolution since it allows particle position to be tracked with a resolution finer than the pixel size of the camera.  Sub-pixel resolution can be used to enhance the dynamic range of PTV measurement, though it is not relied upon for results in this thesis.  The pixel centroid of particle $a$ is defined as the nearest pixel to the particles centroid, and will be denoted as $(x^{(a)},y^{(a)})$.  Finally the size of particle $a$ is simply the 2$^{\rm nd}$ moment of the intensity distribution given by
\[
R^{(a)} \equiv \left( \frac{\sum_{n=1}^{N} ((i(n)-x^{(a)}_c)^2 + (j(n) - y^{(a)}_c)^2)I^{(1)}_{(i(n),j(n))}} {\sum_{n=1}^{N}I^{(1)}_{(i(n),j(n))}} \right) ^{1/2}.
\]
The particle size is used as a filter to discard particles which are too big or too small.

Once particles have been identified the challenge is to track them from one image to the next.  This is done by comparing the regions surrounding particles in the first image with regions around particles in the second to establish how well they correlate.  For particle $a$ in the first image and $b$ in the second image the correlation number $c_{a,b}$ is determined by
\[
c_{a,b} = \frac{\sum_{m=-l}^{l} \sum_{n=-l}^{l} \tilde{I}^{(1)}_{((x^{(a)}+m),(y^{(a)}+n))} \tilde{I}^{(2)}_{((x^{(b)}+m),(y^{(b)}+n))} }{(\sum_{m=-l}^{l} \sum_{n=-l}^{l} (\tilde{I}^{(1)}_{((x^{(a)}+m),(y^{(a)}+n))})^2)^{1/2}(\sum_{m=-l}^{l} \sum_{n=-l}^{l} (\tilde{I}^{(2)}_{((x^{(b)}+m),(y^{(b)}+n))})^2)^{1/2}}.
\]
In the above $2l$ is called the correlation box size.  The intensities, $\tilde{I}^{(z)}_{i,j}$, used in determining the correlation number are the intensities of the images $I^{(z)}_{i,j}$ less their average over their respective correlation boxes so that $c_{a,b}$ assumes a value between $-1$ and $1$.  If the correlation number is close to $1$ then the region around particle $a$ is similar to the region around particle $b$.  The closer to $1$ the more similar the regions.

One need not compare all particles in the first image to all particles in the second.  Only a subset of particles within a certain distance of one another need be considered.  That is if a particle $a$ is found at $(x^{(a)}_c,y^{(a)}_c)$ in the first frame and $b$ is at $(x^{(b)}_c,y^{(b)}_c)$ in the second, their correlation number need only be calculated if $((x^{(a)}-x^{(b)})^2 + (y^{(a)}-y^{(b)})^2)^{1/2} < s$, where $s$ is some reasonable threshold displacement based on the RMS velocity fluctuations and flash spacing.

Once all the particle comparisons have been performed, one need only match particles in the first frame to those in the second.  This is done by an iterative mutual maxima technique.  Look at correlation number $c_{a,b}$ and determine if it is above some initial threshold $c_{limit}$.  If it is, then look to see if $c_{a,b}$ is the maximum correlation value for both particle $a$ and particle $b$.  That is make sure $c_{a,b}>c_{a,x}$ for any $x \neq b$ and $c_{a,b}>c_{y,b}$ for any $y \neq a$.  If so then then $c_{a,b}$ is the mutual maximum for particles $a$ and $b$ and they are considered a match.  Thus particle $a$ has moved from position $(x^{(a)}_c,y^{(a)}_c)$ in the first frame to $(x^{(b)}_c,y^{(b)}_c)$ in the second.  The fact that we have checked that not only is particle $a$ the best fit for $b$, but that $b$ is best fit by $a$ is the self-correction that PTV has that PIV does not (PIV can only check that $a$ is best for $b$), and the reason that PTV can achieve higher vector density without much sacrifice in velocity resolution.

Once matched these particles and their correlation numbers are removed from consideration.  This is done for all correlation numbers above $c_{limit}$, and all mutual maxima are obtained in this way.  The $c_{limit}$ is then lowered slightly and the procedure performed over all particles which have not already been matched.  This is done until $c_{limit}$ hits some specified lowest bound and the particles which have not been matched at this point are discarded.

This leaves a final list of particles which have been tracked from a point in the first image to a point in the second.  The average velocity of the particle is determined by the motion of it's center of mass.  This velocity is assigned to the average particle position.  This yields a field of average velocities which is the final field from the PTV technique.  These fields are generally interpolated to a finite grid (binned) so that derivatives may be taken. This interpolation can be performed by any number of weighted averaging techniques.

\section{Cell Operation}

This section describes the procedure that was used to employ the features of the e-m cell described above.  First a marginally thick film is pulled across the square frame.  The plastic edges of the frame are dried to remove any fluid bridges that might short the electrical current.  A small amount of air is removed from the box until the film is just level to the eye.  This film is then placed above the magnet array and a mixing current, generally around $15$ mA, is turned on.  The feedback loops to inject fluid and pressure balance the film are initiated.  After a balance is reached the magnets are raised (or lowered) to the desired level, and the current is adjusted until the target $u_{rms}$ is reached.  Particle  images are then acquired at a rate on the order of a few Hz until a large number (between $500$-$1000$) image pairs are obtained.  These pairs are then interrogated using the PTV algorithm to obtain velocity information.

\chapter{Modeling Flows in the E-M Cell\label{ebudg}}

A frustrating problem that arises when soap films are used as an experimental system for studying 2D fluid dynamics is the lack of direct evidence that these films obey the 2D incompressible Navier-Stokes equation, 
\begin{eqnarray}
\frac{\partial u_i}{\partial t} + u_s\frac{\partial u_i}{\partial x_s} = -\frac{\partial p}{\partial x_i} + \nu\frac{\partial^2u_i}{\partial x_s^2} + F^{ext}_i,
\label{eq: Navier-Stokes}
\end{eqnarray}
\begin{eqnarray}
\frac{\partial u_s}{\partial x_s} = 0.
\label{eq: incompressible}
\end{eqnarray}
In the above equations $u_i$ is the i$^{\rm th}$ component of the fluid's velocity field, $p$ is the internal pressure field normalized by fluid density, and $F^{ext}_i$ represents the i$^{\rm th}$ component of any external force field (per unit mass) acting on the fluid.  The constant $\nu$ is the fluid's kinematic viscosity.  Einstein summation convention is used, and will be used throughout the thesis unless otherwise noted.  These equations are the governing equations for the time evolution of the velocity field of an incompressible fluid.  If the soap film does not obey this equation then it is not behaving as an incompressible Navier-Stokes fluid and is therefore useless as a system for standard turbulence investigations.  The purpose of this chapter is to demonstrate that the e-m cell does indeed approximate a 2D Navier-Stokes fluid.

\section{Introduction}

\begin{figure}
\hskip 0.5in
\includegraphics*[width=5in,height=4.01in]{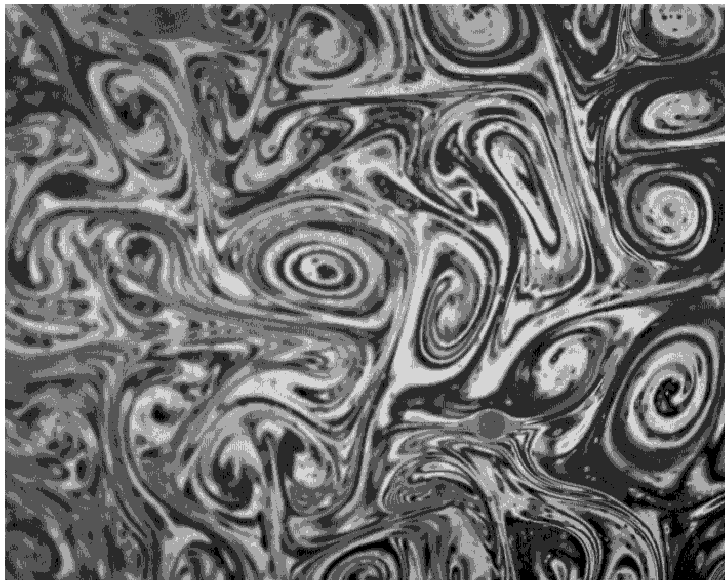}
\caption{Thin film interference fringes demonstrate that the thickness of the soap film in the e-m cell is not constant but varies from point to point in the flow.}
\label{fig: thinfilm}
\end{figure}

There are a number of reasons to be skeptical about soap films behaving as an incompressible Navier-Stokes fluids.  Most of the problems arise from the presence of thickness fluctuations which are caused by the motion of the film.  Such thickness fluctuations can be imaged using thin film interference as in Fig. \ref{fig: thinfilm} and are indicative of compressibility in the soap film.  Compressibility constitutes a failure of Eq. \ref{eq: incompressible} in soap films.  This has already been discussed somewhat in chapter \ref{experimental} where it was implied that keeping the Mach number small eliminates this problem.  This is a little misleading; a small Mach number does not mean that there are no thickness fluctuations.  Rather it means that the thickness fluctuations do not develop shock fronts and vary in a smooth manner from point to point in the flow.  In the range of Mach number present in the e-m cell the thickness fluctuations tend to be around $20$\% the mean thickness of the film \cite{Rivera:PRL98}.  One would like to determine if this is small enough to allow Eq. \ref{eq: incompressible} to hold approximately.

Aside from the incompressibility issue, thickness fluctuations could cause soap films to deviate from a Navier-Stokes fluid in a more sinister way.  Recall from the discussion in chapter \ref{experimental} that the kinematic viscosity of the soap film, $\nu$, is dependent on thickness.  Since the film thickness varies from point to point in the flow, so should the effective viscosity of the fluid.  These thickness fluctuations respond to velocity gradients in the film, therefore the viscosity is dependent on the local shear rate.  A fluid with such a shear dependent viscosity is said to be non-Newtonian and does not obey Eq. \ref{eq: Navier-Stokes}.  Here again one would like to determine if the viscosity fluctuations are small enough to be considered negligible and Eq. \ref{eq: Navier-Stokes} to hold approximately.

Another problem when dealing with soap films is the external frictional coupling to the air.  It's presence is important to attain an energy balance in 2D forced turbulence, as discussed in chapter \ref{intro}.  Indeed its strength and form should dictate the outer scale of the turbulence and affect energy transfer at large scales.  Because of its importance the effects of this coupling must be modeled and tested.  The simplest model of the frictional effects of the air is to assume it acts as a linear drag on the film.  Therefore the external force field $F^{ext}_i$ acting on the e-m cell can be broken into two parts, the Lorentz force caused by current and magnetic field, $F_i$, and the air drag $F^{air}_i = -\alpha u_i$.  The constant $\alpha$ represents the strength of the frictional coupling of the air to the film.  This model must be tested if it is to be used in later investigations.

Though a direct test of the Navier-Stokes equations by inverse methods is not easily performed, it is possible to test an equation known as the Karman-Howarth relationship.  This relationship can be derived from the Navier-Stokes equation with a single assumption, and easily tested with data from the e-m cell.  It's failure or success in describing data from the e-m cell would then constitute an indirect test of the Navier-Stokes equation as well as the linear drag model proposed for the air friction.

\section{The Karman-Howarth Relationship \label{karmanhowarth}}

Although the derivation of the Karman-Howarth relationship can be found in a number of texts on turbulence, it is performed here for two purposes: the relationship is used extensively in later chapters and to present notation which will be used throughout the thesis.  It is also performed with the inclusion of a linear damping term in the Navier-Stokes equation to represent the air friction as discussed earlier. The derivation here, with the exception of the air drag, closely follows that found in Hinze \cite{Hinze}.

The Karman-Howarth relationship governs the time evolution of the two-point velocity correlation, $\langle u_i(\vec{x}) u_j(\vec{x^\prime}) \rangle$, for a fluid in a state of homogenous turbulence.  The brackets $\langle ... \rangle$ represent an ensemble average.  This relationship can be derived from the incompressible Navier-Stokes equation using only the assumption of homogeneity in the following manner.  Multiply Eq. \ref{eq: Navier-Stokes} which is evaluated for the i$^{\rm th}$ component of the velocity field at the point $\vec{x}$ with the j$^{\rm th}$ component of the field at point $\vec{x^{\prime}}$:
\begin{eqnarray}
u_j^{\prime}\frac{\partial u_i}{\partial t} + \frac{\partial}{\partial x_s}(u_i u_s u_j^{\prime}) = -\frac{\partial}{\partial x_i}(pu_j^{\prime}) + \nu\frac{\partial^2}{\partial x_s^2}(u_i u_j^{\prime}) + u_j^{\prime}F_i - \alpha u_i u_j^{\prime}.
\label{eq: khstep1}
\end{eqnarray}
In the above, field quantities at the point $\vec{x^{\prime}}$ are denoted by a $\prime$ and the linear drag model has been explicitly inserted into the equation. The fact that the derivative at point $\vec{x}$ commutes with multiplication by fields evaluated at $\vec{x^{\prime}}$ has been used to move $u_j^{\prime}$ inside spatial derivatives.  Incompressibility has also been used in the second term on the left-hand-side to move $u_s$ into the derivative.

Add Eq. \ref{eq: khstep1} with the corresponding equation evaluated by multiplying  Eq. \ref{eq: Navier-Stokes} evaluated for the j$^{\rm th}$ component of the velocity field at the point $\vec{x^\prime}$ with the i$^{\rm th}$ component of the field at point $\vec{x}$.  This allows both velocity terms to be brought into the time derivative,
\begin{eqnarray}
\frac{\partial}{\partial t}(u_i u_j^{\prime}) &=& -\frac{\partial}{\partial x_s}(u_i u_s u_j^{\prime}) - \frac{\partial}{\partial x_s^\prime}(u_i u_s^{\prime} u_j^{\prime}) - \frac{\partial}{\partial x_i}(p u_j^\prime) - \frac{\partial}{\partial x_j^\prime}(p^\prime u_i) \nonumber \\ & &+\nu \frac{\partial^2}{\partial x_s^2}(u_i u_j^\prime) +\nu \frac{\partial^2}{\partial x_s^{\prime 2}}(u_i u_j^\prime) + u_j^\prime F_i + u_i F_j^\prime - 2\alpha u_i u_j^\prime.
\label{eq: khstep2}
\end{eqnarray}
A coordinate transformation to relative, $r_i \equiv x^\prime_i - x_i$, and absolute, $\xi_i \equiv 1/2(x^\prime_i + x_i)$, coordinates can now be performed.  Grouping the appropriate terms yields:
\begin{eqnarray}
\frac{\partial}{\partial t}(u_i u_j^{\prime}) &=& -\frac{1}{2}\frac{\partial}{\partial \xi_s}(u_i u_s u_j^{\prime} + u_i u_s^{\prime} u_j^{\prime}) -\frac{\partial}{\partial r_s}(u_i u_s^\prime u_j^{\prime} - u_i u_s u_j^{\prime}) \nonumber \\
& &-\frac{1}{2}\frac{\partial}{\partial \xi_i}(p u_j^\prime) - \frac{1}{2}\frac{\partial}{\partial \xi_j}(p^\prime u_i) +\frac{\partial}{\partial r_i}(p u_j^\prime) - \frac{\partial}{\partial r_j}(p^\prime u_i) \nonumber \\
& &+\frac{1}{2}\nu\frac{\partial^2}{\partial \xi_s^2}(u_i u_j^\prime)+2\nu\frac{\partial^2}{\partial r_s^2}(u_i u_j^\prime)
+u_j^\prime F_i + u_i F_j^\prime - 2\alpha u_i u_j^\prime.
\label{eq: khstep3}
\end{eqnarray}

Now an ensemble average is performed.  Using the assumption of homogeneity eliminates the derivative of averages with respect to absolute position, $\xi_i$, leaving only derivatives with respect to relative position, $r_i$.  What remains is called the Karman-Howarth relationship,
\begin{eqnarray}
\frac{\partial}{\partial t}\langle u_i u_j^{\prime} \rangle &=& -\frac{\partial}{\partial r_s}\langle u_i u_s^\prime u_j^{\prime} - u_i u_s u_j^{\prime} \rangle + \frac{\partial}{\partial r_i}\langle p u_j^\prime \rangle  - \frac{\partial}{\partial r_j}\langle p^\prime u_i \rangle  \nonumber \\& &+2\nu\frac{\partial^2}{\partial r_s^2}\langle u_i u_j^\prime \rangle  +\langle u_j^\prime F_i \rangle  + \langle u_i F_j^\prime \rangle  - 2\alpha \langle u_i u_j^\prime \rangle.
\label{eq: Karman-Howarth}
\end{eqnarray}
In the future the n-term two-point velocity correlation functions will be denoted by $b^{(n)}_{ij...,k...} \equiv \langle u_i u_j ... u_k^\prime... \rangle$.  Using this notation, the correlation on the left hand side of Eq. \ref{eq: Karman-Howarth} is given by $b^{(2)}_{i,j}$, while the first term in the first derivative on the right is given by $b^{(3)}_{i,sj}$.

This relationship can also be used to derive the energy balance equation for homogenous turbulence.  Energy balance will be used in what follows as a self consistency check for data that attempts to fit the Karman-Howarth relationship.  Taking the limit of Eq. \ref{eq: Karman-Howarth} as $\vec{r} \rightarrow 0$ (or equivalently as $\vec{x} \rightarrow \vec{x^\prime}$)yields
\begin{eqnarray}
\frac{\partial}{\partial t} \langle u_i u_j \rangle = \Pi_{ij} - \epsilon_{ij} + \langle u_i F_j \rangle - 2\alpha \langle u_i u_j \rangle.
\label{eq: ebudg1}
\end{eqnarray}
In the above the tensors $\Pi_{ij}$ and $\epsilon_{ij}$ are defined as
\begin{eqnarray}
\Pi_{ij} \equiv \lim_{r \rightarrow 0} (\frac{\partial}{\partial r_i}\langle p u_j^\prime \rangle  - \frac{\partial}{\partial r_j}\langle p^\prime u_i \rangle),
\end{eqnarray}
\begin{eqnarray}
\epsilon_{ij} \equiv \lim_{r \rightarrow 0} 2\nu\frac{\partial^2}{\partial r_s^2}\langle u_i u_j^\prime \rangle.
\end{eqnarray}
Taking half the trace of Eq. \ref{eq: ebudg1} eliminates the pressure term $\Pi_{ij}$ and yields the energy balance relationship
\begin{eqnarray}
\frac{1}{2}\frac{\partial}{\partial t} u^2_{rms} = -\nu \omega^2_{rms} + \langle u_s F_s \rangle - \alpha u^2_{rms},
\label{eq: ebudget}
\end{eqnarray}
where the vorticity, $\omega$, is defined as the curl of the velocity field (i.e. $\omega = \vec{\nabla}\times\vec{u}$).  The first term on the right, $\nu \omega^2_{rms} \equiv \epsilon_\nu$, is the amount of energy changed to heat by internal viscous dissipation.  The second, $\langle u_s F_s \rangle \equiv \epsilon_{inj}$, is the work done by the external force.  The final term, $\alpha u^2_{rms} \equiv \epsilon_{air}$, is the energy lost to the linear drag. Equation \ref{eq: ebudget} is the statement that the change in energy in the system is simply the amount gained from the external forcing less the amount lost to dissipative effects.

\section{Testing Karman-Howarth}
\subsection{Experimental Considerations}
Recall that the objective of the experiments presented in this section is to demonstrate that the dynamics of the e-m cell are governed by the Navier-Stokes equation, Eq. \ref{eq: Navier-Stokes}, with the effects of air friction modeled as a linear drag.  This will be demonstrated indirectly by showing that the Karman-Howarth relationship, Eq. \ref{eq: Karman-Howarth}, holds for homogenous turbulence in the e-m cell.  The number of terms which must be measured to check Eq. \ref{eq: Karman-Howarth} can be simplified by using specific characteristics of the e-m cell.  The first is the elimination of the time derivative.  This term can be ignored if the turbulence is in a statistically steady state.  Since the e-m cell was designed specifically to study steady state turbulence it is easy to maintain energy and enstrophy approximately constant.  Elimination of the time derivative in this manner is the main reason that testing of the Karman-Howarth relationship is significantly easier than directly testing the Navier-Stokes equation.

Another term which can be dropped is the viscous term, if one considers length scales, $r$, greater than the viscous scale $r_\nu \approx (\nu^3/\epsilon_{inj})^{1/4}$.  For typical values of energy injection in the e-m cell $r_\nu  = 200 \mu$m.  Since the particle tracking measurements focus on the inverse cascade regime, which occurs at length scales greater than a millimeter in the e-m cell, most of the measurement resolution lies well above this criteria.  Since the viscous term is being ignored these experiments can draw no conclusions about how thickness changes may be affecting changes in viscosity.  Small scale investigations, outside the scope of this thesis, would have to be performed to draw conclusions about this effect.

What remains of the Karman-Howarth relationship after using these assumptions is
\begin{eqnarray}
\frac{\partial}{\partial r_s}(b^{(3)}_{i,sj} - b^{(3)}_{is,j}) - \frac{\partial}{\partial r_i}\langle p u_j^\prime \rangle  + \frac{\partial}{\partial r_j}\langle p^\prime u_i \rangle = \langle u_j^\prime F_i \rangle  + \langle u_i F_j^\prime \rangle - 2\alpha b^{(2)}_{i,j}.
\label{eq: Karman-Howarth_simple}
\end{eqnarray}
Normally the assumption of isotropy would also be made to eliminate the pressure-velocity correlations on the left-hand-side.  Recall from the discussion in chapter \ref{experimental} that this assumption cannot be made in the e-m cell due to unidirectional forcing.  Therefore to check Eq. \ref{eq: Karman-Howarth_simple} a pressure-velocity correlation must be measured, indicating that not only is a velocity field needed for the check but a pressure field as well.

To obtain the pressure field, the divergence operator is applied to Eq. \ref{eq: Navier-Stokes} and Eq. \ref{eq: incompressible} is used.  What is left has the form
\begin{equation}
\nabla^2 p = 2\Lambda + \nabla \cdot \vec{F}
\end{equation}
where $\Lambda \equiv \frac{\partial u_x}{\partial x}\frac{\partial u_y}{\partial y} - \frac{\partial u_x}{\partial y}\frac{\partial u_y}{\partial x}$.  If the Kolmogorov magnet array is used, the divergence of the electromagnetic force on the right may be dropped (see chapter \ref{experimental}).  This allows the pressure to be approximated from the velocity field using Fourier techniques\footnote{This approximation of the pressure field assumes periodic boundary conditions.  Though the velocity fields extracted from the e-m cell do not satisfy this boundary condition it is hoped that the solution for the pressure field will be insensitive to this approximation away from the boundaries}.  For this reason the Kolmogorov array will be used in these experiments.

With Kolmogorov forcing and the assumptions above, all the terms in Eq. \ref{eq: Karman-Howarth_simple} may be measured and tested as an indirect test of the Navier-Stokes equation and linear drag model.  One final simplification can be made.  An exact measure of the external electromagnetic forcing is difficult at best.  However, since the forcing is unidirectional, along the $\hat{x}$ direction, the force-velocity correlation terms can be dropped if the $(i,j) = (y,y)$ component of the tensor equation is considered.  This is easily done leaving the final equation to be tested:
\begin{eqnarray}
\frac{\partial}{\partial r_s}(b^{(3)}_{y,sy} - b^{(3)}_{ys,y}) - \frac{\partial}{\partial r_y}\langle p u_y^\prime \rangle  + \frac{\partial}{\partial r_y}\langle p^\prime u_y \rangle = - 2\alpha b^{(2)}_{y,y}.
\label{eq: Karman-Howarth_yy}
\end{eqnarray}
All of the terms in the above can be measured, and the constant $\alpha$ can be used as a single free fitting parameter.  The quality of the fit will determine if the Karman-Howarth relationship holds in the e-m cell, and therefore by implication the Navier-Stokes equation and linear drag model.  Such a detailed comparison between theory and experiment has not been performed before for 2D soap film systems.

\subsection{The Data}

\begin{figure}
\hskip 1.25in
\includegraphics*[width=3.5in,height=7.31in]{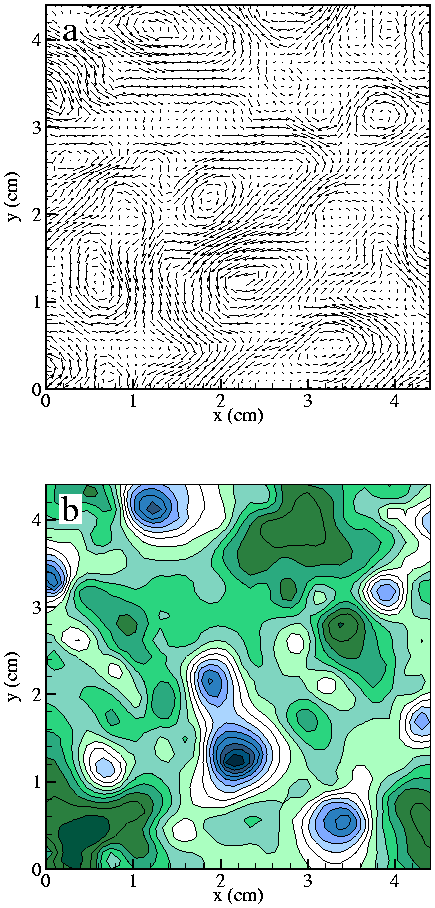}
\caption{Typical velocity (a) and pressure (b) fields obtained from the e-m cell.  In the pressure field green denotes positive and blue negative values.}
\label{fig: typicalfields}
\end{figure}

A single run of the e-m cell using Kolmogorov forcing was performed over a time span of $\sim 300$s during which one thousand vector fields were obtained by PTV.  The cell was driven at a voltage $40$ V with a current of $40$ mA oscillating with a square wave form at $5$ Hz.  The magnet array was placed approximately $1$ mm below the film.  This resulted in velocity fluctuations of around $11$ cm/s over the time of the experiment.  A typical velocity field that is obtained by binning the particle tracks is shown with the associated pressure field derived using the method described above in Fig. \ref{fig: typicalfields}.

\begin{figure}
\hskip 0.25in
\includegraphics*[width=5.5in,height=6.99in]{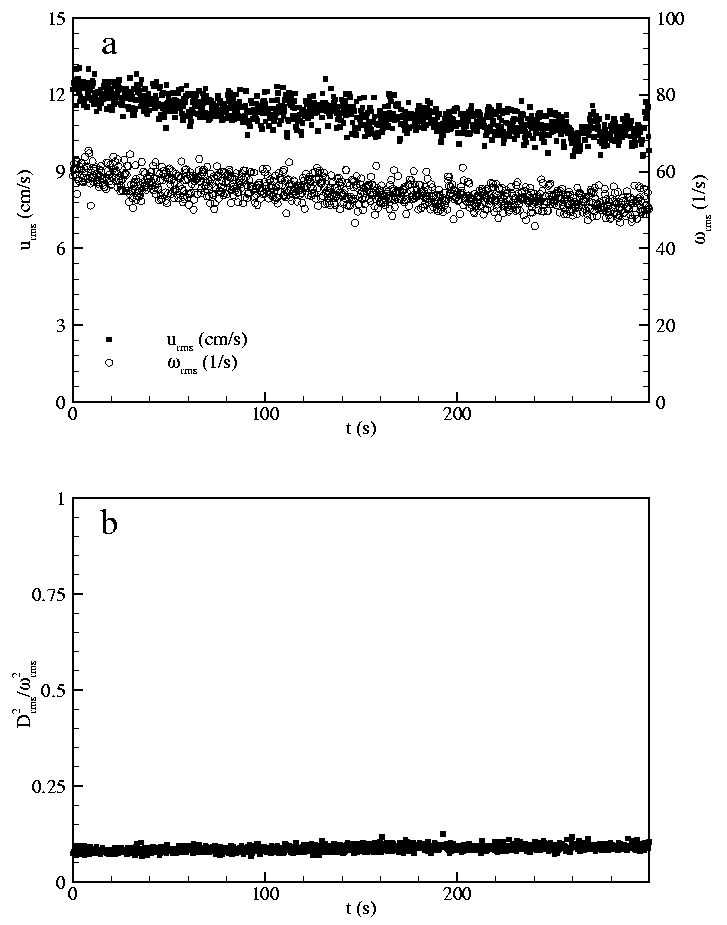}
\caption{(a) Time dependence of $u_{rms}$ and $\omega_{rms}$ for a single run in the e-m cell.  This demonstrates that the e-m cell is in an  approximately steady state. (b) Time dependence of the enstrophy normalized mean square divergence, $D^2/\omega^2_{rms}$, for a single run in the e-m cell.  The fact that $D^2/\omega^2_{rms}$ is small indicates negligible compressibility.}
\label{fig: steadystate}
\end{figure}

The first order of business is to check that the assumptions of incompressibility and that the system is in a steady state are accurate.  Shown in Fig. \ref{fig: steadystate}a is the time dependence of the velocity and vorticity fluctuations for the run.  The fluctuations are not exactly constant, but the change is negligible due to the fact that it happens over a long time, i.e. the average time derivative is approximately zero.  The steady state assumption is therefore approximately correct. The incompressibility assumption can be tested by measuring the divergence of the flow, $D \equiv \vec{\nabla} \cdot \vec{u}$, and normalizing its square by the enstrophy $\Omega \equiv \omega^2_{rms}$.  This forms a dimensionless quantity which must be small if the divergence effect is to be ignorable.  In the e-m cell $D^2/\Omega \approx 0.1$ over the time of the run as shown in Fig. \ref{fig: steadystate}b.  This indicates that the divergence is not overly large and incompressibility can be assumed.  Coincidentally this number is also close to the Mach number of the system, which the reader will recall was kept small for the purpose of minimizing compressibility.
 
\begin{figure}
\hskip 1.25in
\includegraphics*[width=3.5in,height=6.99in]{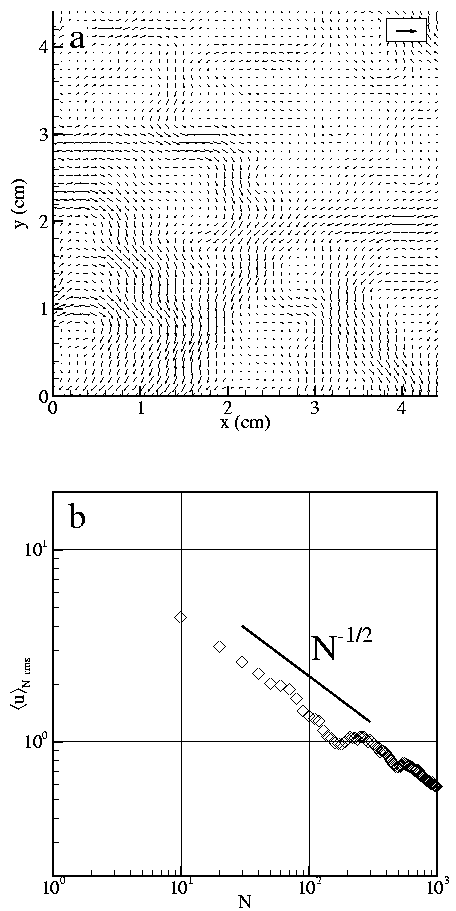}
\caption{(a) The mean flow in the e-m cell averaged over $1000$ vector fields.  The length of the reference vector in the upper right corresponds to $2$ cm/s.  (b)  The decay of the fluctuations in the mean flow as the number of fields, $N$, in the average increases.  The line corresponds to the expected $N^{-1/2}$ decay of a centered Gaussian variable.}
\label{fig: meanflow}
\end{figure}

The final assumption necessary to check before Karman-Howarth is applicable to the system is homogeneity.  That is the average quantities in the turbulence should be invariant with respect to translation.  A crude test of homogeneity is to measure the mean, $\langle \vec{u} (\vec{x})\rangle_N$, and RMS fluctuation, $(\langle |\vec{u}(\vec{x})|^2 \rangle_N)^{\frac{1}{2}}$, of the velocity as a function of position, where $N$ denotes the number of fields the quantity is averaged over.  Both should be independent of position for homogeneity to hold.  Moreover, since the film in the e-m cell does not have a net translation, the mean flow everywhere should be identically zero.  Figure \ref{fig: meanflow}a shows the the mean flow for the run after having averaged over the thousand images ($N=1000$).  One can see that there still exists a small mean.  Though at first this is discouraging, it is misleading since a finite amount of data will almost never converge exactly to zero.  Rather the magnitude of the fluctuations in the mean shear should decrease as $N^{-1/2}$ if one assumes the fluctuations away from the mean are behaving as a centered Gaussian variable.  To this end, the RMS fluctuations of the mean flow, $\langle \vec{u}(\vec{x}) \rangle _N~_{rms}$, is plotted as a function $N$ in Fig. \ref{fig: meanflow}b.  It is clear that the magnitude of the fluctuations in the mean is decreasing almost perfectly as $N^{-1/2}$, indicating that the mean flow as $N \rightarrow \infty$ should go to zero as required by homogeneity.

\begin{figure}
\hskip 1.25in
\includegraphics*[width=3.5in,height=7.31in]{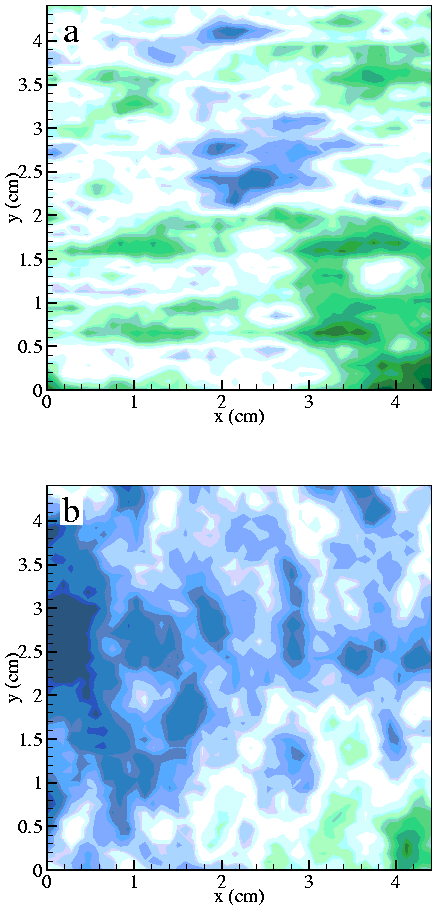}
\caption{The RMS fluctuations of (a) $u_x$ and (b) $u_y$ as a function of position in the e-m cell.  Green denotes large values of the fluctuations while blue denotes small values.}
\label{fig: meanrmsflow}
\end{figure}

Although the mean flow is constant (since it's zero) and satisfies the requirement for homogeneity, the RMS fluctuations, $(\langle |\vec{u}(\vec{x})|^2 \rangle_N)^{\frac{1}{2}}$, does not.  This can be seen by looking at Fig. \ref{fig: meanrmsflow} which shows the RMS fluctuations of the two velocity components averaged over the thousand images.  Although the $\hat{y}$ component of the velocity fluctuations is approximately constant, the $\hat{x}$ fluctuations display strong striations.  These striations reflect the Kolmogorov forcing, as they must if the electromagnetic force is to inject energy into the system.  That is, some part of $u_x$ must be non-random and in phase with the forcing otherwise $\langle \vec{F}\cdot\vec{u} \rangle = 0$ and the e-m cell could not be maintained in an energetically steady state.  Since the forcing is oscillating in time so must this in-phase component; thus it shows up in the RMS fluctuations as a function of position and not in the mean flow.  Fortunately the magnitude of the in phase part of the fluctuations is small, around $1.5$ cm$/$s, compared to the total RMS fluctuations of $12$ cm/s.  Therefore they will be assumed to be ignorable.  Other than these oscillations, the cell appears approximately homogenous in the RMS fluctuations as a function of position.  The assumption of homogeneity can be said to weakly hold for the turbulence in the e-m cell.  This approximation will be refined in later chapters.

\begin{figure}
\includegraphics*[width=6in,height=5.95in]{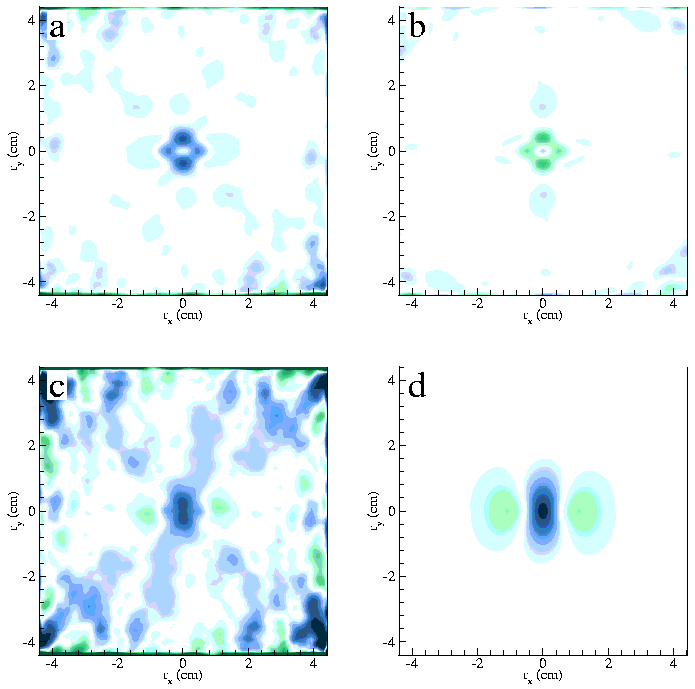}
\caption{Measured values of  (a) $A_{y,y}$, (b) $B_{y,y}$, (c) $A_{y,y} + B_{y,y}$, and (d) $-2 \alpha b^{(2)}_{y,y}$ from Eq. \ref{eq: Karman-Howarth_yy}.  Green denotes positive values and blue denotes negative values.}
\label{fig: karmanhowarth_yy}
\end{figure}

The Karman-Howarth relationship is now in a position to be tested.  For simplicity, define
\begin{eqnarray}
A_{i,j} \equiv \frac{\partial}{\partial r_s}(b^{(3)}_{i,sj} - b^{(3)}_{is,j}),
\end{eqnarray}
\begin{eqnarray}
 B_{i,j} \equiv - \frac{\partial}{\partial r_i}\langle p u_j^\prime \rangle  + \frac{\partial}{\partial r_j}\langle p^\prime u_i \rangle,
\end{eqnarray}
so that the $(y,y)$ component of the Karman-Howarth relationship, Eq. \ref{eq: Karman-Howarth_yy}, may be written $A_{y,y} + B_{y,y} = -2 \alpha b^{(2)}_{y,y}$.  The three separate terms $A_{y,y}$,$B_{y,y}$ and $b^{(2)}_{y,y}$ were measured and a least squares algorithm used to obtain the $\alpha$ value which best fit the measured data to the Karman-Howarth equation.  In this case $\alpha \approx 0.7$ Hz.  The results of these measurements are shown in Fig. \ref{fig: karmanhowarth_yy} a,b,d.  In Fig. \ref{fig: karmanhowarth_yy}c the sum $A_{y,y} + B_{y,y}$ is shown.

\begin{figure}
\hskip 1.5in
\includegraphics*[width=3in,height=6.76in]{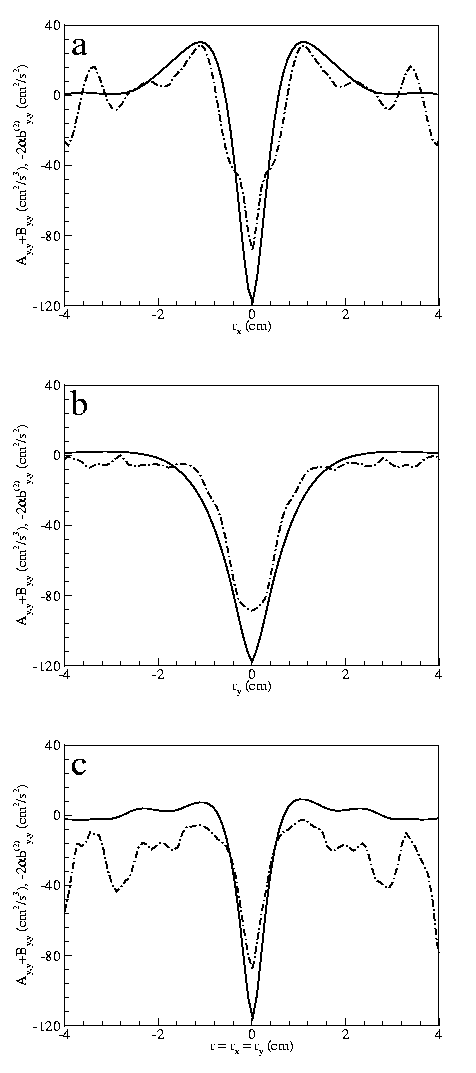}
\caption{Cross sections of $A_{y,y}+B_{y,y}$ ($-\cdot$) and $-2 \alpha b^{(2)}_{y,y}$ ($-$) along the lines (a) $r=r_x$ $(r_y=0)$, (b) $r=r_y$ $(r_x=0)$ and (c) $r = r_x = r_y$.}
\label{fig: karmanhowarth_yy_cross}
\end{figure}

First note that $B_{y,y}$, the term involving pressure velocity correlations is non-zero, as it would be if the turbulence were anisotropic.  This confirms what was earlier assumed to be the case, that the unidirectional forcing does not allow isotropy to be assumed.  Next note that the images in Fig. \ref{fig: karmanhowarth_yy}c and d have very similar forms, namely a central negative trough with two positive peaks on the $r_x$ axis.  This is evidence that the Karman- Howarth relationship is indeed holding in the e-m cell.  To get a better feel for the degree to which there is agreement, several cross-sections of plots c and d are shown in Fig. \ref{fig: karmanhowarth_yy_cross}. The noise in the terms $A_{y,y}$ and $B_{y,y}$ arises from the fact that these terms are derivatives, which are always noisy and converge slowly in experiment.  In spite of the noise there is clearly agreement, and it is therefore concluded that Karman-Howarth, and hence the Navier-Stokes equation with a linear drag, holds for the e-m cell.

A quick check to see if this conclusion is correct is to see if the measured coefficient for the linear drag, $\alpha$, is viable.  Using the discussion in chapter \ref{experimental} the linear drag coefficient can be approximated as $\alpha = \eta/\rho h d$.  Given a $50$ $\mu$m thick film, a magnet-film distance of $1$ mm the coefficient assumes the value $0.36$ Hz.  This value is at least the right order of magnitude, and the difference between this predicted value and the measured one may be accounted for by recalling that the drag on the top surface of the film has been ignored(see chapter \ref{eflux}).

\subsection{Consistency Check: Energy Balance}

The previous section has checked that the $(y,y)$ component of the Karman-Howarth equation is consistent with measurements made in the e-m cell.  However, the fit to the data was somewhat noisy in spite of a thousand fields being used in the average.  What is needed to bolster confidence in the equations is some sort of consistency check.  This is provided by the energy balance relationship, Eq. \ref{eq: ebudget}.

The energy balance statement for the time independent flow simply states that the energy injected into the system by the electromagnetic force, $\epsilon_{inj}$, must be balanced by the energy lost to the air friction, $\epsilon_{air}$, and viscosity, $\epsilon_{\nu}$.  The later two of these can now be measured using the definitions of the various $\epsilon$'s given earlier, the extracted value of $\alpha \approx 0.7$ Hz and the kinematic viscosity of $\nu \approx 0.016$ cm$^2$/s.  The energy dissipated by air is found to be $\epsilon_{air} \approx 85$ cm$^2$/s$^3$, while the energy dissipated by viscous forces is $\epsilon_{\nu} \approx 55$ cm$^2$/s$^3$.  Using energy balance this suggests that the energy injected by the electromagnetic force should be $\epsilon_{inj} \approx 140$ cm$^2$/s$^3$.  An independent measure of $\epsilon_{inj}$ would then yield a consistency check of the measured value of $\alpha$ and the quality of agreement of data to the $(y,y)$ component of the Karman-Howarth relationship.

This check is provided by the $(x,x)$ component of the Karman-Howarth relationship which allows a measure of $\epsilon_{inj}$.  This component of the Karman-Howarth equation has the form
\begin{eqnarray}
A_{x,x} + B_{x,x} = \langle u_x^\prime F_x \rangle  + \langle u_x F_x^\prime \rangle - 2\alpha b^{(2)}_{x,x}.
\label{eq: Karman-Howarth_xx}
\end{eqnarray}
Recall that the Kolmogorov forcing is invariant to translation in the $\hat{x}$ direction, that is, along the forcing.  Let us then restrict the displacement vector $\vec{r}$ to lie along this direction.  Then $\langle u_x^\prime F_x \rangle = \langle u_x^\prime F_x^\prime \rangle$ which by homogeneity equals $\langle u_x F_x \rangle = \langle u_x F_x^\prime \rangle$.  But $\epsilon_{inj} = \langle u_x F_x \rangle$.  Using these relationships in Eq. \ref{eq: Karman-Howarth_xx} yields $A_{x,x} + B_{x,x} = 2\epsilon_{inj} - 2\alpha b^{(2)}_{x,x}$ along the $\vec{r} = r_{x}$ cross-section.  Thus the plots of $A_{x,x} + B_{x,x}$ and $2\alpha b^{(2)}_{x,x}$ should look similar to the ones shown earlier except offset by a constant which is equal to $2\epsilon_{inj}$.

\begin{figure}
\hskip 1.5in
\includegraphics*[width=3in,height=2.16in]{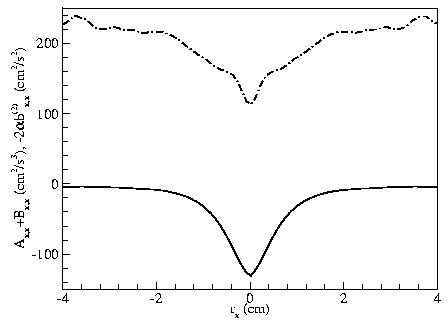}
\caption{Cross section of $A_{x,x} + B_{x,x}$ ($-\cdot$) and $-2 \alpha b^{(2)}_{x,x}$ ($-$) along the line $r = r_x$ ($r_y = 0$).}
\label{fig: karmanhowarth_xx_cross}
\end{figure}

Figure \ref{fig: karmanhowarth_xx_cross} shows the plot of the $r_x$ cross-section of the  $(x,x)$ components of $A_{x,x} + B_{x,x}$ and $-2\alpha b^{(2)}_{x,x}$.  Clearly the plots are offset by  positive constant of $2\epsilon_{inj} \approx 240$ cm$^2$/s$^3$.  Thus $\epsilon_{inj} \approx 120$ cm$^2$/s$^3$ which is close to the expected value of $140$ cm$^2$/s$^3$.  Systematic data discussed in the next chapter will show that the extracted values of $\alpha$ and $\epsilon_{inj}$ fluctuate around $\pm 20$\%, thus the measured value of $\epsilon_{inj}$ is within experimental error of the expected value.  This is further supporting evidence that the Karman-Howarth relationship does indeed work for data extracted from the e-m cell.

\chapter{Energy Distribution and Energy Flow\label{eflux}}

In the last chapter the energy balance relationship, $\epsilon_{inj} - \epsilon_{\nu} - \epsilon_{air} = 0$, was used as a consistency check to determine if the Karman-Howarth relationship was  applicable to data from the e-m cell.  Though it was not discussed there, these measurements are the first indication that an inverse cascade is present.  Recall from the discussion in chapter \ref{intro} that if an inverse cascade exists then energy is moved from small length scales to large ones, away from length scales at which viscous dissipation is effective.  Thus the bulk of the energy is expected to be dissipated by the external dissipation mechanism acting at large scales, which in the case of the e-m cell is the air dissipation.  This is exactly the result that the measured energy rates in the e-m cell demonstrate, i.e. $\epsilon_{air} > \epsilon_{\nu}$.  This chapter begins the investigation of the inverse energy cascade in the e-m cell by quantifying the length scales over which it exists (it's range), measuring how the energy is distributed over these length scales and determining the rate at which energy flows through these length scales.

\section{Distribution of Energy and the Outer Scale}

The energy spectrum, $U(\vec{k})$, provides a means for describing the manner in which turbulent kinetic energy is distributed over different wavenumbers (inverse length scales) in the e-m cell.  It is defined as the average square modulus of the Fourier transform of the velocity fluctuations, $U(\vec{k}) \equiv \langle \tilde{u}(\vec{k}) \tilde{u}^\dagger (\vec{k}) \rangle$\footnote{$\dagger$ denotes a complex conjugate.}, and it's circular integral, $E(k) \equiv \int_0^{2\pi} |k|U(\vec{k})d\theta$, denotes the average amount of kinetic energy stored in wavenumbers of modulus $k$.  If an inverse cascade is present in the e-m cell, the expectation is that energy would build up in wavenumbers smaller than the energy injection wavenumber $k_{inj}$.

\begin{figure}
\includegraphics*[width=6in,height=5.91in]{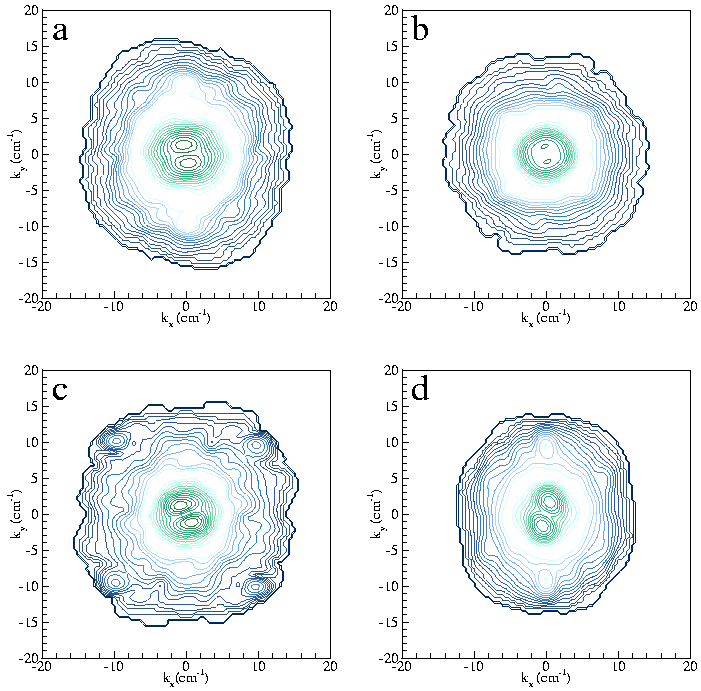}
\caption{The energy spectrum, $U(\vec{k})$, for (a) Kolmogorov forcing, (b) square forcing using $6$ mm round magnets, (c) square forcing using $3$ mm round magnets and (d) stretched hexagonal forcing.  Green denotes large values of $U(\vec{k})$ while blue denotes small values.}
\label{fig: var2Dspectra}
\end{figure}

\begin{figure}
\includegraphics*[width=6in,height=5.92in]{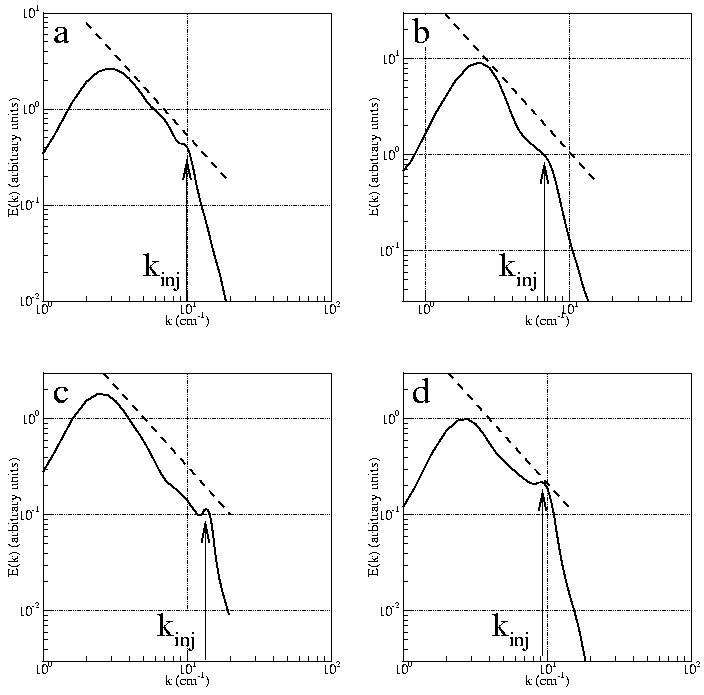}
\caption{The circularly integrated energy spectrum, $E(k)$, for (a) Kolmogorov forcing, (b) square forcing using $6$ mm round magnets, (c) square forcing using $3$ mm round magnets and (d) stretched hexagonal forcing.  The dashed lines correspond to the Kraichnan prediction that $E(k) \propto k^{-5/3}$ \cite{Kraichnan:PFL67}.  The arrows indicate the injection wavenumber $k_{inj}$.}
\label{fig: varspectra}
\end{figure}

This expectation proves to be the case in the e-m cell.  Fig. \ref{fig: var2Dspectra} are plots of $U(\vec{k})$ calculated from transforms of the PTV velocity fields for various types of driving force in the e-m cell.  Note that the peaks corresponding to the injection wavenumber differ as expected for the different types of forcing.  For example the two square arrays produce peaks along the line $k_x = \pm k_y$, with the distance from $0$ being dictated by the size of the magnets used in the array.  Also note that energy contained in small wavenumbers is greater than that contained in large wavenumbers, as expected for an inverse cascade.  This can be better seen in Fig. \ref{fig: varspectra} where the circular integrals have been taken and a build up of energy at wavenumbers smaller than $k_{inj}$ can be seen.

The Kraichnan prediction for the inverse energy cascade range is that $E(k) \sim \epsilon^{2/3}k^{-5/3}$ for $k<k_{inj}$ and $\epsilon$ a typical energy rate (in the case of the e-m cell this is $\epsilon_{air}$) \cite{Kraichnan:PFL67}.  This result is consistent with dimensional predictions for the scaling behavior of $E(k)$.  Lines corresponding to this prediction have been drawn on the plots in Fig. \ref{fig: varspectra}.  These lines should not be interpreted as a fit to the data and are drawn only as a guide.  Only the Kolmogorov data set, (a), appears to be in qualitative agreement with the dimensional prediction over slightly less than half a decade of wavenumbers below the injection wavenumber.  All the remaining data sets have a small range directly below the injection wavenumber which could be interpreted as $k^{-5/3}$, but this is quickly lost to a broad peak in the spectrum at small $k$.  This type of behavior in the spectrum is in agreement with results reported in \cite{Paret:PFL97} where the build up of energy at small $k$ is associated with the saturation of energy in the largest length scales in the system.  Such saturation is not included in the Kraichnan prediction and is therefore not indicative of failure of the theory, but rather a failure of the system to satisfy the assumptions of the theory\footnote{The assumption which is violated in a saturated system is that the velocity fluctuations are homogenous.  A saturated system occurs when the outer scale which is determined by external dissipation exceeds the system size, as it did for the data in Fig. \ref{fig: varspectra}(b)-(d).  When this happens large structures attempt to pack into a small area near the center of the system away from system boundaries.  This will be investigated in some detail later in this chapter.}.  Since case (a) satisfies the assumptions of Kraichnan's theory one might conclude that the $k^{-5/3}$ prediction is correct.  However, it should also be noted that the behavior in the measured $E(k)$ depends on the type of window function used when Fourier transforming the velocity fields.  In the data shown in the Fig. \ref{fig: varspectra}, a three term Blackman-Harris window has been used.  By changing this window one could get up to a $20$\% change in the slope of the energy spectra.  Due to the limitations imposed by windowing no conclusion may be drawn from this data about possible corrections to the Kraichnan scaling prediction.

The low $k$ limit of the inverse cascade range, denoted $k_{out}$, is determined by the size of the largest vortices which result from the inverse cascade, and corresponds to the low $k$ peak in $E(k)$.  The position of this peak should depend on the strength of the air friction since this is the large scale external dissipation mechanism in the e-m cell.  In the chapter \ref{ebudg} the linear damping model for the air friction possessed a coefficient $\alpha$ which determines it's strength.  Using dimensional analysis, $\alpha$, and $\epsilon_{inj}$, a length scale called the outer scale can be calculated by $r_{out} \equiv (\epsilon_{inj}/\alpha^3)^{1/2}$.  The outer scale represents the size of the vortices at which more energy is lost to air friction than is transferred to the next size larger vortices.  The outer scale should be related to the low $k$ peak in $E(k)$ by $k_{out} = 2\pi/r_{out}$.

To check this dimensional prediction, a systematic set of data using the Kolmogorov forcing with various magnet-film distances was taken holding $u_{rms}$ approximately constant.  Kolmogorov forcing was used so that $\alpha$ and $\epsilon_{inj}$ could be extracted using the techniques of chapter \ref{ebudg}.  Between $400$ and $500$ vector fields where obtained for each magnet-film distance.  As in chapter \ref{ebudg}, the energy in the e-m cell remained approximately constant during the data acquisition time for each run.  Table \ref{tab: constants} lists the various constants associated with each of the different data sets.

The first four data sets listed in Table \ref{tab: constants} may be compared for the purpose of error analysis.  The first and second of these were obtained using identical values of the external control parameters and thus the extracted values of $\alpha$ and $\epsilon_{inj}$ should be identical.  This is found to be the case up to two significant figures for the first two data sets.  The third and fourth data sets are also taken under identical control conditions different from those used for the first and second data sets (the magnet-film distance was slightly smaller).  In these data sets the values of $\alpha$ and $\epsilon_{inj}$ vary around the mean by about $10$\%.  Thus, to be conservative, the error in the two quantities $\alpha$ and $\epsilon_{inj}$ will be assumed to be as much as 20\% of the measured value.

\begin{figure}
\hskip 0.5in
\includegraphics*[width=5in,height=3.28in]{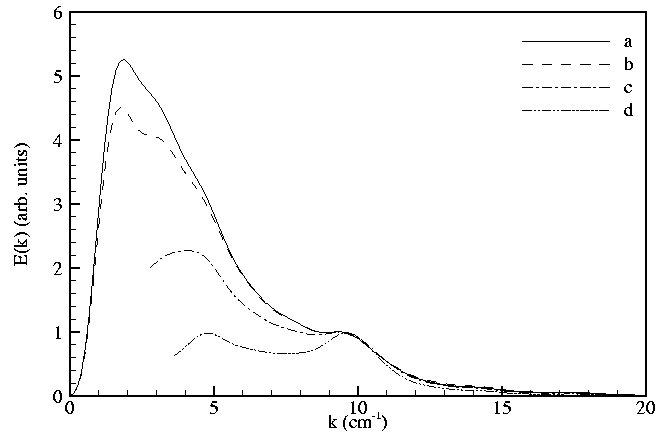}
\caption{The circularly integrated energy spectrum, $E(k)$, for the four cases of Kolmogorov flow labeled in Table \ref{tab: constants}.}
\label{fig: spectra}
\end{figure}

\begin{figure}
\hskip 0.5in
\includegraphics*[width=5in,height=3.41in]{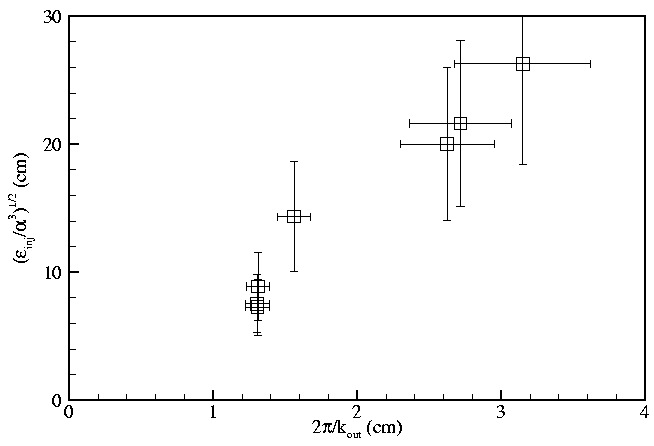}
\caption{Comparison of the outer scale obtained from the energy spectra by $r_{out} = 2\pi/k_{out}$ with that obtained using the dimensional prediction $r_{out} = (\epsilon_{inj}/\alpha^3)^{1/2}$ for all of the data sets in Table \ref{tab: constants}.}
\label{fig: outerscale}
\end{figure}

The $E(k)$ measured from the data sets labeled a,b,c and d are displayed in Fig \ref{fig: spectra}.  The low wavenumber peak, $k_{out}$, moves to smaller and smaller wavenumber as $\alpha$ decreases.  This is in qualitative agreement with the dimensional prediction.  Using the position of this peak the outer scale is calculated by $r_{out} = 2 \pi/k_{out}$ and compared to that obtained from the dimensional prediction using the measured values of $\alpha$ and $\epsilon_{inj}$.  The results of this comparison are shown for all the data sets in Fig. \ref{fig: outerscale}.  The vertical error bars reflect the propagated error of $\epsilon_{inj}$ and $\alpha$ while the horizontal represent the discretization inherent in finite Fourier transforms.  From Fig. \ref{fig: outerscale} one can see that the dimensional prediction of the outer scale is not inconsistent with the measured outer scale using the low $k$ peak in $E(k)$.

\begin{table}
\label{tab: constants}
\begin{tabular}{||c|c|c|c|c|c|c||}\hline\hline
$\epsilon_{inj}$ (cm$^2/$s$^3$) & $\alpha$ (s$^{-1}$) & $u_{rms}$ (cm$^2/$s$^2$)& $\omega^2_{rms}$ (s$^{-2}$) & $2\pi/k_{out} (cm)$& $r_{int} (cm)$ &case\\ \hline
63 & 0.45 & 7.81 & 2049 & 3.14 &0.63&a\\
63 & 0.45 & 7.80 & 2014 & 3.14 &0.64&\\
101 & 0.6 & 8.96 & 2901 & 2.72 &0.59&b\\
110 & 0.65 & 9.18 & 3103 & 2.63 &0.58&\\
150 & 0.9 & 9.18 & 3579 & 1.56 &0.51&c\\
154 & 1.25 & 7.81 & 3211 & 1.31 &0.41&\\
197 & 1.55 & 7.95 & 3567 & 1.31 &0.38&d\\ \hline \hline
\end{tabular}
\caption{Global constants for several runs of the e-m cell using Kolmogorov forcing}
\end{table}

The measurements presented here clearly indicate that an inverse cascade is present in the e-m cell for all types of forcing.  The inverse cascade range is shown to exist over wavenumbers $k$ such that $k_{out}<k<k_{inj}$, where $k_{inj}$ is the energy injection wavenumber determined by the electromagnetic forcing and $k_{out}$ is the outer wavenumber determined by the external dissipation.  $k_{out}$ is found to be not inconsistent with the dimensional prediction, $k_{out} \sim (\epsilon_{inj}/\alpha^3)^{1/2}$.  No strong conclusion can be draw about the manner in which energy is distributed over this range due to windowing difficulties, however the Kraichnan prediction of $E(k) \sim k^{-5/3}$ superficially holds for data sets in agreement with the assumptions of the prediction.

\begin{figure}
\hskip 1.5in
\includegraphics*[width=3in,height=2.89in]{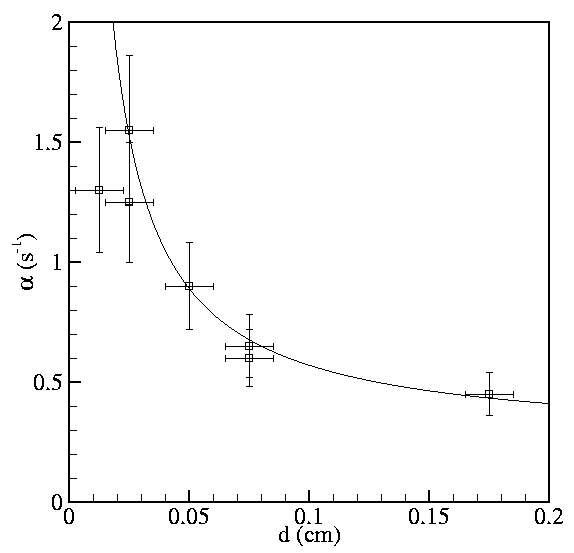}
\caption{The measured linear drag coefficient, $\alpha$, versus the magnet-film distance, $d$, for the data sets reported in Table \ref{tab: constants}.  The dotted line represents the fit $\alpha = \eta_{air}/\rho h d + C$ with $C=0.25$ Hz.}
\label{fig: alphavsd}
\end{figure}

Before leaving this section, the systematic data set used in obtaining the outer scale allow the external dissipation to be compared to the predicted value $\alpha = \eta_{air}/\rho h d$.  The measured values of $\alpha$ versus the magnet-film distance are shown in Fig. \ref{fig: alphavsd}.  The vertical error bars again reflect the $20$\% error in measurement while the horizontal error bars denote the limit of control over magnet-film distance.  A line corresponding to $\eta_{air}/\rho h d + C$, where $C = 0.25$ Hz is also plotted with the data and for the most part is within error of the measured values.  According to this data the prediction for the magnitude of the linear dissipation must be offset by a small positive constant to be accurate.  This constant can be accounted for by recalling that the effect of air friction on the top surface of the film has been ignored.  Approximations of the frictional force on the top surface of the film indicate that the measured value of the offset is appropriate, though more experimentation needs to be done to more accurately account for this offset.

\section{Energy Flow}

A simplified viewpoint of how kinetic energy might be transfered through the inverse cascade range in the e-m cell is to imagine that the energy is flowing like a liquid through a pipe.  Energy produced by the forcing is poured in at one end of the pipe which characterizes the injection scale.  It is then moved along the pipe to larger length scales by the mixing of fluid (i.e. by vortex cannibalization). Finally energy is exhausted from the pipe at the opposite end which characterizes the outer length scale.  Since the total energy in the system is constant, the amount of energy poured into the pipe must be equal to the amount exhausted from the pipe.  It is also expected that the rate of energy being poured into the pipe is equivalent to the rate of energy transferred across any length scale in the middle of the pipe.  That is to say that the energy flux through the pipe is not dependent on the position in the pipe and remains constant.

This viewpoint is a severe simplification of what actually happens in 2D turbulence.  First it ignores losses of energy to viscous dissipation and assumes all energy is lost to external dissipation.  Since viscous losses are presumably smaller than losses to external dissipation, let us accept for now that this simplification is valid.  A more important simplification, the one which is of concern in this section, is that the pipe doesn't leak.  That is there are no holes drilled along the length of the pipe.  That is to say energy cannot be exhausted from the pipe by external dissipation until the outer length scale is reached.  To hope that this is actually the case in the e-m cell, or for that matter any other laboratory 2D turbulence system is quite a stretch.

It is more likely that there exists a range in the pipe, probably close to the injection scale since external dissipation increases at large length scales, over which the amount of energy lost to leaks is negligible compared to the energy flux through the pipe.  This range will be called an ``inertial" range since the energy flow is almost entirely dictated by the fluid's inertia and not the energy dissipation.  The inertial range is of interest due to it's universality.  Presumably two 2D turbulent systems with completely different external dissipation mechanisms will behave identically in their inertial ranges.  To use the pipe analogy, the inertial range of both systems is completely closed and thus fluid mixing and energy flux should be the same in this region.  Outside of this range the pipe leaks, and the energy flux depends on how many holes of what size are drilled at what position in the pipe. Therefore the characteristics of the turbulence may not be universal outside of the inertial range.

To determine if a range is inertial or not, a measurement of energy flow must be made.  One way in which energy flow may be characterized is by the third moment of velocity difference.  The third moment, labeled $S^{(3)}(r)$ for now, can be thought of as the average energy per unit mass advected over a circle of radius $r$ centered at $\vec{x}$ per unit circumference of the circle:
\[
S^{(3)}(r) = \frac{1}{2 \pi r}\int_0^{2\pi} \langle E(\vec{x}+\vec{r}) \vec{u}(\vec{x}+\vec{r})\cdot\vec{r} \rangle d\theta.
\]
Assume also that the reference frame is moving so that the velocity $\vec{u}(\vec{x})=0$.  Assuming homogeneity the circle can be placed anywhere in the turbulence, so that $S^{(3)}$ does not depend on $\vec{x}$.  $E$ in the above is the energy per unit mass.  If the third moment is positive, then on average energy is being advected from inside the circle to outside the circle, and vice versa if the third moment is negative.  It is interesting to note what happens if  $r$ is assumed to be in an inertial range.  If this is the case then no energy is lost to external dissipation at that length scale.  Since the energy held in the turbulence at a scale $r$ is in a steady state then all of the energy injected by the forcing into the circle must be advected over the surface of the circle.  Call $\epsilon$ the rate of energy injection per unit mass into the system.  The rate at which energy is injected into the circle is then $\pi r^2 \epsilon$.  Replacing the integral with this yields $S^{(3)}(r) \approx \epsilon r$.  Thus a linear range in the third moment indicates inertial behavior of the energy transfer.

The above derivation of a linear behavior in the third moment for an inertial range can be put on much more solid foundation.  Indeed the third moment is one of the few quantities for which an exact prediction can be derived from the Navier-Stokes equation.  This derivation was first done by Kolmogorov for 3D homogenous and isotropic turbulence.  To apply to results from the e-m cell Kolmogorov's derivation must be relaxed to the case of 2D homogenous but anisotropic turbulence with an external linear drag.  This relaxation is given below.  Following this are measurement and analysis of the third moment in the e-m cell for the data sets in Table \ref{tab: constants}.

\subsection{The Anisotropic Third Moment}

The starting point for the derivation of the third moment relationship for homogenous anisotropic 2D turbulence is the Karman-Howarth relationship, Eq. \ref{eq: Karman-Howarth}, which was derived in section \ref{karmanhowarth}.  All of the notation and conventions used in that section will be carried over without alteration.  The notation and the first step of the derivation follows that given in a recent paper by Lindborg \cite{Lindborg:JFM96}.

Add Eq. \ref{eq: Karman-Howarth} evaluated for $(i,j)$ to that evaluated for $(j,i)$, and use Eq. \ref{eq: ebudg1} to obtain
\begin{eqnarray}
2\Pi_{ij} - 2\epsilon^{(\nu)}_{ij} & = & \frac{\partial}{\partial r_s}B^{(3)}_{isj} + \frac{\partial}{\partial t}B^{(2)}_{ij} - \frac{\partial}{\partial r_j}P_i - \frac{\partial}{\partial r_i}P_j \nonumber \\ & &- 2\nu\frac{\partial^2}{\partial r_s \partial r_s}B^{(2)}_{ij}  + 2\alpha B^{(2)}_{ij} - W_{ij} - W_{ji}.
\label{eq: KarmanHowarth2}
\end{eqnarray}
In the above, the $n$ term moments $B^{(n)}_{ij...k}(\vec{r}) \equiv \langle \delta u_i \delta u_j ... \delta u_k \rangle$ of velocity difference $\delta u_i \equiv u_i^\prime - u_i$ have been defined. Also defined are $P_i \equiv \langle u_i p^\prime \rangle - \langle u_i^\prime p \rangle$ and $W_{ij} \equiv \langle \delta u_i \delta F_j\rangle$ with $\delta F_j \equiv F_j^\prime - F_j$. Note that the homogeneity assumption has been used in a number of places in the above step. Most notably, it sets
\begin{eqnarray}
\frac{\partial}{\partial r_s}(b^{(3)}_{i,sj} - b^{(3)}_{is,j}) = \frac{\partial}{\partial r_s}B^{(3)}_{isj}.
\label{eq: homogreq1}
\end{eqnarray}

Contracting Eq. \ref{eq: KarmanHowarth2} with the unit vectors $n_i (\equiv r_i/r = r_i/|\vec{r}|)$ and $n_j$ and using the identities
\begin{eqnarray}
n_in_jB^{(2)}_{ij} = B^{(2)}_{rr},
\end{eqnarray}
\begin{eqnarray}
n_i n_j \frac{\partial}{\partial r_s} B^{(3)}_{isj} = \frac{\partial}{\partial r_s}(n_i n_j B^{(3)}_{isj}) - \frac{2}{r}B^{(3)}_{rtt},
\end{eqnarray}
\begin{eqnarray}
n_i n_j \frac{\partial}{\partial r_j}P_i = \frac{\partial}{\partial r_j}(n_i n_j P_i) - \frac{1}{r}n_i P_i,
\end{eqnarray}
\begin{eqnarray}
n_i n_j \frac{\partial^2}{\partial r_s \partial r_s} B^{(2)}_{ij}&=& \frac{2}{r^2}(B^{(2)}_{rr} - B^{(2)}_{tt}) + \frac{4}{r} \frac{\partial}{\partial r} B^{(2)}_{rr} \nonumber \\ & &+ \frac{\partial^2}{\partial r_s \partial r_s}B^{(2)}_{rr},
\end{eqnarray}
where the subscripts $r$ and $t$ denote longitudinal, i.e. along $\vec{r}$, and transverse directional coordinates, we obtain
\begin{eqnarray}
n_i n_j (\Pi_{ij} - \epsilon^{(\nu)}_{ij}) - \frac{1}{2}\frac{\partial}{\partial t}B^{(2)}_{rr} &=& \frac{1}{2}\frac{\partial}{\partial r_s} (n_i n_j B^{(3)}_{isj}) - \frac{1}{r}B^{(3)}_{rtt} - \frac{\partial}{\partial r_j} (n_i n_j P_i) - \frac{1}{r} n_i P_i \nonumber \\ & & -\nu\left( \frac{2}{r^2}(B^{(2)}_{rr} - B^{(2)}_{tt}) + \frac{4}{r}\frac{\partial}{\partial r} B^{(2)}_{rr} + \frac{\partial^2}{\partial r_s \partial r_s}B^{(2)}_{rr}\right)\nonumber \\ & &+\alpha B^{(2)}_{rr} - \frac{1}{2}n_i n_j (W_{ij} + W_{ji}).
\label{eq: KHcontract}
\end{eqnarray}
Eq. \ref{eq: KHcontract} is now in a form which may be easily integrated over a circle of radius $r$. This procedure, along with incompressibility and the assumption of homogeneity eliminates the pressure terms.  Using the divergence theorem and rearranging the terms yields
\begin{eqnarray}
S^{(3)}_{rrr}(r) - \frac{2}{r}\int_0^r dr^\prime S^{(3)}_{rtt}(r^\prime) &=& - \epsilon_{\nu}r - \frac{1}{r}\int_0^r dr^\prime  r^\prime \frac{\partial}{\partial t}S^{(2)}_{rr}(r^\prime)  - \frac{2 \alpha}{r}\int_0^r dr^\prime  r^\prime S^{(2)}_{rr}(r^\prime) \nonumber \\ && + 2\nu \left( \frac{4}{r}S^{(2)}_{rr} + \frac{\partial}{\partial r}S^{(2)}_{rr} + \frac{2}{r}\int_0^r dr^\prime (S^{(2)}_{rr}(r^\prime) - S^{(2)}_{tt}(r^\prime))/r^{\prime}\right) \nonumber \\ &&+ \frac{1}{2 \pi r}\int_0^r dr^\prime \int_0^{2\pi} r^\prime d\theta n_i n_j(W_{ij}(\vec{r^{\prime}})  + W_{ji}(\vec{r^{\prime}})).
\end{eqnarray}
Here the circular averages of velocity moments have been denoted as $S^{(n)}_{ij..k}(r) \equiv \frac{1}{2 \pi r}\int_0^{2 \pi} r d\theta B^{(n)}_{ij...k}(\vec{r})$. Though this equation seems to explicitly contain an $\epsilon r$ term, it is somewhat superficial as this term exactly cancels with  terms contained in the first and final expressions on the right hand side. Removing these terms yields
\begin{eqnarray}
S^{(3)}_{rrr}(r) - \frac{2}{r}\int_0^r dr^\prime S^{(3)}_{rtt}(r^\prime) &=& - \frac{1}{2 \pi r}\int_0^r dr^\prime \int_0^{2\pi} r^\prime d\theta n_i n_j \langle u_i^\prime F_j + u_iF_j^\prime + u_j^\prime F_i + u_j F_i^\prime \rangle \nonumber \\ && + 2\nu \left( \frac{4}{r}S^{(2)}_{rr} + \frac{\partial}{\partial r}S^{(2)}_{rr} + \frac{2}{r}\int_0^r dr^\prime (S^{(2)}_{rr}(r^\prime) - S^{(2)}_{tt}(r^\prime))/r^{\prime}\right) \nonumber \\ && + \left(\frac{1}{\pi r} \frac{\partial}{\partial t} + \frac{2 \alpha}{\pi r}\right) \int_0^r dr^{\prime}\int_0^{2\pi}r^{\prime}d\theta b^{(2)}_{r,r}(\vec{r^{\prime}}).
\label{eq: thirdmom_beforeapprox}
\end{eqnarray}
Note that using the notation introduced here $b^{(2)}_{r,r}(\vec{r})$ is simply the two-point longitudinal velocity correlation.  The terms on the left hand side are defined as the anisotropic third moment of velocity difference, $S^{(3)}_a$, and up to a constant play the role of the third moment of the longitudinal velocity difference in the fully developed isotropic-homogeneous turbulence derived by Kolmogorov \cite{Frisch}. The terms on the right account for the energy flux at some length scale due to external forces. Eq. \ref{eq: thirdmom_beforeapprox} is essentially the scale-by-scale energy balance relationship for the system.  For large $r$ the viscous term in Eq. \ref{eq: thirdmom_beforeapprox} can be ignored, as can the time derivative if the system is in an energetically steady state, leaving the final form which will be used in this thesis
\begin{eqnarray}
S^{(3)}_{rrr}(r) - \frac{2}{r}\int_0^r dr^\prime S^{(3)}_{rtt}(r^\prime) &=& - \frac{1}{2 \pi r}\int_0^r dr^\prime \int_0^{2\pi} r^\prime d\theta n_i n_j \langle u_i^\prime F_j + u_iF_j^\prime + u_j^\prime F_i + u_j F_i^\prime \rangle \nonumber \\ && + \frac{2 \alpha}{\pi r}\int_0^r dr^{\prime}\int_0^{2\pi}r^{\prime}d\theta b^{(2)}_{r,r}(\vec{r^{\prime}}).
\label{eq: thirdmom}
\end{eqnarray}

Limits of this equation are now ready to be taken to establish that a linear range can exist.  First note that the force-velocity correlation term in Eq. \ref{eq: thirdmom} (first term on the right) should decay in magnitude as $1/r$ for $r\gg r_{inj}$ for $F$ periodic in space.  One limit which can be considered is the case where this force-velocity term is neglected.  In this case the only remaining term arises from the linear dissipation.  Assuming $S^{(2)}_{rr}(r) \propto r^{2/3}$ over some range of length scales for $r>r_{inj}$, then $b^{(2)}_{r,r}(r) = u^2_{rms} - S^{(2)}_{rr}(r)/2 \propto u^{2}_{rms} - Ar^{2/3}$ in that range, where A is a constant. Since the longitudinal velocity correlation, $b^{(2)}_{r,r}(r)$, remains positive the first term is dominant.  Thus $b^{(2)}_{r,r}(r) \approx u^{2}_{rms}$ in this range.  Inserting this approximation into Eq. \ref{eq: thirdmom} and integrating yields $S_a^{(3)} = 2\alpha u^2_{rms} r = 2 \epsilon_{air}r$, which is the extension of the earlier mentioned Kolmogorov $4/5$ result for 3D turbulence to 2D anisotropic turbulence.  Another limit of interest is when both the force-velocity and velocity-velocity correlations have disappeared.  In this case the integrals on the right hand side become constant and the third moment decays as $S^{(3)}_a \approx r^{-1}$.

Notice that the $\epsilon$ found in the third moment relationship is $\epsilon_{air}$, the energy dissipated by the e-m cell's external dissipation mechanism of air friction.  One might have expected that this should be $\epsilon_{inj}$, the total energy injection rate.  Recall that in the previous discussion of the third moment, energy lost to viscous forces were ignored so that $\epsilon_{inj} = \epsilon_{air}$.  When viscosity is reintroduced, then the energy flowing over a circle of radius $r$ is the energy injected less the amount dissipated by viscosity, i.e. $\epsilon_{inj} - \epsilon_{\nu}$.  By energy conservation this is just $\epsilon_{air}$, which dictates the remaining energy which must flow over the circle.  This is why $\epsilon_{air}$ determines the third moment and not $\epsilon_{inj}$.

\subsection{Homogeneity}

Before sliding headlong into $S^{(3)}_a$ measurements and blindly searching for positive linear ranges, a word of caution is warranted.  The assumption of homogeneity has been used a number of times in the preceding derivation of $S^{(3)}_a$, not to mention the fact that it was used in deriving the Karman-Howarth equation.  This makes $S^{(3)}_a$ measurements extremely sensitive to inhomogeneity in the e-m cell.\footnote{There might be some concern that results presented in chapter \ref{ebudg} might be inaccurate due to inhomogeneity.  Recall that in that chapter homogeneity was assumed and not exactly checked.  Using the analysis presented in this section on the data of chapter \ref{ebudg} demonstrates that these data sets are approximately homogenous.}

\begin{figure}
\includegraphics*[width=6in,height=6.05in]{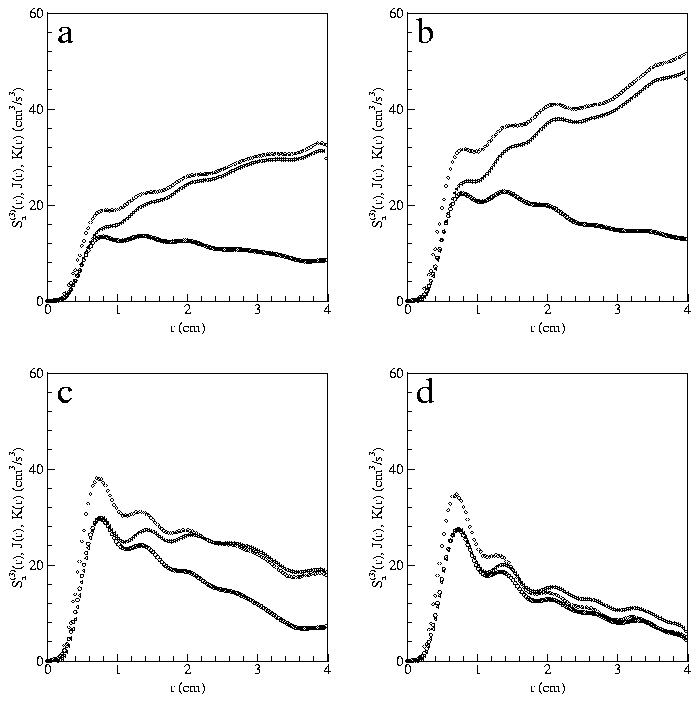}
\caption{Comparison of $S^{(3)}_a(r)$ ($\diamond$),J(r) ($\triangleleft$), and $K(r)$ ($\circ$) for the data sets labeled (a)-(d) in Table \ref{tab: constants}.}
\label{fig: homog1}
\end{figure}

Fortunately the derivation of $S^{(3)}_a$ has provided a simple test of homogeneity.
If the turbulence in the e-m cell is homogenous then the following should be equivalent representations for the anisotropic third moment, $S^{(3)}_a$:
\begin{eqnarray}
S^{(3)}_a(r) &\equiv& S^{(3)}_{rrr}(r) - \frac{2}{r}\int_0^rdr S^{(3)}_{rtt}(r),\\
&=&\frac{1}{2\pi r}\int_0^rdr\int_0^{2 \pi}d\theta  n_i n_j \frac{\partial}{\partial 	x_s}B^{(3)}_{isj}(\vec{r}),\\
&=&\frac{1}{2\pi r}\int_0^r dr\int_0^{2 \pi}d\theta  n_i n_j \frac{\partial}{\partial x_s}(b^{(3)}_{i,sj}(\vec{r}) - b^{(3)}_{is,j}(\vec{r})).
\label{eq: third_equiv}
\end{eqnarray}
The latter two forms will be denoted $J(r)$ and $K(r)$, respectively. Plots of all three of these quantities for the data sets labeled (a)-(d) in Table \ref{tab: constants} are shown in Fig. \ref{fig: homog1}. It is clear that only case (d) of extremely heavy damping (large $\alpha$) produces approximate agreement for all three forms at length scales larger than the injection scale.  It is therefore the only approximately homogenous data set. It is also evident that case (a) and (b) are strongly inhomogeneous, and case (c) is marginal.

\begin{figure}
\includegraphics*[width=6in,height=6in]{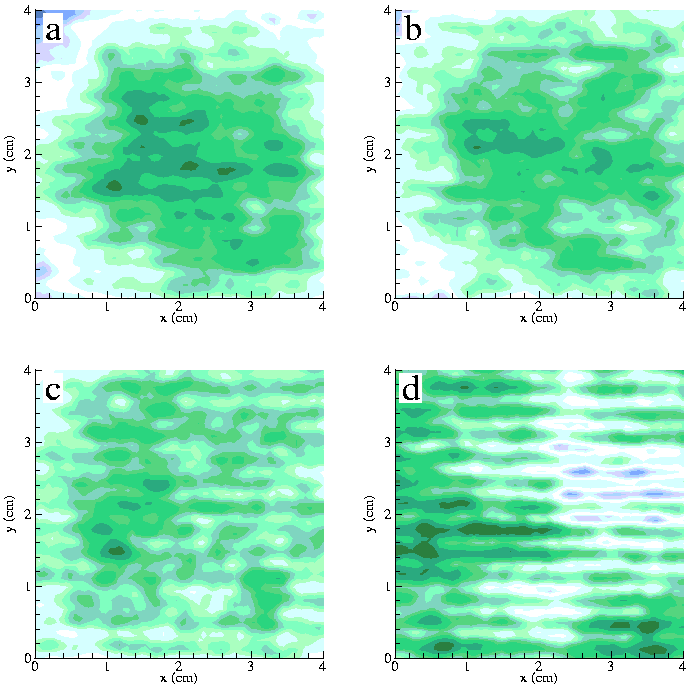}
\caption{Spatial variation of velocity fluctuation for the four data sets labeled in (a)-(d) Table \ref{tab: constants}.  Green denotes large values of the fluctuations while blue denotes small values.}
\label{fig: rmsfluc}
\end{figure}

To understand the nature of the discrepancies, ensemble averages of individual velocity fields were performed over $N$ images. Again the ensemble averaged velocity at any given point in the flow, $\langle \vec{u} (\vec{x})\rangle_N$,  though not identically zero was found to decrease in magnitude as $N^{-1/2}$ for all of the data sets indicating negligible mean flow in the system.  The inhomogeneity, then, stems from the spatial variation of the velocity fluctuations, $(\langle |\vec{u}(\vec{x})|^2 \rangle_N)^{\frac{1}{2}}$,  which is shown in Fig. \ref{fig: rmsfluc} for the four data sets (a)-(d). The oscillating light and dark bands, corresponding to high and low $u_{rms}$ and inhomogeneity at the injection scale are again present in all four sets to a greater or lesser extent.  As before these oscillations will be ignored.  A closer inspection of the $u_{rms}$ fields for the four data sets also reveals a large-scale inhomogeneity which increases in magnitude as $\alpha$ decreases. Note that for the most weakly damped case (a), the fluctuations near the corners are weak compared to those near the box center. This large-scale inhomogeneity is the main source of the discrepancies for the three different forms of $S^{(3)}_a$.

\begin{figure}
\includegraphics*[width=6in,height=5.95in]{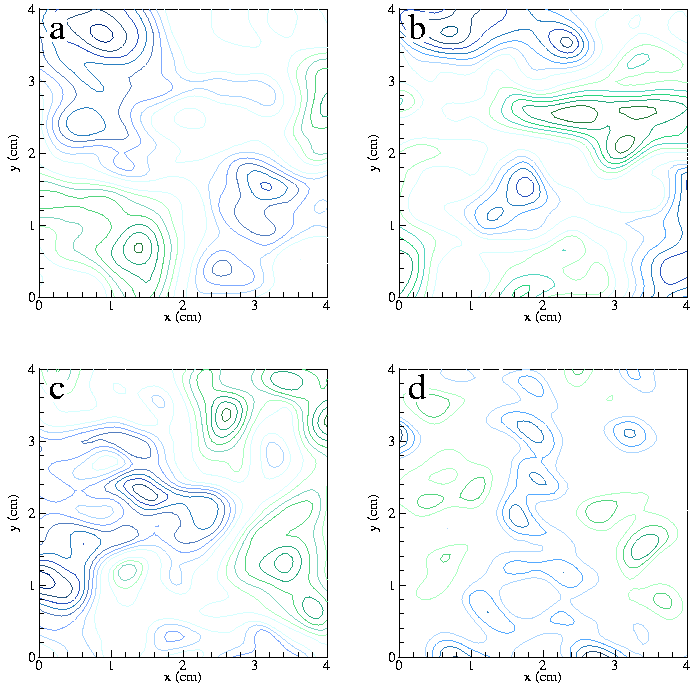}
\caption{Typical streamlines for the four cases labeled (a)-(d) in Table \ref{tab: constants}.}
\label{fig: outervelocity}
\end{figure}

The reason that the velocity fluctuations begin to form this large scale inhomogeneity for weak damping is connected to the growth of the outer scale of the turbulence.  As discussed before the outer scale indicates the largest sized vortices present in the turbulence.  This can be seen in Fig. \ref{fig: outervelocity} which shows typical streamlines for the four cases (a)-(d).  Clearly the heavy damping, case (d), has many small vortices but few large ones compared to the case of weak damping, (a).  These large vortices prefer to exist in regions removed from solid boundaries, otherwise a large shear builds up between the vortex and the boundary and quickly dissipates the vortex.  Since the largest vortices are of diameter $r_{out}$, this  preference causes an absence of fluctuations for distances smaller than $r_{out}$ from the wall.  Should this boundary region invade the PTV measurement area homogeneity is sacrificed.

For all of the data sets in Table \ref{tab: constants} the distance between the PTV measurement area and the boundary was approximately $1.5$ cm, this sets the largest outer scale possible before sacrificing homogeneity.  Table \ref{tab: constants} reveals that the spectrally measured outer scale, $2\pi/k_{out}$, of the case (a) and (b) are well above this value explaining their strong inhomogeneity.  Case (d) is below this value therefore the measurement volume is homogenous.  Case (c) has an outer scale just exceeding this limit, which explains it's marginal homogeneity.

\subsection{The Inertial Range and The Integral Scale}

In the last section only one of the data sets which were analyzed strictly satisfied the homogeneity condition, case (d).  Case (c) was marginal in it's homogeneity, so it can be considered as well.  Considering only these two cases it is apparent that there is no inertial range in the e-m cell.  Recall that from Eq. \ref{eq: thirdmom} an inertial range is indicated by a linear range in $S^{(3)}_a(r)$.  Neither case (c) or (d) shows such a range in $S^{(3)}_a$.  Since both (c) and (d) have an inverse cascade range, this is the first indication that the inertial range is behaving distinctly from the inverse cascade range.  However, since neither case (c) or (d) has an extensive inverse cascade range, it would be helpful to determine if either case (a) or (b), both of which exhibit large inverse cascade ranges, has an inertial range in spite of their lack of homogeneity.

\begin{figure}
\includegraphics*[width=6in,height=5.94in]{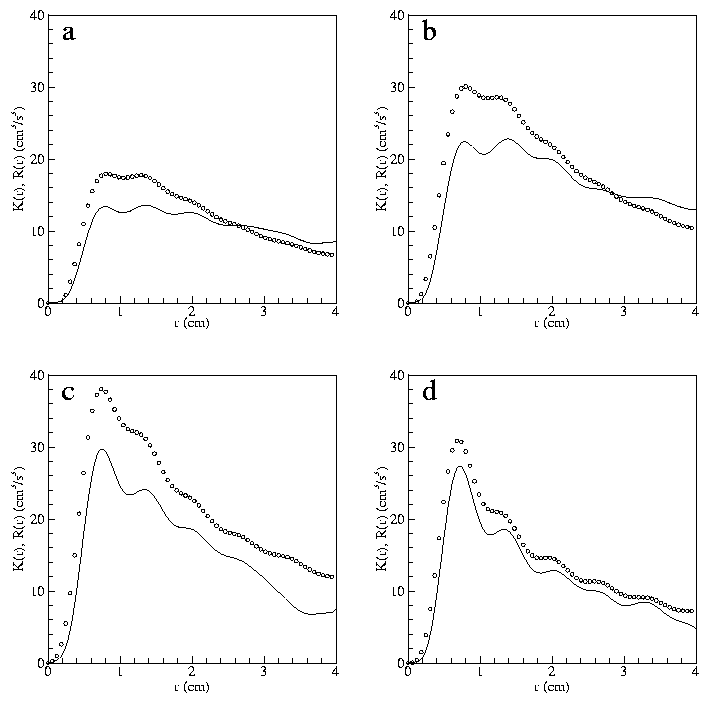}
\caption{Comparison of $K(r)$ ($\circ$) with the independently measured right hand side of Eq. \ref{eq: thirdmom} ,$R(r)$ ($-$),for the data sets labeled (a)-(d) in Table \ref{tab: constants}.}
\label{fig: thirdmom}
\end{figure}

To that end consider Eq. \ref{eq: thirdmom}. The left-hand side of this equation can be represented by any of the three forms $S^{(3)}(r)$, $J(r)$, or $K(r)$ from the last section if the turbulence is homogenous. For the inhomogeneous case it would be useful to find if any of these three forms satisfy Eq. \ref{eq: thirdmom}, effectively relaxing the condition of homogeneity on the left of the equation. To test this idea the right-hand side of Eq. \ref{eq: thirdmom}, denoted $R(r)$, was independently measured and is displayed in Fig. \ref{fig: thirdmom} along with $K(r)$ the third representation of the anisotropic third moment for the four cases discussed earlier. The strongly and moderately damped cases produce moderate agreement between $K(r)$ and $R(r)$ , while for the weakly damped cases there is some disagreement which increases for large scales.  The agreement for all cases, however, is better for $K(r)$ than if $S^{(3)}(r)$ had been used to represent the left hand side.  One can weakly conclude then that the homogeneity assumption used to obtain Eq. \ref{eq: thirdmom} can be relaxed on the left for cases of moderate inhomogeneity if the $K(r)$ representation of the anisotropic third moment is used. For this reason we call $K(r)$ the quasi-homogenous part of the third moment.

Now the determination of the inertial range for weakly inhomogeneous turbulence boils down to whether or not a linear range exists in the quasi-homogenous part of the anisotropic third moment for $r>r_{inj}$. Clearly, even for the cases of weak damping, this does not happen. Indeed over the majority of the range displaying inverse cascade, the quasi-homogenous part of the anisotropic third moment decays as $r^{-1}$, which is indicative of a linear dissipation dominated regime. Therefore, none of the inverse cascade ranges produced in the e-m cell are inertial, they are all dissipation dominated.

\begin{figure}
\hskip 1.25in
\includegraphics*[width=3.5in,height=3.55in]{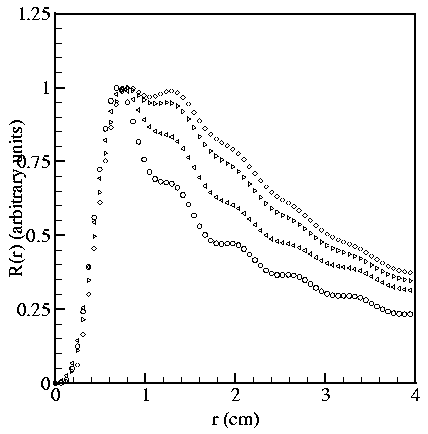}
\caption{The right hand side of Eq. \ref{eq: thirdmom} ,$R(r)$,for the data sets labeled (a)-(d) in Table \ref{tab: constants}: (a) $\diamond$, (b) $\triangleright$, (c) $ \triangleleft $, (d) $\circ$.  $R(r)$ has been normalized so that the peak value just after $r_{inj}$ is unity.}
\label{fig: thirdmom_norm}
\end{figure}

In the previous section, the outer scale of turbulence was determined to behave as $r_{out}\propto (\frac{\epsilon_{inj}}{\alpha^3})^{1/2}$. This length scale, obviously, does not determine the upward extent of the range over which energy transfer is inertial, otherwise, a few of the data sets should display a linear range. It is reasonable to ask what condition is required in order for any linearly damped 2D system to have an inertial range. Equation \ref{eq: thirdmom} indicates that if a linear range is to exist it arises from the dominance of the second integral on the right-hand side over the first integral at large scales.  The first integral, the force-velocity correlation, decays (as $1/r$) at length scales larger than the injection scale. Thus, if a typical length scale of the second integral greatly exceeds the injection scale, the third moment should have some linear range. The second integral contains the longitudinal two-point velocity correlation as one of its functional arguments.  This suggests that a measure of the typical length scale of this integral is the so-called integral scale,
\[
r_{int} \equiv \frac{1}{b^{(2)}_{r,r}(0)}\int_0^\infty dr b^{(2)}_{r,r}(r).
\]
The size of the integral scale for the data sets is displayed in Table \ref{tab: constants}. Note that all but the most weakly damped data sets have integral scales smaller than the injection scale ($r_{inj}\approx 0.6$ cm). For the case of very weak damping the integral scale has just exceeded the injection scale.  This is reflected in the quasi-homogenous part of the third moment for the weakly damped cases (a) and (b), which show an extended plateau right after the injection peak. To better visualize this plateau growth, the right hand side of Eq. \ref{eq: thirdmom} for the four cases is displayed in the Fig. \ref{fig: thirdmom_norm} with the plots normalized by the value of the peak after the injection scale.  Such a feature is absent in the more heavily damped cases. Thus, the integral scale seems to govern the upward extent of the inertial range, in contrast to the outer scale that governs the upward extent of the inverse cascade in 2D.

\begin{figure}
\hskip 1.25in
\includegraphics*[width=3.5in,height=3.41in]{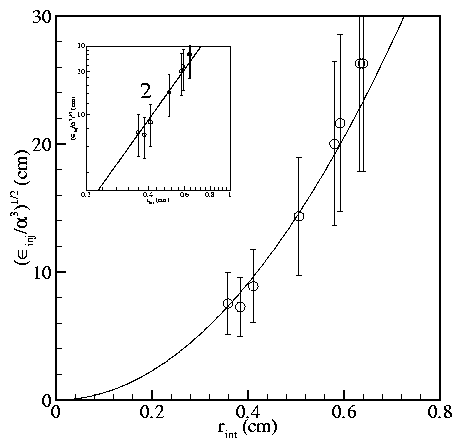}
\caption{The dimensionally predicted outer scale, $(\epsilon_{inj}/\alpha^3)^{1/2}$ vs. integral scale $r_{int}$ (inset is the same plot on log-log scales).  The line corresponds to the power law fit of $r_{int}^{2}$.}
\label{fig: lscales}
\end{figure}

One might believe that the integral scale and the outer scale should be linearly proportional to each other. This may not be the case. Figure \ref{fig: lscales} is a comparison of the the outer scale calculated using the dimensional argument with the integral scale. Though the error in the plot does allow for the possibility of a linear fit, it is more likely a power law growth as shown by the $r_{int}^{2}$ curve drawn in the figure.  Within the error indicated on the plots, the integral scale seems to behave as the geometric average of the outer scale and injection scale (see the inset),
\begin{eqnarray}
r_{int} = \sqrt{r_{inj}r_{out}}  = \left(\frac{r^2_{inj}\epsilon_{inj}}{\alpha^3}\right)^{1/4}.
\label{eq: lengthscales}
\end{eqnarray}
It seems that this relationship should be predictable by inserting finite ranges in the energy spectrum and inverse transforming to get the two-point correlation. At this time such a calculation has not been performed.

\chapter{High Order Moments \label{himom}}

Recall from the introduction that the inverse energy cascade range was expected to have two important properties: locality and scale invariance.  The experiments in the e-m cell do not have enough inverse cascade range for any conclusion to be directly drawn about either of these properties.  However, some indirect conclusions pertaining to scale invariance might be drawn if one assumes that certain measured properties of the moments of velocity difference can be extended to the case of a large inverse cascade range.  Displaying these properties and demonstrating how such conclusions may be drawn is the purpose of this chapter.  First, to facilitate the extension of measured results, a brief discussion which describes the expected behavior for moments of velocity difference assuming scale invariance is presented. This discussion is similar to one in \cite{Frisch}.

\section{Scale Invariance and Moments}

In the introduction the velocity difference over a length scale $r$, $\delta \vec{u}(\vec{r}) \equiv (\vec{u}(\vec{x}+\vec{r}) - \vec{u}(\vec{x}))$, was said to be scale invariant in the inverse cascade range (Note that though $\delta \vec{u}$ depends on the absolute position, $\vec{x}$, any statistical quantities formed from $\delta \vec{u}$ does not if homogeneity is assumed).  Here, scale invariance means that there exists a unique scaling exponent $h$ such that $P(\delta \vec{u}(\lambda \vec{r})) = P(\lambda^h \delta \vec{u}(\vec{r}))$, where $P$ denotes a probability distribution function (PDF).  For the moment, consider homogenous isotropic 2D turbulence.  Instead of using both components of $\delta \vec{u}(\vec{r})$ in $P$, isotropy allows for the simplification to only a single component.  For various reasons this component is usually the longitudinal component, $\delta u_{||}(\vec{r}) = \delta \vec{u}(\vec{r}) \cdot {\hat{r}}$.  Also note that isotropy allows dependence on $\vec{r}$ to be reduced to only dependence on $r = |\vec{r}|$.  From this the scaling relationship becomes $P(\delta u_{||}(\lambda r)) = P(\lambda^h u_{||}(r))$. 

The $n^{\rm th}$ order moment of longitudinal velocity difference is defined as
\begin{eqnarray}
S^{(n)}_{rrr...}(r) \equiv \langle (\delta u_{||}(r))^n \rangle = \int_{-\infty }^{+\infty }d\delta u_{||}(r) ~(\delta u_{||}(r))^nP(\delta u_{||}(r)). 
\label{eq: pdfmoms}
\end{eqnarray}
Note that these functions were already defined in chapter \ref{eflux}.  Since longitudinal fluctuations will only be considered here the notation can be simplified by defining $S_n(r) = S^{(n)}_{rrr...}(r)$. The scale invariance of the PDF's directly results in the scale invariance of the moments of velocity difference, with the scaling exponent $nh$ for the $n^{\rm th}$ order moment.  That is
\begin{eqnarray}
S_n(\lambda r) &=& \int_{-\infty}^{+\infty}d\delta u_{||}(\lambda r) ~(\delta u_{||}(\lambda r))^n P(\delta u_{||}(\lambda r)) \nonumber \\ &=& \int_{-\infty}^{+\infty}d\lambda^h\delta u_{||}(r) ~(\lambda^h\delta u_{||}(r))^n P(\lambda^h\delta u_{||}(r)) \nonumber \\&=& \lambda^{nh} \int_{-\infty}^{+\infty}\lambda^{h}d\delta u_{||}(r) ~(\delta u_{||}(r))^n \lambda^{-h}P(\delta u_{||}(r)) \nonumber\\ &=&\lambda^{nh}S_n(r).
\end{eqnarray}
Thus scale invariance implies that the moments of longitudinal velocity difference behave as $S_n(r) \propto r^{nh}$.

Recall from the discussion in chapter \ref{eflux} that there exists an exact result for the third moment of velocity difference in an inertial range, $S_3(r) = \frac{3}{2} \epsilon r$.  If an inertial range exists, and if it is scale invariant, then this exact result fixes the scaling exponent $h = 1/3$ and the final expectation for the scaling behavior of the $n^{\rm th}$ order moment is that $S_{n}(r) \propto r^{n/3}$.  This result was first reached by Kolmogorov in his 1941 theory of homogenous isotropic turbulence and will be called the K41 theory.  The result can be fleshed out a little more if one assumes that the only variables of importance in the inertial range are the constant energy flow rate, $\epsilon$, and the length scale, $r$.  Under these assumptions dimensional analysis predicts $S_{n}(r) \propto C_n(\epsilon r)^{n/3}$, where the dimensionless $C_n$ are assumed to be universal constants.  Incompressibility sets $C_1 = 0$ and the previously mentioned exact result sets $C_3 = 3/2$.

This analysis yields a simple way in which the scale invariance of the turbulent fields can be verified.  Simply make sure that the PDF's of longitudinal velocity difference for various $r$ in the inertial range, when properly normalized, collapse to a single curve.  Equivalently, make sure that the $n^{\rm th}$ order moments of longitudinal velocity difference scale as $r^{n/3}$ for all $n$.  It should be pointed out that this analysis holds for homogenous isotropic 3D turbulence, except for a change in coefficients $C_n$.  Although one might expect the velocity differences in 3D turbulence to be scale invariant as expressed above, it is an experimental fact that the moments steadily deviate from the expected scaling behavior of the K41 theory as the order of the moment is increased.  The possibility of such deviations happening in 2D turbulence is still an open experimental question, though what follows in this chapter might be considered clues as to what the answer may be.

\section{Disclaimer}

The analysis in this chapter must begin by pointing out a couple of reasons {\em not} to extend certain conclusion presented in it to cases of 2D turbulence beyond the e-m cell.  The first of these reasons stems from the type of statistical quantities which will be used in the determination of scale invariance.  The analysis in the previous section was phrased solely in terms of longitudinal velocity differences.  In order to go from full velocity differences to longitudinal differences without loss of information it is necessary that the turbulence be fully isotropic.  The fact that the forcing is unidirectional, as discussed in chapter \ref{experimental} does not allow isotropy to be assumed in the e-m cell.  This means that the analysis given above must be done for fully anisotropic turbulence to be absolutely correct.

Unfortunately, this analysis becomes prohibitive for anisotropic turbulence as the order of the moments increase since the number of  moments containing transverse velocity differences that need to be measured grows.  Earlier experiments show that the deviations from scaling prediction in 3D become apparent only for moments of large order.  Assuming that 2D might be similar means that a large number of quantities would have to be accurately measured and compared, making fully anisotropic analysis exceedingly difficult.  The fact that anisotropy does not allow the dependence on the difference vector $\vec{r}$ to be simplified to dependence on $r=\vec{r}$ exacerbates this difficulty.  From an experimental point of view this is devastating since dependence only on $r$ allows statistics to be averaged over circles, increasing the number of points used in evaluating the PDF's by $2 \pi r$ for each scale $r$.  Without this buffering of the statistics one cannot hope to obtain enough data to measure high order moments of the PDF.  For these reasons a fully anisotropic analysis must be abandoned.  All results in this chapter use only longitudinal differences, and it is hoped that the anisotropy will not seriously affect the final results.

The second reason was hinted at in chapter \ref{eflux}.  The results that will be discussed will be coming from an inverse cascade range that is not inertial.  This means that the form of the external damping may be affecting the results.  Since the external damping in the e-m cell is known to have a linear form, the results can strictly be said to only apply to other linearly damped 2D turbulent fluids.  This fact, along with the e-m cells anisotropy, raises serious questions about the universality of certain results which are obtained.  In spite of this, the author feels that many of the results presented are robust due to similarities with data obtained from other experiments \cite{Paret:PFL98} and simulation \cite{Boffetta:PRE00}.

\section{The PDF of Longitudinal Velocity Difference}

\begin{figure}
\hskip 1.25in
\includegraphics*[width=3.5in,height=3.12in]{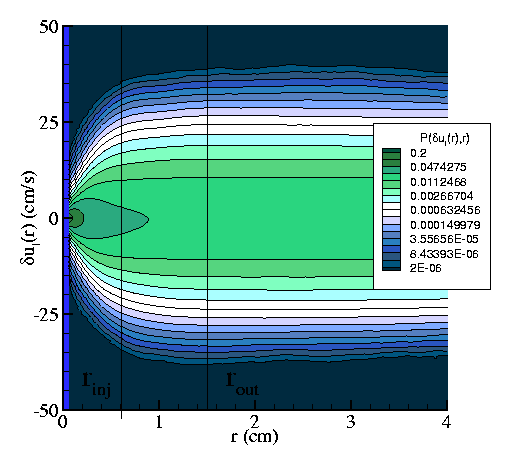}
\caption{$P(\delta u_l(r),r)$ calculated from data set (c) in Table \ref{tab: constants}.  Divisions in the coloration increase on an exponential scale.  The injection and outer scale are marked by lines, in between which is the inverse energy cascade range.}
\label{fig: longitudinalpdf1}
\end{figure}

\begin{figure}
\hskip 1.25in
\includegraphics*[width=3.5in,height=3.39in]{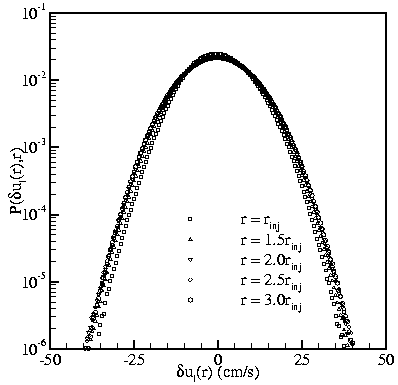}
\caption{Cross sections at various $r$ for $P(\delta u_l(r),r)$ shown in Fig \ref{fig: longitudinalpdf1}.}
\label{fig: longitudinalpdf2}
\end{figure}

Of the data sets presented in Table \ref{tab: constants} the only two to display homogeneity or marginal homogeneity by the analysis of chapter \ref{eflux} are cases (c) and (d).  These will be the main data sets from which conclusions in this chapter will be drawn.  In particular, focus will be given to (c) since it has the largest inverse cascade range of the two as shown in Fig. \ref{fig: spectra}.  The PDF of longitudinal velocity difference, $P=P(\delta u_{||}(r))$ was calculated in a straightforward manner from data set (c) and is presented in the color plot of Fig. \ref{fig: longitudinalpdf1}.  Several cross-sections of the plot for various $r$ are shown in  Fig. \ref{fig: longitudinalpdf2}.

It is clear from the cross-sections that $P$ is approximately Gaussian at all $r$.  Of course it cannot be perfectly Gaussian since there must be some third moment as was measured in chapter \ref{eflux}.  The magnitude of the odd moments, such as the third, must be small compared to the even order moments for $P$ to have such a strongly Gaussian character.  The tails of $P$ do not seem to strongly deviate from Gaussian decay into either exponential or algebraic decay at any $r$ in the inverse cascade range.  In 3D, deviations of the high order moments from the K41 theory is associated with slower than Gaussian decay in the PDF tails.  This behavior is commonly termed ``intermittency" since it indicates intermittent bursty behavior in the velocity field.  That no such deviation in the PDF tails is seen in these experiments is an indication that the scale invariant result may hold.  The approximately Gaussian PDF's  measured here are in agreement with both recent experiments \cite{Paret:PFL98} and simulations \cite{Boffetta:PRE00}.

\begin{figure}
\includegraphics*[width=6in,height=8.46in]{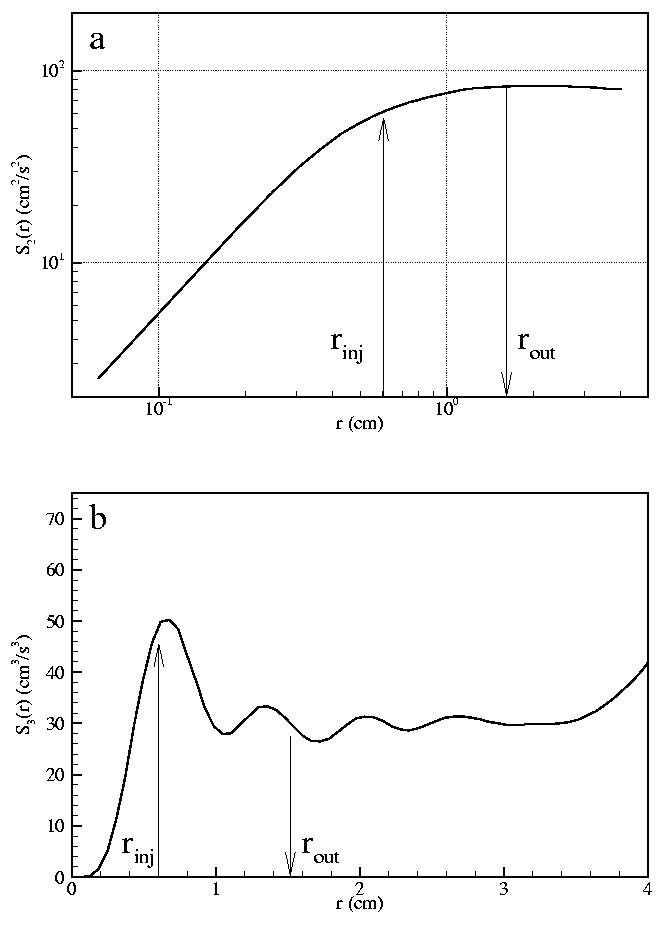}
\caption{(a) $S_2(r)$ (log-log) and (b) $S_3(r)$ (lin-lin) calculated from data set (c) in Table \ref{tab: constants}.}
\label{fig: lowmoms}
\end{figure}

The moments of the longitudinal velocity difference are calculated from $P$ using Eq. \ref{eq: pdfmoms}.  Since the two lowest order moments in 3D do not measurably deviate from the K41 theory, and it is expected that this will be the case in 2D, the low order moments are analyzed in the e-m cell first.  Displayed in Fig: \ref{fig: lowmoms} are $S_2(r)$ and $S_3(r)$ for the data set (c).  Clearly there is no range in the second moment which scales as $r^{2/3}$  in between the injection and outer scale as K41 predicts.  There is no linear behavior of $S_3(r)$ in this range either.  This later result is hardly surprising considering that a linear range is indicative of an inertial range which has already been shown not to exist in the e-m cell (see chapter \ref{eflux}).

\begin{figure}
\includegraphics*[width=6in,height=8.10in]{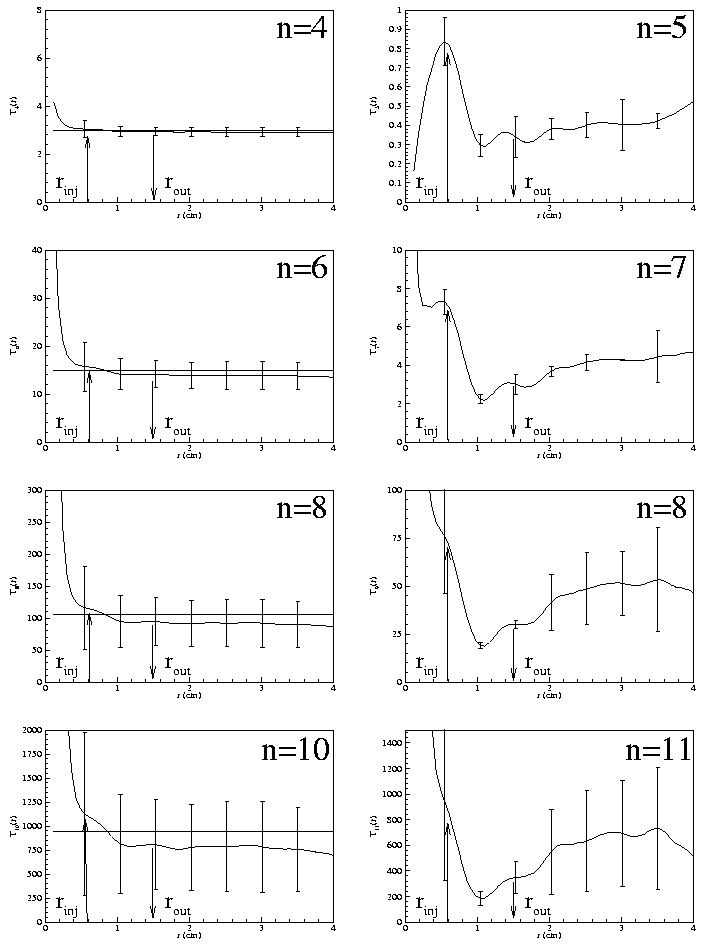}
\caption{The normalized high order moments, $T_n(r)$, evaluated from data set (c) of Table \ref{tab: constants} for $4 \leq n \leq 11$.}
\label{fig: highmoms}
\end{figure}

These graphs clearly violate the scaling prediction in the K41 theory, therefore the e-m cell's inverse cascade range is not scale invariant.  It might become scale invariant if an inertial range was allowed to build in the cell.  To better understand what might happen, the higher order moments (i.e. $n > 3$) of the longitudinal velocity difference are evaluated.  These moments are more easily displayed if they are made dimensionless by dividing out the appropriate power of the second moments.  Define $T_n(r) = S_n(r)/(S_2(r))^{n/2}$, so that $T_3$ is the skewness, $T_4$ is the flatness etc.  These normalized moments are shown in Fig. \ref{fig: highmoms} for $4 \leq n \leq 11$.  The error bars are calculated by truncating the PDF wherever noise dominates the PDF measurement.

First consider the moments of even order.  With the exception of a blow up at small $r$, the even $T_n(r)$ are essentially constant for $r > r_{inj}$ with the exception of logarithmically small fluctuations between $r_{inj}$ and $r_{out}$.  The constant values that the even $T_n$ assume at large $r$ are only slightly less than those of a pure Gaussian distribution, namely
\begin{eqnarray}
a_n = \frac{n !}{2^{(n/2)}(n/2)!}, 
\label{eq: gaussval}
\end{eqnarray}
for $n \geq 2$.  These are shown on the even plots of Fig. \ref{fig: highmoms} as dotted lines.  This confirms the earlier visual observation that $P$ was approximately Gaussian.  The large blow up at small scales $r<r_{inj}$ is due to poor statistics.  The fluctuations of the even $T_n$ at scales in between $r_{inj}$ and $r_{out}$ seems to force the value slightly higher than the Gaussian value.  It should be pointed out that a similar rise is seen in \cite{Paret:PFL98} though without the fluctuating character.  Fluctuations are also seen in \cite{Boffetta:PRE00} near the outer scale, though these settle back down to the Gaussian values once the inertial range is reached.  Since the fluctuations are small the Gaussian values will be assumed to hold for all $r>r_{inj}$.  From the measured constant values of the even $T_n$ the higher order even moments in the e-m cell can be approximated as $S_n(r) \approx a_n(S_2(r))^{n/2}$ with the $a_n$ assuming the values of a Gaussian distribution. 

\begin{figure}
\hskip 0.5in
\includegraphics*[width=5in,height=3.38in]{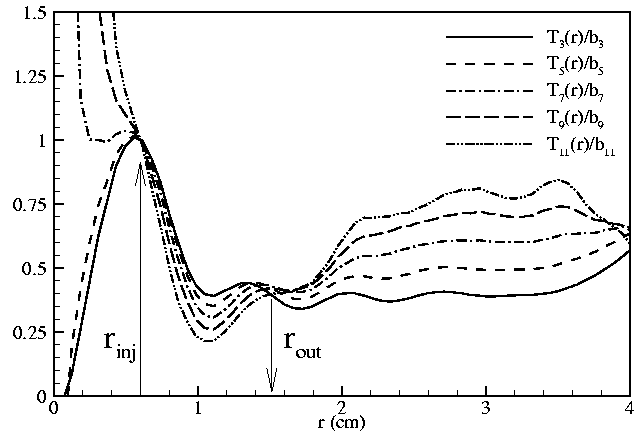}
\caption{$T_n(r)/b_n$ for odd $n \geq 3$ evaluated using the data set (c) in Table \ref{tab: constants}.}
\label{fig: normoddmom}
\end{figure}

\begin{figure}
\hskip 0.5in
\includegraphics*[width=5in,height=3.46in]{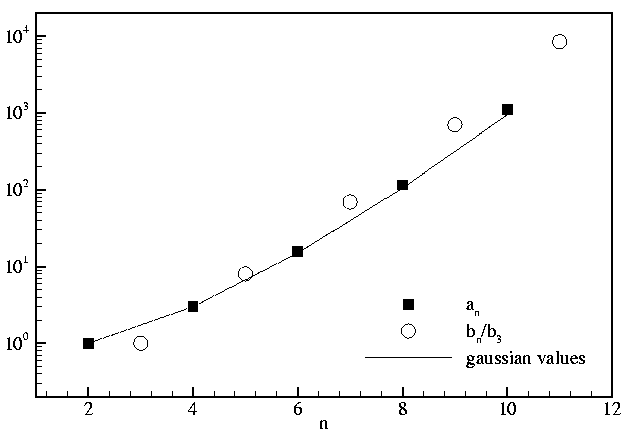}
\caption{The multiplicative constants $a_n$ and $b_n/b_3$ for the data set (c) in Table \ref{tab: constants}.  The dotted line corresponds to the exact values of a purely Gaussian distribution given by Eq. \ref{eq: gaussval}.}
\label{fig: consts}
\end{figure}

Now consider the odd moments.  Unlike the even moments, the odd $T_n$ display a complicated behavior for $r > r_{inj}$.  However, up to a multiplicative constant, which will be denoted $b_n$, the odd $T_n$ have similar behaviors.  To see this the multiplicative constant has been removed and all odd $T_n$ for $n \geq 3$ have been plotted in Fig. \ref{fig: normoddmom}.  The $b_n$ where chosen to be the value of $T_n$ at $r_{inj}$.  This was completely arbitrary and more complicated procedures could be performed to get better fits.  Though the plots are not identical, they clearly have the same trends, and agreement is within measurement error.  Using the $T_3$ as a base function this experimental data indicates that $T_n(r) \approx \frac{b_n}{b_3}T_3(r)$.  Or, in terms of the unnormalized moments $S_n(r) \approx \frac{b_n}{b_3} S_3(S_2)^{(n-3)/2}$.  The coefficients $\frac{b_n}{b_3}$ behave in a different manner from their even counterparts, the $a_n$.  The two sets of coefficients are plotted in Fig. \ref{fig: consts} along with a dotted line representing Gaussian values.  Where the $a_n$ are following the Gaussian prediction quite nicely, the $b_n/b_3$ behave as an almost perfect exponential.

The experimental results lead to the conclusion that the higher order moments can be expressed approximately in terms of the two lowest order moments as
\begin{eqnarray}
S_n(r) &=& a_n(S_2(r))^{\frac{n}{2}}~:~n~\mbox{even},\\
S_m(r) &=& d_m S_3(r)(S_2(r))^{\frac{m-3}{2}}~:~m~\mbox{odd}.
\end{eqnarray}
where $d_m \equiv b_m/b_3$.  We are now in a position to draw some conclusions about scale invariance in the inertial range.  In the inertial range it is clear that $S_3(r) \propto r$ as the exact result from chapter \ref{eflux} shows.  Further, the range $S_2(r)$ is not expected to deviate in an inertial range from it's $r^{2/3}$ behavior.  Assuming these two scaling laws hold in an inertial range and assuming also that the above results can be extended into the inertial range yields
\begin{eqnarray}
S_n(r) &\propto& r^{\frac{n}{3}}~:~\forall n.
\end{eqnarray}
This is precisely the statement of the K41 theory for $r$ in an inertial range.  Assuming, then, that the extensions and assumptions made are correct, the inertial range of 2D turbulence should behave in a scale invariant manner.

\begin{figure}
\hskip 0.5in
\includegraphics*[width=5in,height=7.18in]{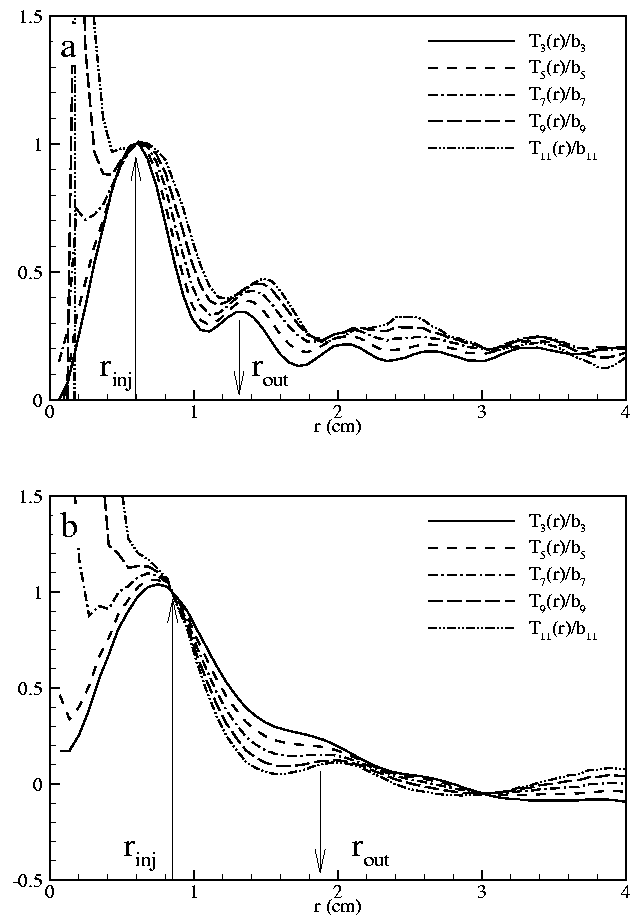}
\caption{$T_n(r)/b_n$ for odd $n$ for (a) case (d) in Table \ref{tab: constants} and (b) a run of the e-m cell with a square array.}
\label{fig: universality1}
\end{figure}

\begin{figure}
\hskip 0.5in
\includegraphics*[width=5in,height=3.45in]{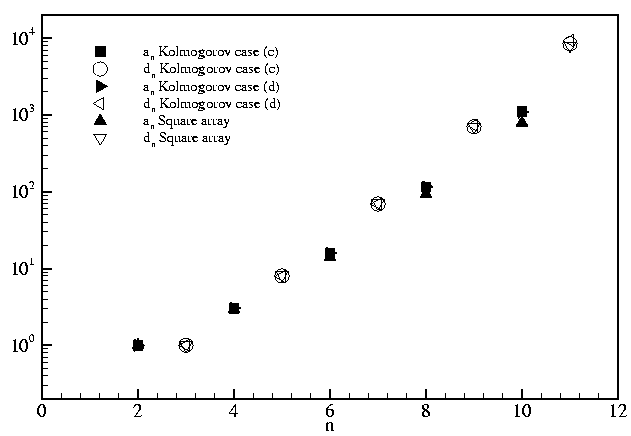}
\caption{The measured $a_n$ and $d_n (=b_n/b_3)$ for three data sets using different $\alpha$ and different types of forcing.}
\label{fig: universality2}
\end{figure}

One might ask about the universality of the coefficients $a_n$ and $d_n$.  Recall that one of the predictions of $K41$ is that these dimensionless numbers should be independent of any external parameters, such as $\alpha$ or $\vec{F}$ which govern the turbulence.  To test this the procedure presented above is performed on data set (d) of Table \ref{tab: constants} as well as a set of data taken using a square magnet array instead of a Kolmogorov array.  Similar to the earlier data sets, the even moments display strong Gaussian characteristics.  The underlying structure of the odd normalized moments has changed, but they are still marginally collapsible by extracting an arbitrary constant.  The odd $T_n/b_n$ are shown in Fig. \ref{fig: universality1}(a) and (b) for the strongly damped Kolmogorov flow (case (d) in Table \ref{tab: constants}) and square array respectively.  In Fig. \ref{fig: universality2} the coefficients for all of the data sets are displayed simultaneously.  Note that the measured values do not significantly differ from one data set to the next, the $a_n$ remain close to the Gaussian values given by Eq. \ref{eq: gaussval}, and the $d_n$ are exponential.  This indicates that the coefficients in the K41 theory are indeed universal as predicted.

\chapter{Conclusion}

The inverse energy cascade of 2D turbulence as it occurs in the e-m cell has been measured and quantified in the preceding chapters.  Clearly an inverse cascade exists and it's energy distribution and extent behave as predicted by dimensional analysis.  The energy flow, when homogenous enough to be accurately compared with theory agrees almost perfectly with exact predictions made using the 2D incompressible Navier-Stokes equation.  However, at no time is this energy flow inertial.

That the inertial range and inverse cascade range are not coincident for a linearly damped 2D fluid is perhaps the most important result in the thesis, from an experimental and numerical stand point.  This is because previous 2D turbulence experiments and simulations did not attempt to differentiate between the two ranges, and merely assumed that a $-5/3$ range in $E(k)$ must be accompanied by the necessary linear range in $S_3(r)$ (or $S^{(3)}_a$ depending on if the turbulence was isotropic or not).  This is now known not to be the case and casts doubt on some of the earlier experimental results.  Finally, the inverse cascade range measured in the e-m cell is not scale invariant, though there is some evidence that it might become so if an inertial range ever developed.

This is perhaps as far as experimental results on the inverse cascade can be taken in the current incarnation of the e-m cell.  The lack of an inertial range being the most significant limiting feature.  To get an idea of what would be needed to eliminate this feature we can use the knowledge that has been gained that the inertial range grows as the geometric average of the injection and outer scale (Eq. \ref{eq: lengthscales}).  If a decade of inertial range is desired for accurate measurements of scaling behavior then by  Eq. \ref{eq: lengthscales} the system size would have to be two decades larger than the injection scale, that is $r_{out} = 100 \times r_{inj}$ .  To get an idea what this would mean in the e-m cell, recall that the Kolmogorov magnets produced an injection scale of $0.6$ cm.  Thus the e-m cell frame would need to be $60$ cm across to support a large enough system to have a decade of inertial range while maintaining homogeneity.  Moreover, the outer scale would have to be forced higher by either increasing $\epsilon_{inj}$, which is risky because velocity fluctuations would grow and sacrifice incompressibility, or by reducing $\alpha$.  Since $\alpha$ has almost reached it's asymptotic limit for the weakly damped cases considered in this thesis, a partial vacuum must be employed to achieve lower $\alpha$.  The author need not emphasize the difficulty in creating a half meter sized soap bubble, balancing it with respect to gravity, and finally putting the whole system in a partial vacuum.

Some sort of happy medium might be reached by marginally increasing the system size, say by a factor of two, and reducing the magnet size a bit.  Some inertial range would exist, though hardly enough to draw conclusions about scaling behavior.  However, if use of the e-m cell in inverse cascade investigations is to continue such compromises must be made to obtain an inertial range.

\appendix
\chapter{Particle Tracking Velocimetry: Program Listing\label{program}}

Included in this appendix is the program code, piv\_chk.cpp, for the particle tracking routine presented in chapter \ref{experimental}.  It has been written in the C program language for no other reason than personal preference.  The program has been successfully compiled with both the GNU C compiler and the Microsoft compiler.  No performance enhancement was found through use of different compilers.  Use of the code after compilations is done from the command line by\\
\begin{center}
piv\_chk~~foo1.tif~~foo2.tif~~outfoo.vec
\end{center}
where ``foo1.tif" and ``foo2.tif" are the first and second tif images to be compared respectively and ``outfoo.vec" is the output file holding the particle positions and displacements.  In ``outfoo.vec" the origin of the coordinate system is assumed to be at the upper left corner of the tif image, with the $\hat{x}$ direction denoting the vertical coordinate in the image and the $\hat{y}$ direction denoting the horizontal coordinate.  Values of $x$ increase as one goes down the tif image and values of $y$ increase as one moves right.  If a matched particle set is found at position $(x_1,y_1)$ in the first image and $(x_2,y_2)$ in the second image it is saved in ``outfoo.vec" as:\\
\begin{center}
$\xi_y$, $\xi_x$, $u_y$, $u_x$\\
\end{center}
where $\xi_x = 1/2(x_1 + x_2)$, $\xi_y = 1/2(y_1+y_2)$, $u_x = x_2-x_1$ and $u_y=y_2-y_1$.  All units in the output file are in pixels.

The various parameters in the routine that determine the characteristics of the particles, the search radius, the correlation box size and other parameters are set by \#define statements at the beginning of the routine.  These are commented in the source code for clarity.\\

{\small \tt
\noindent
{\tt piv\_chk.cpp:}
\begin{flushleft}
\verb%#include <stdio.h>%
\\* \verb%#include <stdlib.h>%
\\* \verb%#include <conio.h>%
\\* \verb%#include <math.h>%
\end{flushleft}
\verb%/*****Control Parameters*****/%
\\* \verb%/*The three #define statements below determine the size and properties%
\\* \verb%of the pictures.  For example, if your picture is (640x480) with 30%
\\* \verb%header bits prefacing the contiguous data block in the .tif file, then%
\\* \verb%header is 30, row is 480, and col is 640. (use 8 bit .tif)*/%
\begin{flushleft}
\verb%#define col 768%
\\* \verb%#define row 480%
\\* \verb%#define header 234%
\end{flushleft}
\verb%/*The two variables below set your background subtraction properties.%
\\*\verb%abox is the size of the box to take a local average over and must%
\\*\verb%be odd.  thresh is the multiple of the background to be subtracted%
\\*\verb%from the picture for the purposes of particle identification. Once%
\\*\verb%the background has been subtracted, any contiguous group of points %
\\*\verb%in the picture with pixel values greater than zero is a candidate %
\\*\verb%to be a particle.*/%
\begin{flushleft}
\verb%#define abox 21%
\\* \verb%#define thresh 1.1%
\end{flushleft}
\verb%/*minsize (maxsze) is the minimum (maximum) rms of a candidate group%
\\* \verb%of points intensity distribution for that group to be labeled a%
\\* \verb%particle.  One can think of this as the particle size in pixels.*/%
\begin{flushleft}
\verb%#define minsize .25%
\\* \verb%#define maxsze 600%
\end{flushleft}
\verb%/*The three variables below are the meat of the routine.%
\\* \verb%srad is the distance in pixels to search for the particle from%
\\* \verb%frame one to frame two.  Set this as low as is reasonable.%
\\* \verb%If you know the particles travel no farther than 5 pixels from frame%
\\* \verb%one to frame two set srad to 5.  cbox is the correlation box size.%
\\* \verb%This box must be big enough to contain at least 4 neighbors of%
\\* \verb%a particle and must be odd.  cthresh is the lower bound of the%
\\* \verb%correlation.  Any correlation above cthresh will be considered for%
\\* \verb%a match, and saved.*/%
\begin{flushleft}
\verb%#define srad 15%
\\* \verb%#define cbox 17%
\\* \verb%#define cthresh 0.5%
\end{flushleft}
\verb%/*Most bad interogations happen near the boundaries.%
\\* \verb%Set brdr to approx cbox/2 to eliminate these.*/%
\begin{flushleft}
\verb%#define brdr 12%
\end{flushleft}
\verb%/*The below just allocates memory.  maxnum is the maximum # of%
\\* \verb%particles in a picture and maxsize is the maximum number of%
\\* \verb%contiguous pixels in the candidate block.  These are set%
\\* \verb%unreasonably high since computer memory is cheap.*/%
\begin{flushleft}
\verb%#define maxnum 30000%
\\* \verb%#define maxsize 10000%
\end{flushleft}
\verb%/*Do not go below here unless you have strongly correlated particle%
\\* \verb%motion.  If you do not have strongly correlated motion cvstat should%
\\* \verb%be zero, which turns the below variables off.*/%
\begin{flushleft}
\verb%#define cvstat 0%
\\* \verb%#define nbrstat 2%
\\* \verb%#define nrad 15%
\\* \verb%#define cvthresh 1.25%
\\* \verb%#define lbnd .6%
\end{flushleft}
\noindent
\verb%/*****Declare Global Variables*****/%
\begin{flushleft}
\verb%struct connection{%
\\* \verb%  float val;%
\\* \verb%  int plc;%
\\* \verb%};%
\end{flushleft}

\begin{flushleft}
\verb%struct object{%
\\* \verb%  float size;%
\\* \verb%  float x,y;%
\\* \verb%  float theta;%
\\* \verb%  int numnbr,numcan;%
\\* \verb%  struct connection *nbr;%
\\* \verb%  struct connection *can;%
\\* \verb%  int status;%
\\* \verb%};%
\end{flushleft}
\verb%/*****Declare Functions*****/%
\begin{flushleft}
\verb%void sort(struct connection *,int);%
\\* \verb%int mutmax(struct object *,struct object *,int,int);%
\\* \verb%void background(unsigned char **,float **,int);%
\\* \verb%struct object findparts(unsigned char **,unsigned char **,int,int);%
\\* \verb%void connect(struct object *,struct object *,int,int,int);%
\\* \verb%float correl(unsigned char **, unsigned char **, int);%
\\* \verb%void getbox(unsigned char **, unsigned char **, int,int,int);%
\\* \verb%void clean(struct object *,struct object *,int,int);%
\\* \verb%int check_vec(struct object *,struct object *,int,int);%
\end{flushleft}
\verb%/*****Main Function*****/%
\begin{flushleft}
\verb%void main(int argv, char *argc[])%
\\* \verb%{%
\\* \verb%  FILE *fin1, *fin2, *fout;%
\end{flushleft}

\begin{flushleft}
\verb%  int i,j,k,l,m,n;%
\\* \verb%  int x0,y0,r,s,t;%
\\* \verb%  int n1,n2,flag;%
\\* \verb%  float val;%
\\* \verb%  float x,y,u,v;%
\\* \verb%  float theta0,theta1,thetadiff;%
\\* \verb%  float vorticity;%
\\* \verb%  float dist1,dist2,xdiff,ydiff;%
\\* \verb%  double rem,pos;%
\\* \verb%  unsigned char **pic1,**pic2,**bfr1,**bfr2,*hdr;%
\\* \verb%  float **mean;%
\\* \verb%  struct object *list1, *list2;%
\end{flushleft}

\begin{flushleft}
\verb%  if(argv<4){%
\\* \verb%    printf("\nsyntax: piv <tif file #1> <tif file #2> <vec file>");%
\\* \verb%    exit(0);%
\\* \verb%  }%
\end{flushleft}
\verb%/*****Open tif and output files*****/%
\begin{flushleft}
\verb%  if((fin1=fopen(argc[1],"rb"))==NULL){%
\\* \verb%    printf("Could not open %\verb-%-\verb%s",argc[1]);%
\\* \verb%    exit(0);%
\\* \verb%  }%
\end{flushleft}

\begin{flushleft}
\verb%  if((fin2=fopen(argc[2],"rb"))==NULL){%
\\* \verb%    printf("Could not open %\verb-%-\verb%s",argc[2]);%
\\* \verb%    exit(0);%
\\* \verb%  }%
\end{flushleft}

\begin{flushleft}
\verb%  if((fout=fopen(argc[3],"w"))==NULL){%
\\* \verb%    printf("Could not open %\verb-%-\verb%s",argc[3]);%
\\* \verb%    exit(0);%
\\* \verb%  }%
\end{flushleft}

\begin{flushleft}
\verb%/*****Allocate some memory*****/%
\\* \verb%  hdr = new(unsigned char[header]);%
\\* \verb%  pic1 = new(unsigned char *[row]);%
\\* \verb%  pic2 = new(unsigned char *[row]);%
\\* \verb%  bfr1 = new(unsigned char *[row]);%
\\* \verb%  bfr2 = new(unsigned char *[row]);%
\\* \verb%  mean = new(float *[row]);%
\end{flushleft}

\begin{flushleft}
\verb%  for(i=0;i<row;i++){%
\\* \verb%    pic1[i] = new(unsigned char[col]);%
\\* \verb%    pic2[i] = new(unsigned char[col]);%
\\* \verb%    bfr1[i] = new(unsigned char[col]);%
\\* \verb%    bfr2[i] = new(unsigned char[col]);%
\\* \verb%    mean[i] = new(float[col]);%
\\* \verb%  }%
\end{flushleft}

\begin{flushleft}
\verb%/*****Read picture files*****/%
\\* \verb%  printf("Reading files\n");%
\end{flushleft}

\begin{flushleft}
\verb%  if(fread(hdr,sizeof(unsigned char),header,fin1)!=header){%
\\* \verb%    printf("Error stripping header from %\verb-%-\verb%s",argc[1]);%
\\* \verb%    exit(0);%
\\* \verb%  }%
\end{flushleft}

\begin{flushleft}
\verb%  if(fread(hdr,sizeof(unsigned char),header,fin2)!=header){%
\\* \verb%    printf("Error stripping header from %\verb-%-\verb%s",argc[2]);%
\\* \verb%    exit(0);%
\\* \verb%  }%
\end{flushleft}

\begin{flushleft}
\verb%  for(i=0;i<row;i++){%
\\* \verb%    if(fread(pic1[i],sizeof(unsigned char),col,fin1)!=col){%
\\* \verb%      printf("Error reading %\verb-%-\verb%s",argc[1]);%
\\* \verb%      exit(0);%
\\* \verb%    }%
\\* \verb%    if(fread(pic2[i],sizeof(unsigned char),col,fin2)!=col){%
\\* \verb%      printf("Error reading %\verb-%-\verb%s",argc[2]);%
\\* \verb%      exit(0);%
\\* \verb%    }%
\\* \verb%  }%
\end{flushleft}

\begin{flushleft}
\verb%/*****Begin background subtraction*****/%
\\* \verb%  printf("\nSubtracting background");%
\\* \verb%  background(pic1,mean,abox);%
\\* \verb%  for(i=0;i<row;i++){%
\\* \verb%    for(j=0;j<col;j++){%
\\* \verb%      if(thresh*mean[i][j] >= pic1[i][j]) bfr1[i][j] = 0;%
\\* \verb%      else{%
\\* \verb%        val = mean[i][j];%
\\* \verb%        rem = modf(val,&pos);%
\\* \verb%        if(rem > .5) pos++;%
\\* \verb%        val =(float)pos;%
\\* \verb%        bfr1[i][j] = pic1[i][j] - (unsigned char)val;%
\\* \verb%      }%
\\* \verb%    }%
\\* \verb%  }%
\end{flushleft}

\begin{flushleft}
\verb%  background(pic2,mean,abox);%
\\* \verb%  for(i=0;i<row;i++){%
\\* \verb%    for(j=0;j<col;j++){%
\\* \verb%      if(thresh*mean[i][j] >= pic2[i][j]) bfr2[i][j] = 0;%
\\* \verb%      else{%
\\* \verb%        val = mean[i][j];%
\\* \verb%        rem = modf(val,&pos);%
\\* \verb%        if(rem > .5) pos++;%
\\* \verb%        val =(float)pos;%
\\* \verb%        bfr2[i][j] = pic2[i][j] - (unsigned char)val;%
\\* \verb%      }%
\\* \verb%    }%
\\* \verb%  }%
\end{flushleft}
\verb%/*****A little housekeeping*****/%
\begin{flushleft}
\verb%  delete[] mean;%
\\* \verb%  list1 = new(struct object[maxnum]);%
\\* \verb%  list2 = new(struct object[maxnum]);%
\\* \verb%  n1 = 0;%
\\* \verb%  n2 = 0;%
\end{flushleft}
\begin{flushleft}
\verb%/*****Begin Finding particles*****/%
\\* \verb%  printf("\nFinding particles");%
\\* \verb%  for(i=0;i<row;i++){%
\\* \verb%    for(j=0;j<col;j++){%
\\* \verb%      if(bfr1[i][j] > 0){%
\\* \verb%        list1[n1]=findparts(bfr1,pic1,i,j);%
\\* \verb%        if(list1[n1].size < maxsze && list1[n1].size > minsize/2 ) n1++;%
\\* \verb%      }%
\\* \verb%      if(bfr2[i][j] > 0){%
\\* \verb%        list2[n2]=findparts(bfr2,pic2,i,j);%
\\* \verb%        if(list2[n2].size < maxsze && list2[n2].size > minsize/2 ) n2++;%
\\* \verb%      }%
\\* \verb%    }%
\\* \verb%  }%
\end{flushleft}

\verb%/*****A little housekeeping*****/%
\begin{flushleft}
\verb%  delete[] bfr1;%
\\* \verb%  delete[] bfr2;%
\end{flushleft}

\verb%/*****Start finding neighbors and candidates*****/%
\begin{flushleft}
\verb%  printf("\n%\verb-%-\verb%d\t%\verb-%-\verb%d\nEstablishing connections",n1,n2);%
\\* \verb%  connect(list1,list2,n1,n2,srad);%
\\* \verb%  bfr1 = new(unsigned char *[cbox]);%
\\* \verb%  bfr2 = new(unsigned char *[cbox]);%
\\* \verb%  for(i=0;i<cbox;i++){%
\\* \verb%    bfr1[i] = new(unsigned char[cbox]);%
\\* \verb%    bfr2[i] = new(unsigned char[cbox]);%
\\* \verb%    for(j=0;j<cbox;j++){%
\\* \verb%      bfr1[i][j] = 0;%
\\* \verb%      bfr2[i][j] = 0;%
\\* \verb%    }%
\\* \verb%  }%
\end{flushleft}

\begin{flushleft}
\verb%/*****Calculate correlations for the candidates*****/%
\\* \verb%  printf("\nBeginning correlations");%
\\* \verb%  for(i=0;i<n1;i++){%
\\* \verb%    if(i%\verb-%-\verb%100 == 0) printf(".");%
\\* \verb%    m = list1[i].numcan;%
\\* \verb%    if(m==0) continue;%
\\* \verb%    if(kbhit()) break;%
\end{flushleft}

\begin{flushleft}
\verb%    val = list1[i].x;%
\\* \verb%    rem = modf(val,&pos);%
\\* \verb%    if(rem > .5) pos++;%
\\* \verb%    x0 =(int)pos;%
\\* \verb%    val = list1[i].y;%
\\* \verb%    rem = modf(val,&pos);%
\\* \verb%    if(rem > .5) pos++;%
\\* \verb%    y0 =(int)pos;%
\\* \verb%    getbox(pic1,bfr1,x0,y0,cbox);%
\end{flushleft}

\begin{flushleft}
\verb%    for(j=0;j<m;j++){%
\\* \verb%      n=list1[i].can[j].plc;%
\\* \verb%      val = list2[n].x;%
\\* \verb%      rem = modf(val,&pos);%
\\* \verb%      if(rem > .5) pos++;%
\\* \verb%      x0 = (int)pos;%
\\* \verb%      val = list2[n].y;%
\\* \verb%      rem = modf(val,&pos);%
\\* \verb%      if(rem > .5) pos++;%
\\* \verb%      y0 = (int)pos;%
\\* \verb%      getbox(pic2,bfr2,x0,y0,cbox);%
\end{flushleft}

\begin{flushleft}
\verb%      val = correl(bfr1,bfr2,cbox);%
\\* \verb%      list1[i].can[j].val = val;%
\\* \verb%      x0 = 0;%
\\* \verb%      while(list2[n].can[x0].plc != i) x0++;%
\\* \verb%      list2[n].can[x0].val = val;%
\\* \verb%    }%
\\* \verb%  }%
\end{flushleft}

\verb%/*****Sort by size of correlations*****/%
\begin{flushleft}
\verb%  printf("\nsorting");%
\\* \verb%  for(i=0;i<n1;i++){%
\\* \verb%    m = list1[i].numcan;%
\\* \verb%    sort(list1[i].can,m);%
\\* \verb%    m = list1[i].numnbr;%
\\* \verb%    sort(list1[i].nbr,m);%
\\* \verb%  }%
\end{flushleft}

\begin{flushleft}
\verb%  for(i=0;i<n2;i++){%
\\* \verb%    m = list2[i].numcan;%
\\* \verb%    sort(list2[i].can,m);%
\\* \verb%    m = list2[i].numnbr;%
\\* \verb%    sort(list2[i].nbr,m);%
\\* \verb%  }%
\end{flushleft}

\begin{flushleft}
\verb%/*****Start matching*****/%
\\* \verb%  printf("\nmatching");%
\\* \verb%  delete[] pic1;%
\\* \verb%  delete[] pic2;%
\\* \verb%  for(val=.9;val>cthresh;val-=0.01){%
\\* \verb%    for(i=0;i<n1;i++){%
\\* \verb%      if(list1[i].numcan==0) continue;%
\\* \verb%      if(list1[i].status==1) continue;%
\\* \verb%      m = list1[i].numcan - 1;%
\\* \verb%      n = list1[i].can[m].plc;%
\end{flushleft}

\begin{flushleft}
\verb%      if(mutmax(list1,list2,i,n)==1 && list1[i].can[m].val>=val){%
\\* \verb%        if(cvstat == 1 && val < lbnd){%
\\* \verb%          if(check_vec(list1,list2,i,n)==1){%
\\* \verb%            list1[i].status = 1;%
\\* \verb%            list2[n].status = 1;%
\\* \verb%            clean(list1,list2,i,n);%
\\* \verb%          }%
\end{flushleft}
\begin{flushleft}
\verb%                else if(check_vec(list1,list2,i,n)==0){%
\\* \verb%            list1[i].numcan--;%
\\* \verb%            list2[n].numcan--;%
\\* \verb%          }%
\\* \verb%        }%
\end{flushleft}

\begin{flushleft}
\verb%        else{%
\\* \verb%          list1[i].status = 1;%
\\* \verb%          list2[n].status = 1;%
\\* \verb%          clean(list1,list2,i,n);%
\\* \verb%        }%
\\* \verb%      }%
\\* \verb%    }%
\\* \verb%  }%
\\* \verb%  n=0;%
\end{flushleft}

\begin{flushleft}
\verb%/*****Print results*****/%
\\* \verb%  for(i=0;i<n1;i++){%
\\* \verb%    if(list1[i].status == 1){%
\\* \verb%      j = list1[i].numcan-1;%
\\* \verb%    m = list1[i].can[j].plc;%
\\* \verb%    if(list1[i].size < minsize || list2[m].size < minsize) continue;%
\\* \verb%      x = (list2[m].x + list1[i].x)/2;%
\\* \verb%      y = (list2[m].y + list1[i].y)/2;%
\\* \verb%      u = list2[m].x - list1[i].x;%
\\* \verb%      v = list2[m].y - list1[i].y;%
\end{flushleft}
\begin{flushleft}
\verb%      if(x > brdr && x <= row-brdr && y > brdr && y<= col-brdr){%
\\* \verb%        fprintf(fout,"%\verb-%-\verb%f\t%\verb-%-\verb%f\t%\verb-%-\verb%f\t%\verb-%-\verb%f\n",y,x,v,u);%
\\* \verb%        n++;%
\\* \verb%      }%
\\* \verb%    }%
\\* \verb%  }%
\\* \verb%  printf("\n%\verb-%-\verb%d",n);%
\end{flushleft}

\verb%/*****Some final housekeeping*****/%
\begin{flushleft}
\verb%  delete[] list1;%
\\* \verb%  delete[] list2;%
\\* \verb%}%
\end{flushleft}

\verb%/*This function attempts to loosen the correlation limit by comparing%
\\* \verb%candidate motions with previously matched neighbors.*/%
\begin{flushleft}
\verb%int check_vec(struct object *list1,struct object *list2,int n1, int n2)%
\\* \verb%{%
\\* \verb%  int num1,num2;%
\\* \verb%  int nbrs,matched;%
\\* \verb%  int i,j,m,n;%
\\* \verb%  float x,y,u,v,val;%
\\* \verb%  float xm,ym,um,vm;%
\\* \verb%  float xp,yp,up,vp;%
\\* \verb%  float jtr;%
\end{flushleft}

\begin{flushleft}
\verb%  matched=0;%
\\* \verb%  nbrs = list1[n1].numnbr;%
\\* \verb%  xp = (list2[n2].x+list1[n1].x)/2;%
\\* \verb%  yp = (list2[n2].y+list1[n1].y)/2;%
\\* \verb%  up = (list2[n2].x-list1[n1].x);%
\\* \verb%  vp = (list2[n2].y-list1[n1].y);%
\end{flushleft}

\begin{flushleft}
\verb%  for(i=0;i<nbrs;i++){%
\\* \verb%    m=list1[n1].nbr[i].plc;%
\\* \verb%    if(list1[m].status == 1){%
\\* \verb%      j=list1[m].numcan-1;%
\\* \verb%      n=list1[m].can[j].plc;%
\\* \verb%      x = (list2[n].x+list1[m].x)/2 - xp;%
\\* \verb%      y = (list2[n].y+list1[m].y)/2 - yp;%
\\* \verb%      val = sqrt(x*x + y*y);%
\\* \verb%    if(val < nrad && x+xp > brdr && x+xp<=row-brdr%
\\* \verb%                      && y+yp > brdr && y+yp <= col-brdr) matched++;%
\\* \verb%    }%
\\* \verb%  }%
\end{flushleft}

\begin{flushleft}
\verb%  if(matched < nbrstat) return(2);%
\end{flushleft}

\begin{flushleft}
\verb%  xm=0;ym=0;um=0;vm=0;%
\\* \verb%  for(i=0;i<nbrs;i++){%
\\* \verb%    m=list1[n1].nbr[i].plc;%
\end{flushleft}

\begin{flushleft}
\verb%    if(list1[m].status == 1){%
\\* \verb%      j=list1[m].numcan-1;%
\\* \verb%      n=list1[m].can[j].plc;%
\\* \verb%      x = (list2[n].x+list1[m].x)/2 - xp;%
\\* \verb%      y = (list2[n].y+list1[m].y)/2 - yp;%
\\* \verb%      val = sqrt(x*x + y*y);%
\end{flushleft}

\begin{flushleft}
\verb%    if(val < nrad && x+xp > brdr && x+xp<=row-brdr%
\\* \verb%                  && y+yp > brdr && y+yp <= col-brdr){%
\\* \verb%        xm += (list2[n].x+list1[m].x)/(2*(float)matched);%
\\* \verb%        ym += (list2[n].y+list1[m].y)/(2*(float)matched);%
\\* \verb%        um += (list2[n].x-list1[m].x)/(float)matched;%
\\* \verb%        vm += (list2[n].y-list1[m].y)/(float)matched;%
\\* \verb%      }%
\end{flushleft}

\begin{flushleft}
\verb%    }%
\\* \verb%  }%
\end{flushleft}

\begin{flushleft}
\verb%  jtr=0;%
\\* \verb%  for(i=0;i<nbrs;i++){%
\\* \verb%    m=list1[n1].nbr[i].plc;%
\end{flushleft}

\begin{flushleft}
\verb%    if(list1[m].status == 1){%
\\* \verb%      j=list1[m].numcan-1;%
\\* \verb%      n=list1[m].can[j].plc;%
\\* \verb%      x = (list2[n].x+list1[m].x)/2 - xp;%
\\* \verb%      y = (list2[n].y+list1[m].y)/2 - yp;%
\\* \verb%      val = sqrt(x*x + y*y);%
\end{flushleft}

\begin{flushleft}
\verb%    if(val < nrad && x+xp > brdr && x+xp<=row-brdr%
\\* \verb%                  && y+yp > brdr && y+yp <= col-brdr){%
\\* \verb%        u = (list2[n].x-list1[m].x) - um;%
\\* \verb%        v = (list2[n].y-list1[m].y) - vm;%
\\* \verb%        jtr += (u*u+v*v)/(float)matched;%
\\* \verb%      }%
\end{flushleft}

\begin{flushleft}
\verb%    }%
\\* \verb%  }%
\end{flushleft}

\begin{flushleft}
\verb%  jtr=sqrt(jtr);%
\\* \verb%  u = up-um;%
\\* \verb%  v = vp-vm;%
\\* \verb%  val = sqrt(u*u + v*v);%
\\* \verb%  if(val < cvthresh*jtr) return(1);%
\\* \verb%  else return(0);%
\\* \verb%}%
\end{flushleft}
\noindent
\verb%/*Cleans the objects i and j from any candidate list since they%
\\* \verb%have presumably been matched.*/%
\begin{flushleft}
\verb%void clean(struct object *list1,struct object *list2,int n1, int n2)%
\\* \verb%{%
\\* \verb%  int num1,num2;%
\\* \verb%  int i,j,k,l,m;%
\\* \verb%  struct connection temp;%
\\* \verb%  num1 = list1[n1].numcan - 1;%
\end{flushleft}

\begin{flushleft}
\verb%  if(num1 > 0){%
\\* \verb%    for(i=0;i<num1;i++){%
\\* \verb%      j = list1[n1].can[i].plc;%
\\* \verb%      m = list2[j].numcan;%
\\* \verb%      temp = list2[j].can[m-1];%
\\* \verb%      k=0;%
\\* \verb%      while(list2[j].can[k].plc != n1) k++;%
\\* \verb%      list2[j].can[k] = temp;%
\\* \verb%      list2[j].numcan--;%
\\* \verb%      m--;%
\\* \verb%      sort(list2[j].can,m);%
\\* \verb%    }%
\\* \verb%  }%
\end{flushleft}

\begin{flushleft}
\verb%  num2 = list2[n2].numcan - 1;%
\end{flushleft}

\begin{flushleft}
\verb%  if(num2 > 0){%
\\* \verb%    for(i=0;i<num2;i++){%
\\* \verb%      j = list2[n2].can[i].plc;%
\\* \verb%      m = list1[j].numcan;%
\\* \verb%      temp = list1[j].can[m-1];%
\\* \verb%      k=0;%
\\* \verb%      while(list1[j].can[k].plc != n2) k++;%
\\* \verb%      list1[j].can[k] = temp;%
\\* \verb%      list1[j].numcan--;%
\\* \verb%      m--;%
\\* \verb%      sort(list1[j].can,m);%
\\* \verb%    }%
\end{flushleft}

\begin{flushleft}
\verb%  }%
\\* \verb%}%
\end{flushleft}
\noindent
\verb%/*Returns a 1 if the maximum correlation of list1[n1] and list2[n2]%
\\* \verb%point to one another.  0 otherwise. This routine assumes we%
\\* \verb%have already sorted the connections using "sort".*/%
\begin{flushleft}
\verb%int mutmax(struct object *list1,struct object *list2,int n1,int n2)%
\\* \verb%{       %
\\* \verb%  int num1,num2;%
\\* \verb%  int i,j;%
\\* \verb%  num1 = list1[n1].numcan - 1;%
\\* \verb%  num2 = list2[n2].numcan - 1;%
\\* \verb%  i = list1[n1].can[num1].plc;%
\\* \verb%  j = list2[n2].can[num2].plc;%
\\* \verb%  if(i == n2 && j == n1) return(1);%
\\* \verb%  else return(0);%
\\* \verb%}%
\end{flushleft}
\noindent
\verb%/*Sorting routine used in object lists.  This is used to sort the%
\\* \verb%nbr connections in order from closest to farthest away and the can%
\\* \verb%connection from lowest correlation to highest.*/%
\begin{flushleft}
\verb%void sort(struct connection *ptr,int num)%
\\* \verb%{%
\\* \verb%  int i,j;%
\\* \verb%  struct connection temp;%
\\* \verb%  for(j=1;j<num;j++){%
\\* \verb%    temp = ptr[j];%
\\* \verb%    i=j-1;%
\\* \verb%    while(i>=0 && ptr[i].val > temp.val){%
\\* \verb%      ptr[i+1]=ptr[i];%
\\* \verb%      i--;%
\\* \verb%    }%
\\* \verb%    ptr[i+1] = temp;%
\\* \verb%  }%
\\* \verb%}%
\end{flushleft}
\noindent
\verb%/*Finds all the parts of an particle given that there is a bright spot%
\\* \verb%at x0,y0.  Return the value of the centroid and the rms.*/%
\begin{flushleft}
\verb%struct object findparts(unsigned char **ptr1,unsigned char **ptr2,%
\\* \verb%                                        int x0,int y0)%
\\* \verb%{%
\\* \verb%  int i,m=1,n=1;%
\\* \verb%  int *x,*y;%
\\* \verb%  int j,k,val;%
\\* \verb%  unsigned char *intensity;%
\\* \verb%  int brght=0;%
\\* \verb%  float tx,ty,std;%
\\* \verb%  struct object out;%
\end{flushleft}

\begin{flushleft}
\verb%  x = new(int[maxsize]);%
\\* \verb%  y = new(int[maxsize]);%
\\* \verb%  x[0]=x0;%
\\* \verb%  y[0]=y0;%
\\* \verb%  ptr1[x0][y0]=0;%
\end{flushleft}

\begin{flushleft}
\verb%  while(n<maxsize){%
\end{flushleft}

\begin{flushleft}
\verb%    for(i=0;i<m;i++){%
\\* \verb%      if(x[i] - 1 >= 0){%
\\* \verb%        if(ptr1[x[i]-1][y[i]] > 0){%
\\* \verb%          x[n] = x[i] - 1;%
\\* \verb%          y[n] = y[i];%
\\* \verb%          ptr1[x[n]][y[n]]=0;%
\\* \verb%          n++;%
\\* \verb%        }%
\\* \verb%      }%
\end{flushleft}

\begin{flushleft}
\verb%      if(x[i] + 1 < row){%
\\* \verb%        if(ptr1[x[i]+1][y[i]] > 0){%
\\* \verb%          x[n] = x[i] + 1;%
\\* \verb%          y[n] = y[i];%
\\* \verb%          ptr1[x[n]][y[n]]=0;%
\\* \verb%          n++;%
\\* \verb%        }%
\\* \verb%      }%
\end{flushleft}

\begin{flushleft}
\verb%      if(y[i] + 1 < col){%
\\* \verb%        if(ptr1[x[i]][y[i]+1] > 0){%
\\* \verb%          x[n] = x[i];%
\\* \verb%          y[n] = y[i]+1;%
\\* \verb%          ptr1[x[n]][y[n]]=0;%
\\* \verb%          n++;%
\\* \verb%        }%
\\* \verb%      }%
\end{flushleft}

\begin{flushleft}
\verb%      if(y[i] - 1 >= 0){%
\\* \verb%        if(ptr1[x[i]][y[i]-1] > 0){%
\\* \verb%          x[n] = x[i];%
\\* \verb%          y[n] = y[i]-1;%
\\* \verb%          ptr1[x[n]][y[n]]=0;%
\\* \verb%          n++;%
\\* \verb%        }%
\\* \verb%      }%
\\* \verb%    }%
\end{flushleft}

\begin{flushleft}
\verb%    if(n==m) break;%
\\* \verb%    else m=n;%
\\* \verb%  }%
\end{flushleft}

\begin{flushleft}
\verb%  intensity = new(unsigned char[n]);%
\end{flushleft}

\begin{flushleft}
\verb%  for(i=0;i<n;i++){%
\\* \verb%    intensity[i] = ptr2[x[i]][y[i]];%
\\* \verb%    brght += (int)intensity[i];%
\\* \verb%  }%
\end{flushleft}

\begin{flushleft}
\verb%  out.x=0;                            %
\\* \verb%  out.y=0;%
\end{flushleft}

\begin{flushleft}
\verb%  for(i=0;i<n;i++){%
\\* \verb%    out.x += (float)intensity[i]*(float)x[i]/(float)brght;%
\\* \verb%    out.y += (float)intensity[i]*(float)y[i]/(float)brght;%
\\* \verb%  }%
\end{flushleft}

\begin{flushleft}
\verb%  std = 0;%
\end{flushleft}

\begin{flushleft}
\verb%  for(i=0;i<n;i++){%
\\* \verb%    tx = (float)x[i]-out.x;%
\\* \verb%    ty = (float)y[i]-out.y;%
\\* \verb%    std += (float)intensity[i]*(tx*tx+ty*ty)/(float)brght;%
\\* \verb%  }%
\end{flushleft}

\begin{flushleft}
\verb%  std = sqrt(std);%
\\* \verb%  out.size = std;%
\\* \verb%  out.numnbr = 0;%
\\* \verb%  out.numcan = 0;%
\\* \verb%  out.status = 0;%
\end{flushleft}

\begin{flushleft}
\verb%  delete[] intensity;%
\\* \verb%  delete[] x;%
\\* \verb%  delete[] y;%
\\* \verb%  return(out);%
\\* \verb%}%
\end{flushleft}
\noindent
\verb%/*Creates float ** mean which contains the value of the mean%
\\* \verb%of a box of sizexsize around each point in pic.*/%
\begin{flushleft}
\verb%void background(unsigned char **pic,float **back,int size)%
\\* \verb%{%
\\* \verb%  int i,j;%
\\* \verb%  int x,y,val=0;%
\\* \verb%  int half = (size-1)/2;%
\\* \verb%  float mean;%
\end{flushleft}

\begin{flushleft}
\verb%  mean=0;%
\end{flushleft}

\begin{flushleft}
\verb%  for(i=0;i<=half;i++){%
\\* \verb%    for(j=0;j<=half;j++){%
\\* \verb%      mean = (val*mean +(float)pic[i][j])/((float)val+1);%
\\* \verb%      val++;%
\\* \verb%    }%
\\* \verb%  }%
\end{flushleft}

\begin{flushleft}
\verb%  back[0][0]=mean;%
\end{flushleft}

\begin{flushleft}
\verb%  for(j=0;j<row;j+=2){%
\\* \verb%    if(j%\verb-%-\verb%64==0) printf(".");%
\\* \verb%    for(i=1;i<col;i++){%
\\* \verb%      y=i-half-1;%
\\* \verb%      for(x=j-half;x<=j+half;x++){%
\\* \verb%        if(y < 0 || x < 0 || x >= row) continue;%
\\* \verb%        mean = (mean*val - (float)pic[x][y])/((float)val-1);%
\\* \verb%        val--;%
\\* \verb%      }%
\end{flushleft}

\begin{flushleft}
\verb%      y=i+half;%
\\* \verb%      for(x=j-half;x<=j+half;x++){%
\\* \verb%        if(y >= col || x < 0 || x >= row) continue;%
\\* \verb%        mean = (mean*val + (float)pic[x][y])/((float)val+1);%
\\* \verb%        val++;%
\\* \verb%      }%
\end{flushleft}

\begin{flushleft}
\verb%      back[j][i]=mean;%
\\* \verb%    }%
\end{flushleft}

\begin{flushleft}
\verb%    x = j-half;%
\\* \verb%    for(y=col-1-half;y<col;y++){%
\\* \verb%      if(x < 0) continue;%
\\* \verb%      mean =(mean*val - (float)pic[x][y])/((float)val-1);%
\\* \verb%      val--;%
\\* \verb%    }%
\end{flushleft}

\begin{flushleft}
\verb%    x = j+1+half;%
\\* \verb%    for(y=col-1-half;y<col;y++){%
\\* \verb%      if(x >= row) continue;%
\\* \verb%      mean =(mean*val + (float)pic[x][y])/((float)val+1);%
\\* \verb%      val++;%
\\* \verb%    }%
\end{flushleft}

\begin{flushleft}
\verb%    back[j+1][col-1] = mean;%
\\* \verb%    for(i=col-2;i>=0;i--){%
\\* \verb%      y=i+half+1;%
\\* \verb%      for(x=j+1-half;x<=j+1+half;x++){%
\\* \verb%        if(y >= col || x < 0 || x >= row) continue;%
\\* \verb%        mean = (mean*val - (float)pic[x][y])/((float)val-1);%
\\* \verb%        val--;%
\\* \verb%      }%
\end{flushleft}

\begin{flushleft}
\verb%      y=i-half;%
\\* \verb%      for(x=j+1-half;x<=j+1+half;x++){%
\\* \verb%        if(y < 0 || x < 0 || x >= row) continue;%
\\* \verb%        mean = (mean*val + (float)pic[x][y])/((float)val+1);%
\\* \verb%        val++;%
\\* \verb%      }%
\end{flushleft}

\begin{flushleft}
\verb%      back[j+1][i]=mean;%
\\* \verb%    }%
\end{flushleft}

\begin{flushleft}
\verb%    x = j+1-half;%
\\* \verb%    for(y=0;y<=half;y++){%
\\* \verb%      if(x < 0) continue;%
\\* \verb%      mean =(mean*val - (float)pic[x][y])/((float)val-1);%
\\* \verb%      val--;%
\\* \verb%    }%
\end{flushleft}

\begin{flushleft}
\verb%    x = j+2+half;%
\\* \verb%    for(y=0;y<=half;y++){%
\\* \verb%      if(x >= row) continue;%
\\* \verb%      mean =(mean*val + (float)pic[x][y])/((float)val+1);%
\\* \verb%      val++;%
\\* \verb%    }%
\end{flushleft}

\begin{flushleft}
\verb%    if((j+2) >= row) continue;%
\\* \verb%    back[j+2][0] = mean;%
\\* \verb%  }%
\\* \verb%}%
\end{flushleft}
\noindent
\verb%/*Finds all the neighbors and candidates for a particle and then%
\\* \verb%stores this info in the appropriate spot in the lists.*/%
\begin{flushleft}
\verb%void connect(struct object *list1,struct object *list2,%
\\* \verb%                                         int n1,int n2,int maxdist)%
\\* \verb%{%
\\* \verb%  int x0,y0;%
\\* \verb%  int i,j;%
\\* \verb%  int x,y;%
\\* \verb%  int m1,m2;%
\\* \verb%  int **pic1,**pic2;%
\\* \verb%  int *temp1,*temp2;%
\\* \verb%  float val,dist,xdiff,ydiff;%
\\* \verb%  double rem,pos;%
\end{flushleft}

\begin{flushleft}
\verb%  pic1 = new(int *[row]);%
\\* \verb%  pic2 = new(int *[row]);%
\end{flushleft}

\begin{flushleft}
\verb%  for(i=0;i<row;i++){%
\\* \verb%    pic1[i] = new(int[col]);%
\\* \verb%    pic2[i] = new(int[col]);%
\\* \verb%    for(j=0;j<col;j++){%
\\* \verb%      pic1[i][j] = -1;%
\\* \verb%      pic2[i][j] = -1;%
\\* \verb%    }%
\\* \verb%  }%
\end{flushleft}

\begin{flushleft}
\verb%  for(i=0;i<n1;i++){%
\\* \verb%    val = list1[i].x;%
\\* \verb%    rem = modf(val,&pos);%
\\* \verb%    if(rem >.5) pos++;%
\\* \verb%    x = (int)pos;%
\\* \verb%    val = list1[i].y;%
\\* \verb%    rem = modf(val,&pos);%
\\* \verb%    if(rem >.5) pos++;%
\\* \verb%    y = (int)pos;%
\\* \verb%    pic1[x][y] = i;%
\\* \verb%  }%
\end{flushleft}

\begin{flushleft}
\verb%  for(i=0;i<n2;i++){%
\\* \verb%    val = list2[i].x;%
\\* \verb%    rem = modf(val,&pos);%
\\* \verb%    if(rem >.5) pos++;%
\\* \verb%    x = (int)pos;%
\\* \verb%    val = list2[i].y;%
\\* \verb%    rem = modf(val,&pos);%
\\* \verb%    if(rem >.5) pos++;%
\\* \verb%    y = (int)pos;%
\\* \verb%    pic2[x][y] = i;%
\\* \verb%  }%
\end{flushleft}

\begin{flushleft}
\verb%  temp1 = new(int[maxdist*maxdist]);%
\\* \verb%  temp2 = new(int[maxdist*maxdist]);%
\end{flushleft}

\begin{flushleft}
\verb%  for(x0=0;x0<row;x0++){%
\\* \verb%    if(x0 %\verb-%-\verb% 64 == 0) printf(".");%
\\* \verb%    for(y0=0;y0<col;y0++){%
\\* \verb%      if(pic1[x0][y0] == -1 && pic2[x0][y0]==-1) continue;%
\\* \verb%      if(pic1[x0][y0] != -1){%
\\* \verb%        m1=0;%
\\* \verb%        m2=0;%
\end{flushleft}

\begin{flushleft}
\verb%        for(i=-maxdist;i<=maxdist;i++){%
\\* \verb%          x = x0 + i;%
\\* \verb%          if(x < 0 || x >= row) continue;%
\end{flushleft}

\begin{flushleft}
\verb%          for(j=-maxdist;j<=maxdist;j++){%
\\* \verb%            y = y0 + j;%
\\* \verb%            if(y < 0 || y >= col) continue;%
\\* \verb%            if((i*i + j*j)> maxdist*maxdist) continue;%
\\* \verb%            if(pic2[x][y] != -1){%
\\* \verb%              temp2[m2] = pic2[x][y];%
\\* \verb%              m2++;%
\\* \verb%            }%
\end{flushleft}

\begin{flushleft}
\verb%            if(pic1[x][y] != -1){%
\\* \verb%              if(x==x0 && y==y0) continue;%
\\* \verb%              temp1[m1] = pic1[x][y];%
\\* \verb%              m1++;%
\\* \verb%            }%
\end{flushleft}

\begin{flushleft}
\verb%          }%
\\* \verb%        }%
\end{flushleft}

\begin{flushleft}
\verb%        list1[pic1[x0][y0]].nbr = new(struct connection[m1]);%
\end{flushleft}

\begin{flushleft}
\verb%        for(i=0;i<m1;i++){%
\\* \verb%          xdiff = list1[temp1[i]].x - list1[pic1[x0][y0]].x;%
\\* \verb%          ydiff = list1[temp1[i]].y - list1[pic1[x0][y0]].y;%
\\* \verb%          dist = sqrt(xdiff*xdiff+ydiff*ydiff);%
\\* \verb%          list1[pic1[x0][y0]].nbr[i].plc = temp1[i];%
\\* \verb%          list1[pic1[x0][y0]].nbr[i].val = dist;%
\\* \verb%        }%
\end{flushleft}

\begin{flushleft}
\verb%        list1[pic1[x0][y0]].numnbr = m1;%
\\* \verb%        list1[pic1[x0][y0]].can = new(struct connection[m2]);%
\end{flushleft}

\begin{flushleft}
\verb%        for(i=0;i<m2;i++){%
\\* \verb%          list1[pic1[x0][y0]].can[i].plc = temp2[i];%
\\* \verb%        }%
\end{flushleft}

\begin{flushleft}
\verb%        list1[pic1[x0][y0]].numcan = m2;%
\\* \verb%      }%
\end{flushleft}

\begin{flushleft}
\verb%      if(pic2[x0][y0] != -1){%
\\* \verb%        m1=0;%
\\* \verb%        m2=0;%
\end{flushleft}

\begin{flushleft}
\verb%        for(i=-maxdist;i<=maxdist;i++){%
\\* \verb%          x = x0 + i;%
\\* \verb%          if(x < 0 || x >= row) continue;%
\end{flushleft}

\begin{flushleft}
\verb%          for(j=-maxdist;j<=maxdist;j++){%
\\* \verb%            y = y0 + j;%
\\* \verb%            if(y < 0 || y >= col) continue;%
\\* \verb%            if((i*i + j*j)> maxdist*maxdist) continue;%
\\* \verb%            if(pic1[x][y] != -1){%
\\* \verb%              temp2[m2] = pic1[x][y];%
\\* \verb%              m2++;%
\\* \verb%            }%
\end{flushleft}

\begin{flushleft}
\verb%            if(pic2[x][y] != -1){%
\\* \verb%              if(x==x0 && y==y0) continue;%
\\* \verb%              temp1[m1] = pic2[x][y];%
\\* \verb%              m1++;%
\\* \verb%            }%
\\* \verb%          }%
\\* \verb%        }%
\end{flushleft}

\begin{flushleft}
\verb%        list2[pic2[x0][y0]].nbr = new(struct connection[m1]);%
\end{flushleft}

\begin{flushleft}
\verb%        for(i=0;i<m1;i++){%
\\* \verb%          xdiff = list2[temp1[i]].x - list2[pic2[x0][y0]].x;%
\\* \verb%          ydiff = list2[temp1[i]].y - list2[pic2[x0][y0]].y;%
\\* \verb%          dist = sqrt(xdiff*xdiff + ydiff*ydiff);%
\\* \verb%          list2[pic2[x0][y0]].nbr[i].plc = temp1[i];%
\\* \verb%          list2[pic2[x0][y0]].nbr[i].val = dist;%
\\* \verb%        }%
\end{flushleft}

\begin{flushleft}
\verb%        list2[pic2[x0][y0]].numnbr = m1;%
\\* \verb%        list2[pic2[x0][y0]].can = new(struct connection[m2]);%
\end{flushleft}

\begin{flushleft}
\verb%        for(i=0;i<m2;i++){%
\\* \verb%          list2[pic2[x0][y0]].can[i].plc = temp2[i];%
\\* \verb%        }%
\\* \verb%        list2[pic2[x0][y0]].numcan = m2;%
\\* \verb%      }%
\\* \verb%    }%
\\* \verb%  }%
\end{flushleft}

\begin{flushleft}
\verb%  delete[] pic1;%
\\* \verb%  delete[] pic2;%
\\* \verb%  delete[] temp1;%
\\* \verb%  delete[] temp2;%
\\* \verb%}%
\end{flushleft}
\noindent
\verb%/*Returns the correlation number between two arrays of size x size.*/%
\begin{flushleft}
\verb%float correl(unsigned char **ptr1,unsigned char **ptr2,int size)%
\\* \verb%{%
\\* \verb%  int i,j;%
\\* \verb%  float mean1,mean2;%
\\* \verb%  float std1,std2,cor;%
\\* \verb%  mean1=0;%
\\* \verb%  mean2=0;%
\end{flushleft}

\begin{flushleft}
\verb%  for(i=0;i<size;i++){%
\\* \verb%    for(j=0;j<size;j++){%
\\* \verb%      mean1 += (float)ptr1[i][j]/(float)(size*size);%
\\* \verb%      mean2 += (float)ptr2[i][j]/(float)(size*size);%
\\* \verb%    }%
\\* \verb%  }%
\end{flushleft}

\begin{flushleft}
\verb%  for(i=0;i<size;i++){%
\\* \verb%    for(j=0;j<size;j++){%
\\* \verb%      std1 += (((float)ptr1[i][j] - mean1)*((float)ptr1[i][j] - mean1))%
\\* \verb%                   /(float)(size*size);%
\\* \verb%      std2 += (((float)ptr2[i][j] - mean2)*((float)ptr2[i][j] - mean2))%
\\* \verb%                   /(float)(size*size);%
\\* \verb%      cor += (((float)ptr1[i][j] - mean1)*((float)ptr2[i][j] - mean2))%
\\* \verb%                   /(float)(size*size);%
\\* \verb%    }%
\\* \verb%  }%
\end{flushleft}

\begin{flushleft}
\verb%  cor /=(sqrt(std1)*sqrt(std2));%
\\* \verb%  return(cor);%
\\* \verb%}%
\end{flushleft}
\noindent
\verb%/*Gets a box from pic1 centered at x0,y0 and stores it in ptr1.%
\\* \verb%ptr1 should be at least size x size and size must be odd!!!.*/%
\begin{flushleft}
\verb%void getbox(unsigned char **pic1, unsigned char **ptr1,%
\\* \verb%                             int x0,int y0,int size)%
\\* \verb%{%
\\* \verb%  int i,j,x,y;%
\\* \verb%  int half = (size-1)/2;%
\end{flushleft}

\begin{flushleft}
\verb%  for(i=0;i<size;i++){%
\\* \verb%    x = x0 + i - half;%
\\* \verb%    if(x<0 || x>=row){%
\\* \verb%      for(j=0;j<size;j++){%
\\* \verb%        ptr1[i][j] = 0;%
\\* \verb%      }%
\\* \verb%      continue;%
\\* \verb%    }%
\end{flushleft}

\begin{flushleft}
\verb%    for(j=0;j<size;j++){%
\\* \verb%      y = y0 + j - half;%
\\* \verb%      if(y<0 || y>=col){%
\\* \verb%        ptr1[i][j] = 0;%
\\* \verb%        continue;%
\\* \verb%      }%
\\* \verb%      ptr1[i][j] = pic1[x][y];%
\\* \verb%    }%
\\* \verb%  }%
\\* \verb%}%
\end{flushleft}
}

\pagebreak
\pagestyle{empty}
\addcontentsline{toc}{chapter}{Bibliography}
\begin{center}
\mbox{}
\vskip 3in
{\Huge \bf Bibliography}
\end{center}

\nocite{*}
\bibliographystyle{unsrt}
\bibliography{thesis}
\end{document}